\documentclass[11 pt]{article}
\usepackage{amssymb,latexsym,amsmath,amsfonts}
\usepackage{subfigure,color}
\usepackage{graphicx, epsf}
\usepackage[active]{srcltx}
\usepackage[active]{srcltx}
\usepackage{bm}
\numberwithin{equation}{section}

\renewcommand{\theta}{\vartheta}
\newcommand{\mat}{\operatorname{\begin{pmatrix}0&0\\0&1\end{pmatrix}}}
\newcommand{\vet}{\operatorname{\begin{pmatrix}0\\M\end{pmatrix}}}

\newcommand{\kk}{\operatorname{\bar{k}_c^2}}
\newcommand{\tr}{\operatorname{tr}}

\newcommand{\ro}{\operatorname{\mathbf{r}}}

\newcommand{\ww}{\operatorname{\mathbf{w}}}
\newcommand{\wuno}{\operatorname{\mathbf{w}_1}}
\newcommand{\wdue}{\operatorname{\mathbf{w}_2}}
\newcommand{\wtre}{\operatorname{\mathbf{w}_3}}
\newcommand{\wqua}{\operatorname{\mathbf{w}_4}}
\newcommand{\wcin}{\operatorname{\mathbf{w}_5}}
\newcommand{\wdz}{\operatorname{\mathbf{w}_{20}}}

\newcommand{\wdd}{\operatorname{\mathbf{w}_{22}}}
\newcommand{\wtu}{\operatorname{\mathbf{w}_{31}}}
\newcommand{\wtd}{\operatorname{\mathbf{w}_{32}}}
\newcommand{\wtt}{\operatorname{\mathbf{w}_{33}}}
\newcommand{\wqz}{\operatorname{\mathbf{w}_{40}}}
\newcommand{\wqu}{\operatorname{\mathbf{w}_{41}}}
\newcommand{\wqd}{\operatorname{\mathbf{w}_{42}}}
\newcommand{\wqt}{\operatorname{\mathbf{w}_{43}}}
\newcommand{\wqq}{\operatorname{\mathbf{w}_{44}}}

\newcommand{\half}{\operatorname{\frac{1}{2}}}

\newcommand{\quart}{\operatorname{\frac{1}{4}}}

\newcommand{\duno}{\operatorname{d^{(1)}}}
\newcommand{\ddue}{\operatorname{d^{(2)}}}

\def\bfpsi{\mbox{\boldmath$\psi$}}
\def\kcb{\bar{k}_c}

\def\SS{Schnakenberg }
\begin{document}

\title{ Effects of cross-diffusion on Turing patterns in a reaction-diffusion
\SS model.}

\author{G. Gambino, S. Lupo, M. Sammartino}

\maketitle

\begin{abstract}
In this paper the Turing pattern formation mechanism of a two component reaction-diffusion system modeling the \SS chemical reaction
coupled to linear cross-diffusion terms is studied.
The linear cross-diffusion terms favors the destabilization of
the constant steady state and the mechanism of pattern formation with respect to the standard linear diffusion case, as
shown in \cite{Madzvamuse}. Since the subcritical Turing bifurcations of reaction-diffusion systems lead to spontaneous onset of
robust, finite-amplitude localized patterns, here a detailed investigation of
the Turing pattern forming region is performed to show how the diffusion coefficients for both species (the activator and the inhibitor)
influence the occurrence of supercritical or subcritical bifurcations. \\
The weakly nonlinear (WNL) multiple scales analysis is employed to derive the equations
for the amplitude of the Turing patterns and to distinguish the supercritical and the
subcritical pattern region, both in 1D and 2D domains. Numerical simulations are employed to confirm the
WNL theoretical predictions through which a classification of the patterns (squares, rhombi, rectangle and hexagons) is obtained. \\
In particular, due to the hysteretic nature of the subcritical bifurcation, we observe the phenomenon of pattern transition from rolls to hexagons, in agreement with the bifurcation diagram.
\end{abstract}

\section{Introduction}\label{Sec1}

\noindent
Self-organized patterning in reaction-diffusion system driven by linear diffusion
has been extensively studied since the seminal paper of Turing.
The interaction between diffusion and reaction has shown to yield rich and unexpected phenomena in different contexts as biological sciences, geology, geography, chemistry, industrial process, networks of electrical circuits and, of course, mathematics \cite{BGLLSS15,BLS13,bozzini_EJAM, CWKC14,SBB10,BEM11,ZLW11,MRW11,GR10,ASAST13,BPS07,BDSP07}.\\
In this paper the following reaction-diffusion system, firstly introduced in \cite{Madzvamuse} is studied:


\begin{equation}
\label{model}
    \begin{array}{ll}
     \displaystyle \frac{\partial u}{\partial t} = \nabla^2 u + d_v\nabla^2 v +\gamma f(u,v), \\
   \ & \ \\
   \displaystyle \frac{\partial v}{\partial t} = d \nabla^2 u + d_u\nabla^2 u +\gamma g(u,v),
\end{array}
\end{equation}
where $\nabla^2$ is the bidimensional Laplacian operator, $d$ is the ratio of the linear diffusion coefficients, $d_u$ and $d_v$ are respectively the ratios of the cross-diffusion and the diffusion coefficients, and $\gamma$ is a positive constant. The nonlinear kinetics:

\begin{equation}
\label{reactions}
\begin{array}{ll}
f(u,v)= a-u+u^2v, \\
g(u,v)= b-u^2v,
\end{array}
\end{equation}

\noindent
describe the \SS chemical reaction.

We also require (\ref{model})-(\ref{reactions}) to be equipped with the following initial conditions:
$$
u(x,y, 0)=u_0(x,y), \hspace{.2in} v(x,y, 0)=v_0(x,y), \hspace{.2in} (x, y) \in \left[0, L_x \right] \times
\left[0, L_y \right],
$$
where $L_x$ and $L_y$ are characteristic lengths.

Cross-diffusion, the phenomenon in which a gradient in the concentration of one species induces a flux of another species, is very significant in generating spatial structures. Recently, a large number of papers have appeared in literature, which investigate pattern formation in reaction-diffusion systems including linear and nonlinear cross-diffusion \cite{LRT14, R-BT13, GS14, GV14, ZKHH13, GLS09, GLSS13, GLS14, BC13, ZY15, TLP10, TLL14, Tian11}. A typical example of nonlinear cross-diffusion system is the well known Shigesada-Kawasaky-Teramoto cross-diffusion system, which describes segregation effects for competing species \cite{SKT79}. More recently, studies on the same model have shown that cross-diffusion is the responsible of Turing instability \cite{GLS12,GLS13,TLS14}.
In \cite{Vanag, Chatt} has been shown that linear cross-diffusion coefficients, even though they are relatively small or negative (in this last case
one species tends to move in the direction of higher concentration of the other species), lead and favor pattern formation.
In the above mentioned papers, the diffusion is coupled with nonlinear kinetic terms. To stress the role of cross-diffusion in pattern formation, in \cite{Wang, Vanag} suitable cross-diffusion coefficients have been coupled to linear reaction terms and showing they are sufficient to assure pattern formation.

Madzvamuse et al. in \cite{Madzvamuse} have shown that the introduction of linear cross-diffusion in the \SS model enhances the process of pattern formation and generalizes the classical diffusion-driven instability (i.e. without cross-diffusion). In fact, in the absence of cross-diffusion terms, to obtain pattern occurrence the inhibitor must diffuse faster than the activator,  $d\gg1$. When the cross-diffusion terms are involved, it is no longer necessary to choose $d$ much greater than one. To show the role of cross-diffusion in pattern formation, the authors in \cite{Madzvamuse2} choose the parameter values of the system in such a way they do not belong to the classical linear diffusion-driven instability region, but they lie on the cross-diffusion driven parameter space.
In \cite{Xua}, sufficient conditions to guarantee the occurrence of Hopf bifurcation have been derived for the one-dimensional \SS reaction-diffusion model. In \cite{Liu}, the occurrence of
oscillating patterns due to the Hopf bifurcation is shown.\\
In this paper we discuss the role of the cross-diffusion coefficients into
influencing the occurrence of supercritical or subcritical Turing bifurcations.
In fact, pattern formation is more robust via subcritical bifurcation than via supercritical Turing instability, see \cite{Medina}.
When the bifurcation is supercritical, the pattern is spatially extended, it is born from zero amplitude and, it
is subject to further instabilities in large domains due to the presence of different unstable modes
which interact. In contrast, when the Turing
bifurcation is subcritical, the arising spatial structure jumps to
a finite amplitude pattern  (due to the large branch amplitude into the bifurcation diagram), localized in the spatial
domain, and robust to small fluctuations in the bifurcation parameter
values, therefore it is more difficult to
destroy the localized pattern. It is therefore important to investigate what is the mechanism which helps the subcritical
Turing instability. The weakly nonlinear (WNL) multiple scales analysis is employed to derive the equations
for the amplitude of the Turing patterns and to distinguish the supercritical and the
subcritical pattern region.
We have observed that cross-diffusion in the inhibitor component only (i.e. $d_u$) helps the growth of the subcritical region, while the cross-diffusion in the activator component only (i.e. $d_v$) reduces its size. \\

We extend our analysis to the two-dimensional case, where the study, although much more complex, gives rise to a rich variety of super- and subcritical patterns. In particular, we recover that the case when the eigenvalue corresponding to the most unstable mode has double multiplicity and the resonance condition occurs is always subcritical. In this case we have observed the phenomenon of hexagons-rolls transition: due to the hysteretic nature of subcritical bifurcation, roll patterns lose their stability and "jump" to hexagonal structures.
Our paper is organized as follows: in Sec.~\ref{Sec2} we perform the linear stability analysis around the uniform steady state to obtain the cross-diffusion driven instability conditions and find the corresponding Turing diffusively-driven instability parameter space.
We illustrate the role of cross-diffusion and, once performed a WNL analysis, we point out how cross-diffusion coefficients influence the occurrence of supercritical or subcritical bifurcations.
In addition, we show how the Stuart-Landau equation captures the behaviour of the pattern amplitude close to the marginal instability of the equilibrium $P_0$.\\
In Sec.~\ref{Sec3} we focus on the pattern formation in a two-dimensional domain. In this case the weakly nonlinear analysis is able to predict the formation of a large number of patterns since the bifurcation can occur via a simple eigenvalue or a multiple one. On the first case, as in the one-dimensional domain, we derive the Stuart-Landau equation and the patterns emerging are squares or rolls. When the bifurcation occurs via a multiple eigenvalue - in particular with double multiplicity - more complex patterns arise, due to the interaction of different modes, such as rhombic, mixed-mode or hexagonal patterns, characterized by different amplitudes governed by a system of coupled Stuart-Landau equations.
Finally, the interesting phenomenon of hysteresis is observed in correspondence od subcritical bifurcations, when different stable states for one single valued of the control parameter coexist. \\


\section{The role of cross-diffusion in the \SS model}\label{Sec2}
\setcounter{equation}{0}

First of all we perform the linear stability analysis. The unique positive steady state is given by
\begin{equation}
\label{equilibrium}
P_0=(u_0,v_0)=\left(a+b,\frac{b}{(a+b)^2}\right).
\end{equation}
Linearizing system (\ref{model})-(\ref{reactions}) in the neighborhood of $P_0$, one gets:
\begin{equation}
\dot{\textbf{w}}=J(P_0)\textbf{w}+D^d\nabla^2 \textbf{w}\;,
\qquad \qquad\textrm{}\qquad
\textbf{w}\equiv\left(\begin{array}{c}{u-u_0}\\
{v-v_0}\end{array}\right) \; ,
\label{sistema_rd}
\end{equation}
where
$$D^d=\begin{pmatrix} 1 & d_v\\ d_u & d \end{pmatrix} \qquad \mbox{and} \qquad
J(P_0)=
\begin{pmatrix}
\frac{b-a}{a+b} & (a+b)^2\\
-\frac{2b}{a+b} & -(a+b)^2\\
\end{pmatrix}.$$\\
In Appendix \ref{linear} we derived the necessary conditions for cross-diffusion driven instability:
\begin{subequations}
\begin{eqnarray}\label{cond_inst1}
J_{11}+J_{22}<0,&\\
\label{cond_inst2}
J_{11}J_{22}-J_{12}J_{21}>0,&\\
\label{cond_inst3}
d-d_ud_v>0,&\\
\label{cond_inst4}
dJ_{11}+J_{22}-d_uJ_{12}-d_vJ_{21}>0,&\\
\label{cond_inst5}
(dJ_{11}+J_{22}-d_uJ_{12}-d_vJ_{21})^2-4(d-d_ud_v)(J_{11}J_{22}-J_{12}J_{21})>0,&
\end{eqnarray}
\label{conditions_instability}
\end{subequations}
while the critical value of the bifurcation parameter $d$ is:
\begin{equation}\label{dc}
d_c=\frac{\det J-J_{12}J_{21}+J_{11}(d_vJ_{21}+d_uJ_{12})+2\sqrt{\det J(d_vJ_{11}-J_{12})(J_{21}-d_uJ_{11})}}{J_{11}^2}.
\end{equation}
Here we study the Turing bifurcation in the presence of cross-diffusion, where the diffusive flux of a given species is also affected by the gradients of other species, and the diffusion  matrix is no longer a diagonal matrix, as in the standard case.
In absence of cross-diffusion, condition \eqref{cond_inst4} becomes $dJ_{11}+J_{22}>0$. This condition together with \eqref{cond_inst1} leads to $d>-\frac{J_{22}}{J_{11}}>1$, that means:
\begin{itemize}
\item[(i)] the coefficients $J_{11}$ and $J_{22}$ do not have the same sign ($J_{11}>0$ and $J_{22}<0$),
\item[(ii)] the diffusion coefficients of the two species are not equal and so in the absence of cross-diffusion, the inhibitor $v$ must diffuse faster than the activator $u$.
\end{itemize}
The presence of cross diffusive effects renders the necessary conditions less restrictive: neither (i) nor (ii) need to be fulfilled.\\
It is important to observe that equation \eqref{dc} corresponds to a Turing threshold condition if $d_c$ is a real number, i.e., the discriminant $\det J(d_vJ_{11}-J_{12})(J_{21}-d_uJ_{11})$ is non-negative. Because of $\det J>0$, according to \eqref{cond_inst2}, this imply that $(d_vJ_{11}-J_{12})$ and $(J_{21}-d_uJ_{11})$ have the same sign.\\ In the \SS model both of the factors are negative and so we can define the following set for the cross-diffusion coefficients:
\begin{equation}
S=\left\{(d_u,d_v)|d_v\leq \frac{(a+b)^3}{b-a}, d_u\geq \frac{2b}{a-b}\right\}.
\label{set_S}
\end{equation}
To emphasize the effects of cross-diffusion on the Turing instability in the \SS model, in the following we consider the case in which one of $d_u$ and $d_v$ are equal to zero.\\
\vskip 0.2cm
\begin{figure}[t]
\centering
\subfigure[]{\includegraphics[height=4.5cm, width=7cm]{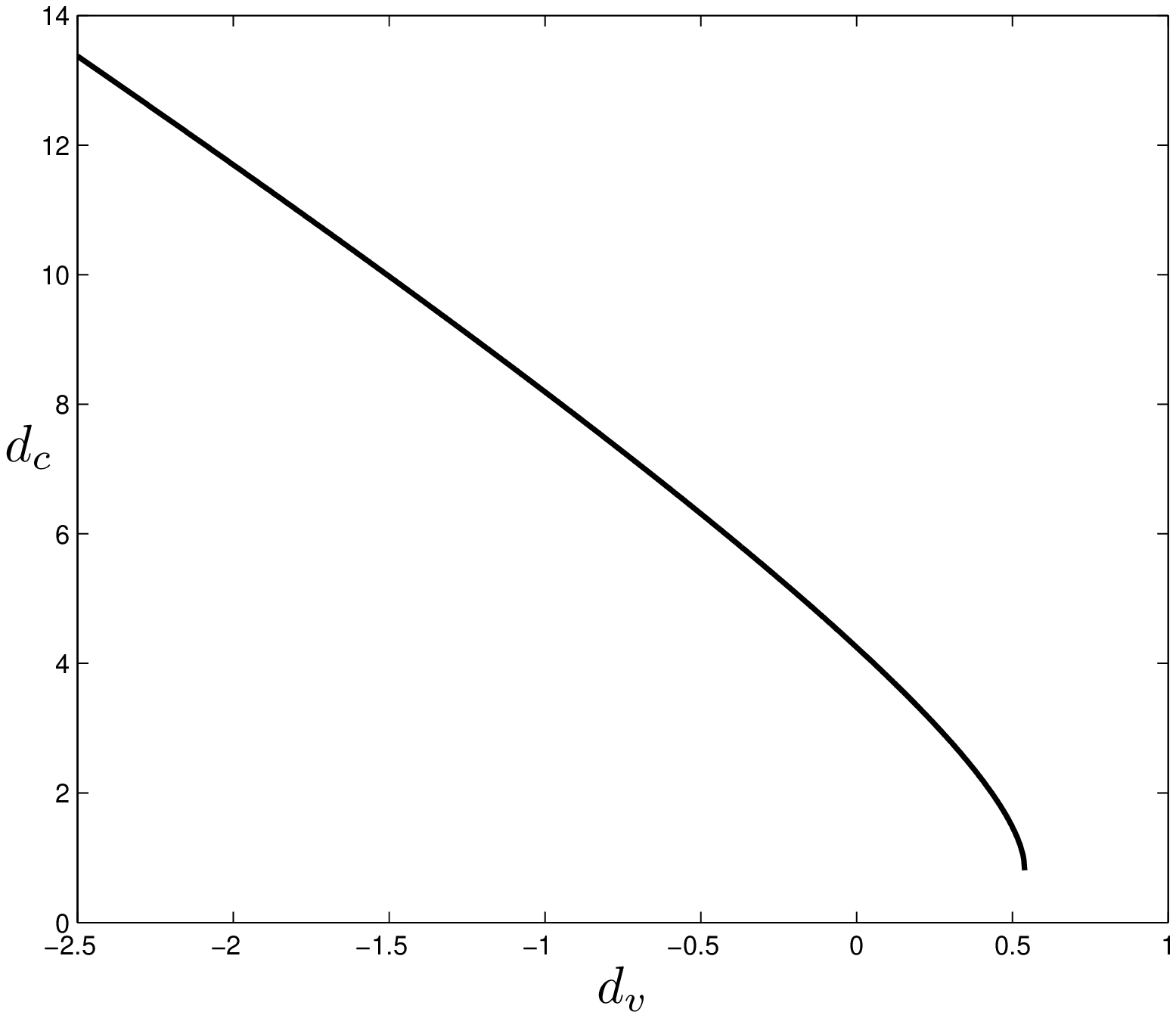}}
\subfigure[]{\includegraphics[height=4.5cm, width=7cm]{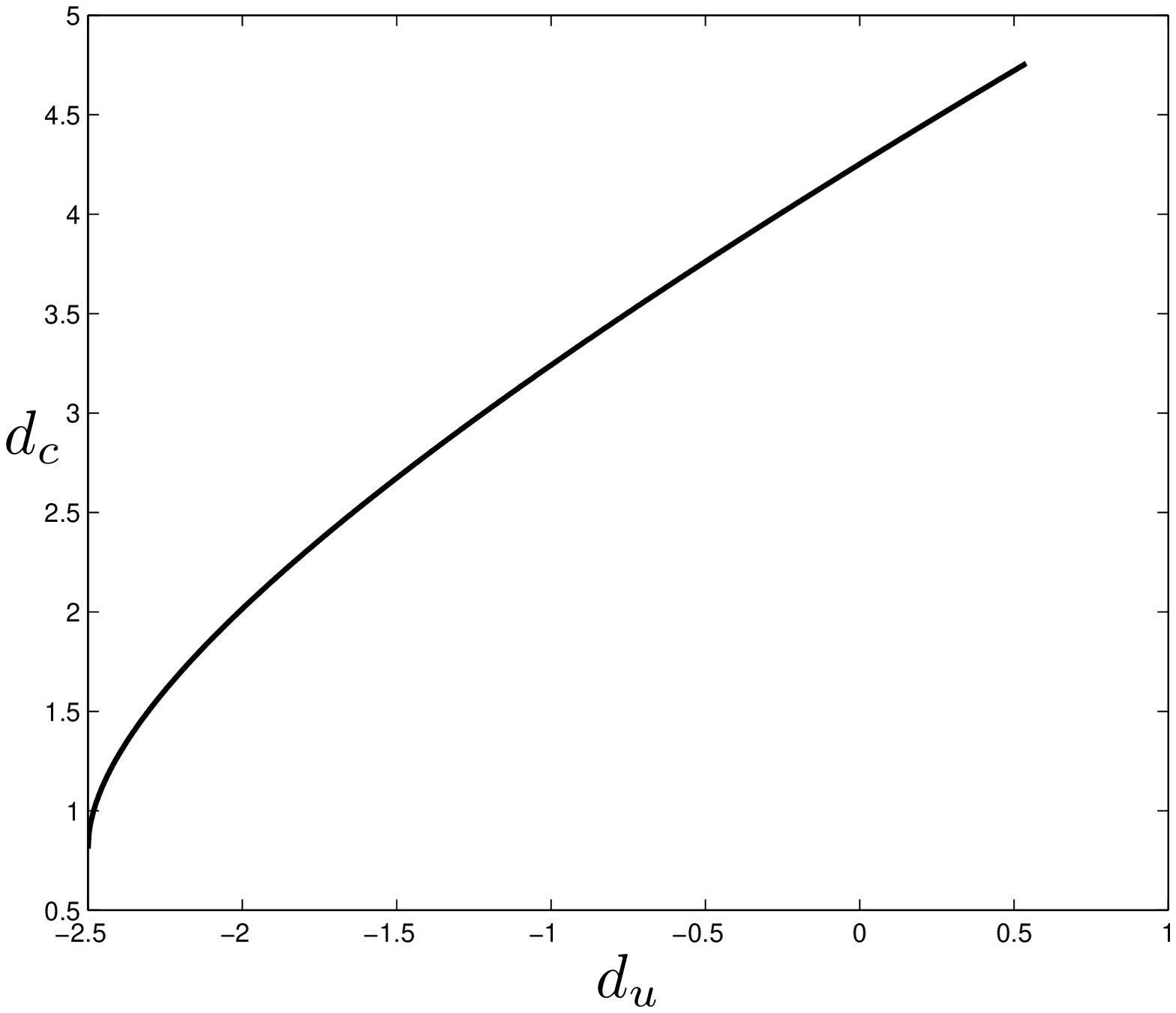}}
\caption{\label{du_dv} (a) $d_c$ vs $d_v$; (b) $d_c$ vs $d_u$. The parameters chosen for the \SS model are: $a=0.1$ and $b=0.6$}
\end{figure}
\begin{itemize}
\item $d_u=0$.\\
In this case the threshold condition \eqref{cond_inst5} reduces to
$$d_c=\left(\frac{a+b}{b-a}\right)^2\left[(a+b)^2+2b(a+b)+\frac{2b}{(a+b)^2}(a-b)d_v+2\sqrt{2b[d_v(a-b)+(a+b)^3}\right].$$
The derivative of $d_c$ with respect to $d_v$ and evaluated at $d_v=0$, i.e., at the point corresponding to the standard Turing condition, is given by
$$\frac{\partial d_c}{\partial d_v}|_{d_v=0}=-\frac{2b}{b-a}-\frac{2b(a+b)^2}{\sqrt{2b\left[d_v(a-b)+(a+b)^3\right]}},$$
which is negative. \\
This implies that conditions for the Turing instability in the \SS model become more favorable, i.e., the threshold value $d_c$ decrease as $d_v$ increases.\\
Note, however, that the Turing instability will be suppressed if the cross-diffusion coefficient becomes too large, i.e., exceeds $\frac{(a+b)^3}{b-a}$.
In Fig. \ref{du_dv}(a) we illustrate the results choosing $a=0.1$ and $b=0.5$. With this choice of the parameters we have that ${d_v}_{\max}=0.54$ and ${d_u}_{\min}=-2.5$.
\vskip.2cm
\item $d_v=0$.\\
In this case the threshold condition \eqref{cond_inst5} reduces to
$$d_c=\left(\frac{a+b}{b-a}\right)^2\left[(a+b)^2+2b(a+b)+d_u(b^2-a^2)+2\sqrt{(a+b)^3[2b+d_u(b-a)]}\right].$$
The derivative of $d_c$ with respect to $d_u$ and evaluated at $d_u=0$, i.e., at the point corresponding to the standard Turing condition, is given by
$$\frac{\partial d_c}{\partial d_u}|_{d_u=0}= \frac{(a+b)^3}{b-a}\left[1+\frac{a+b}{\sqrt{2b(a+b)}}\right],$$
which is positive. \\
This implies that conditions for the Turing instability in the \SS model become less favorable, i.e., the threshold value $d_c$ increase as $d_u$ increases.\\
Note, however, that the Turing instability will be suppressed if the cross-diffusion coefficient drops below $\frac{2b}{a-b}$.
In Fig. \ref{du_dv}(b) we plot the dependence of the Turing threshold value $d_c$ against the cross-diffusion coefficient $d_u$.\\
\end{itemize}
  \begin{figure}[h!]
\begin{center}
\subfigure[]{\includegraphics[height=5.3cm, width=7.5cm]{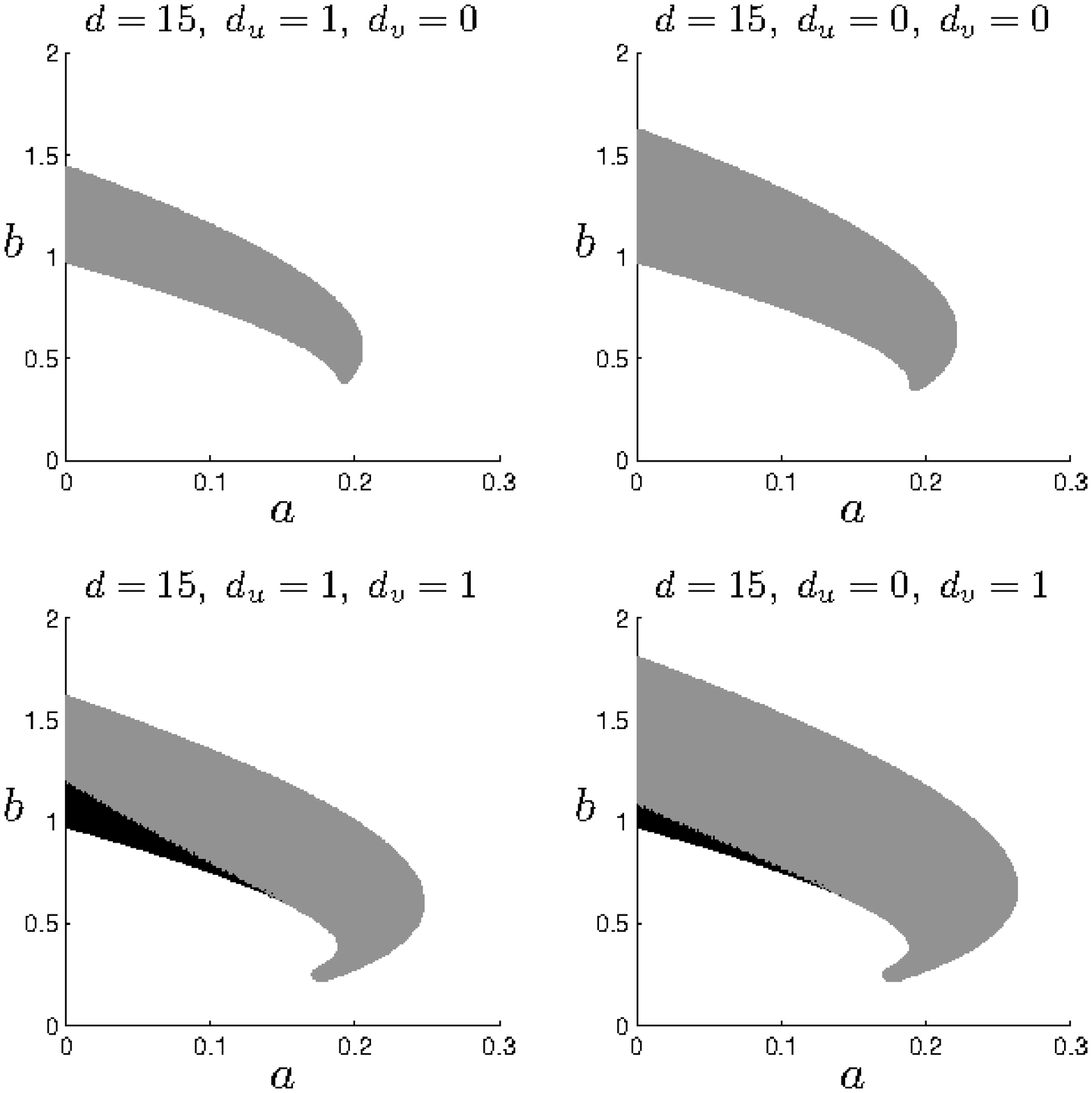}}
\subfigure[]{\includegraphics[height=5.3cm, width=7.5cm]{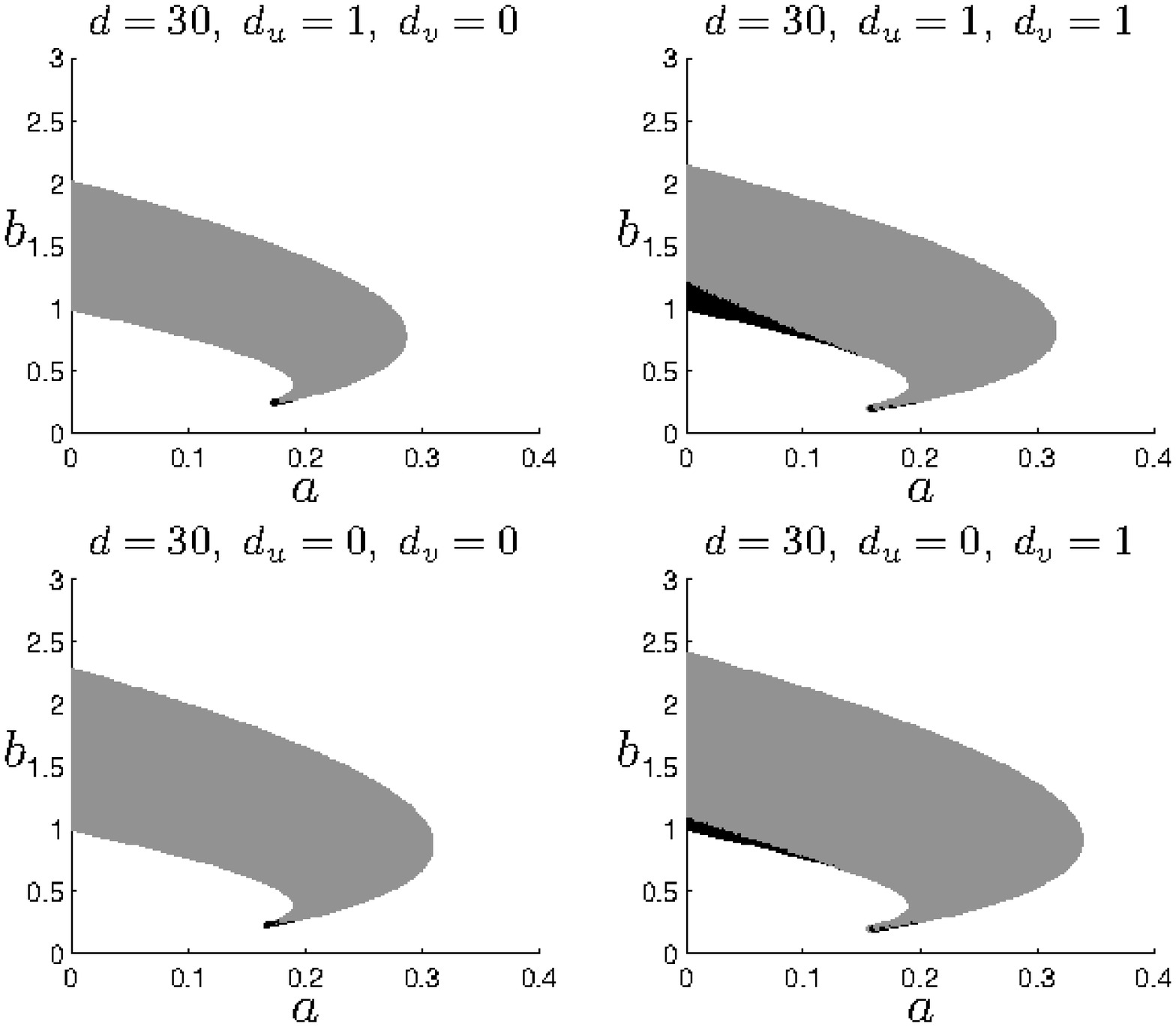}}
\subfigure[]{\includegraphics[height=5.3cm, width=7.5cm]{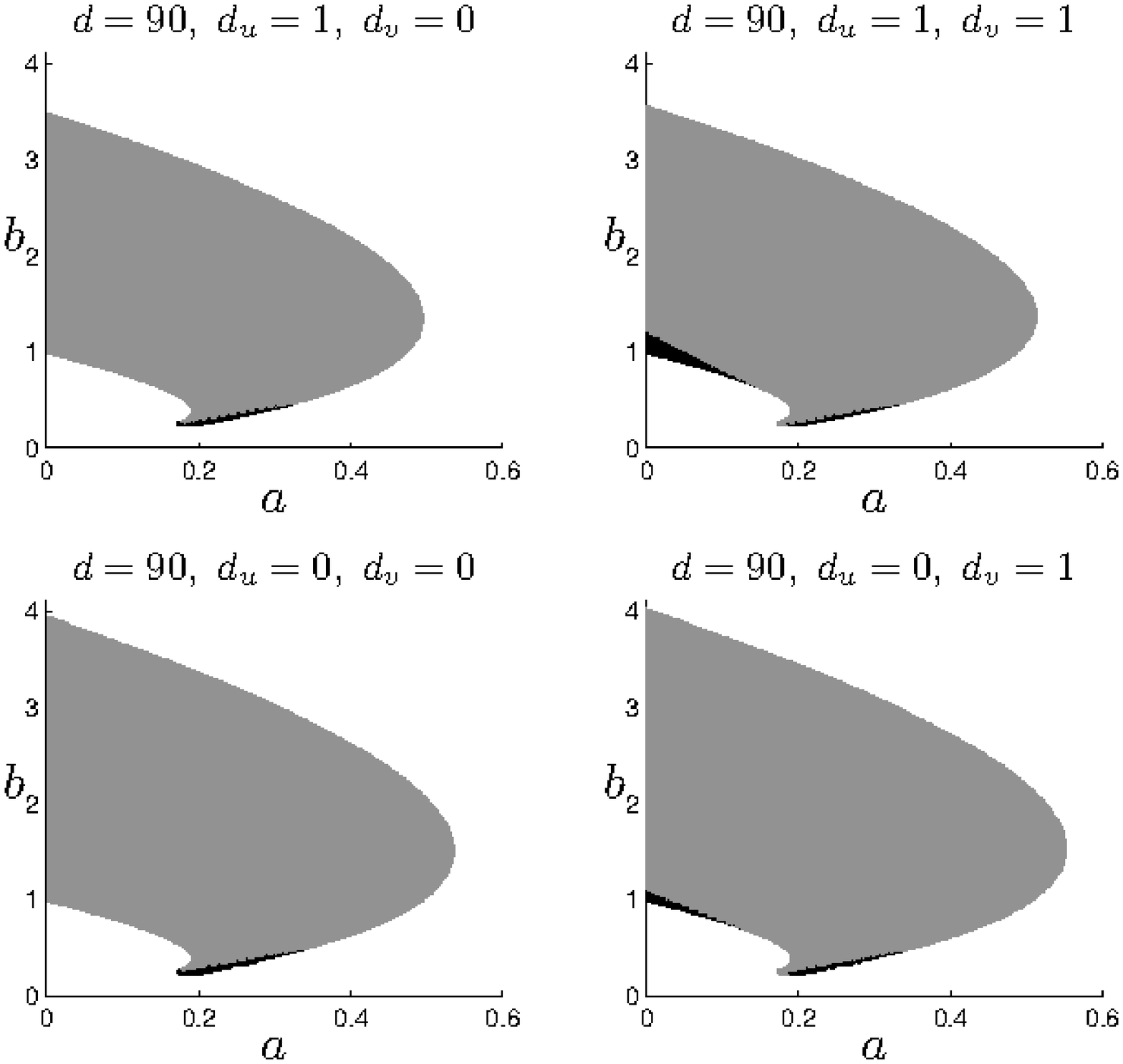}}
\subfigure[]{\includegraphics[height=5.3cm, width=7.5cm]{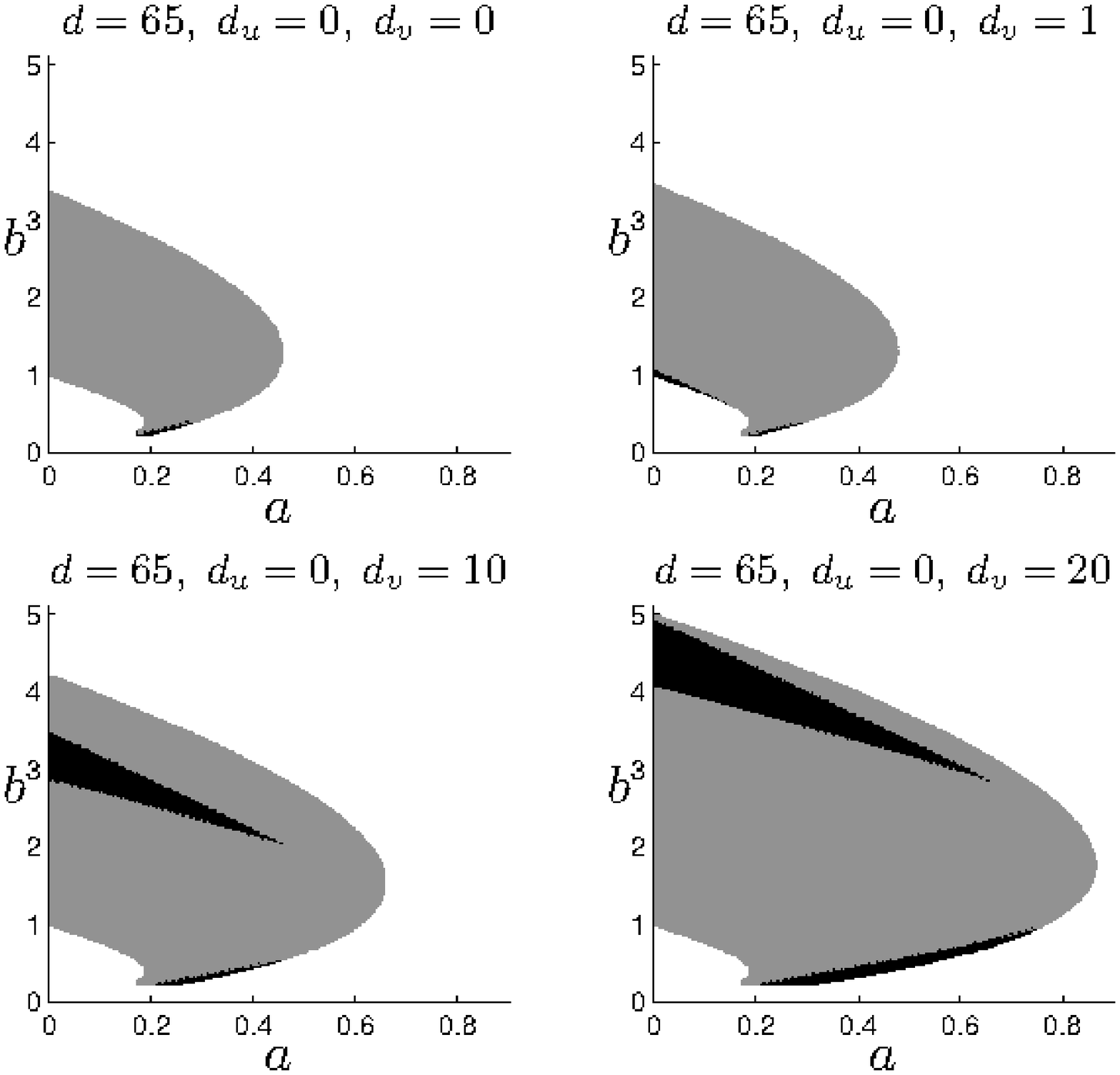}}
\subfigure[]{\includegraphics[height=5.3cm, width=7.5cm]{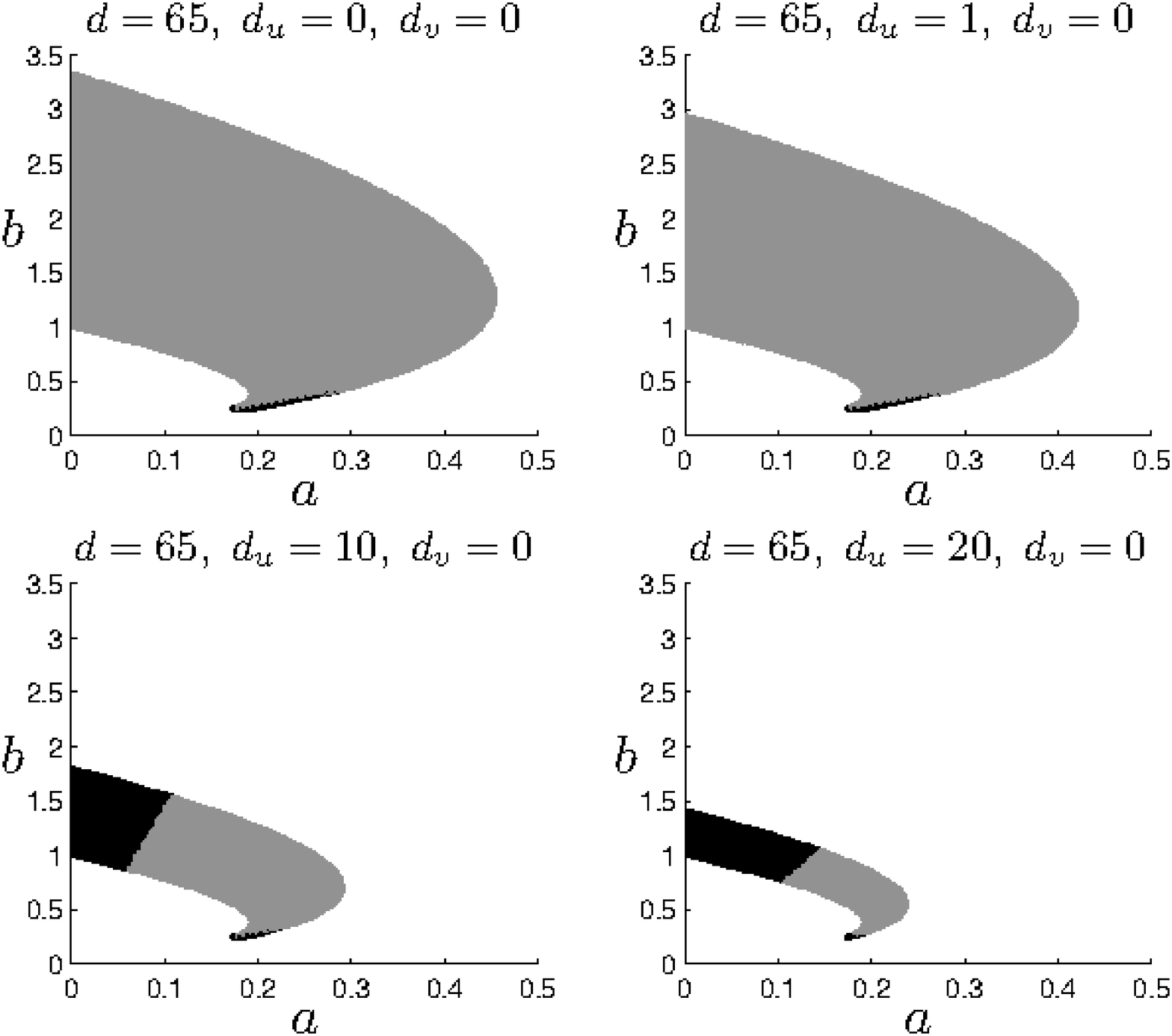}}
\subfigure[]{\includegraphics[height=5.3cm, width=7.5cm]{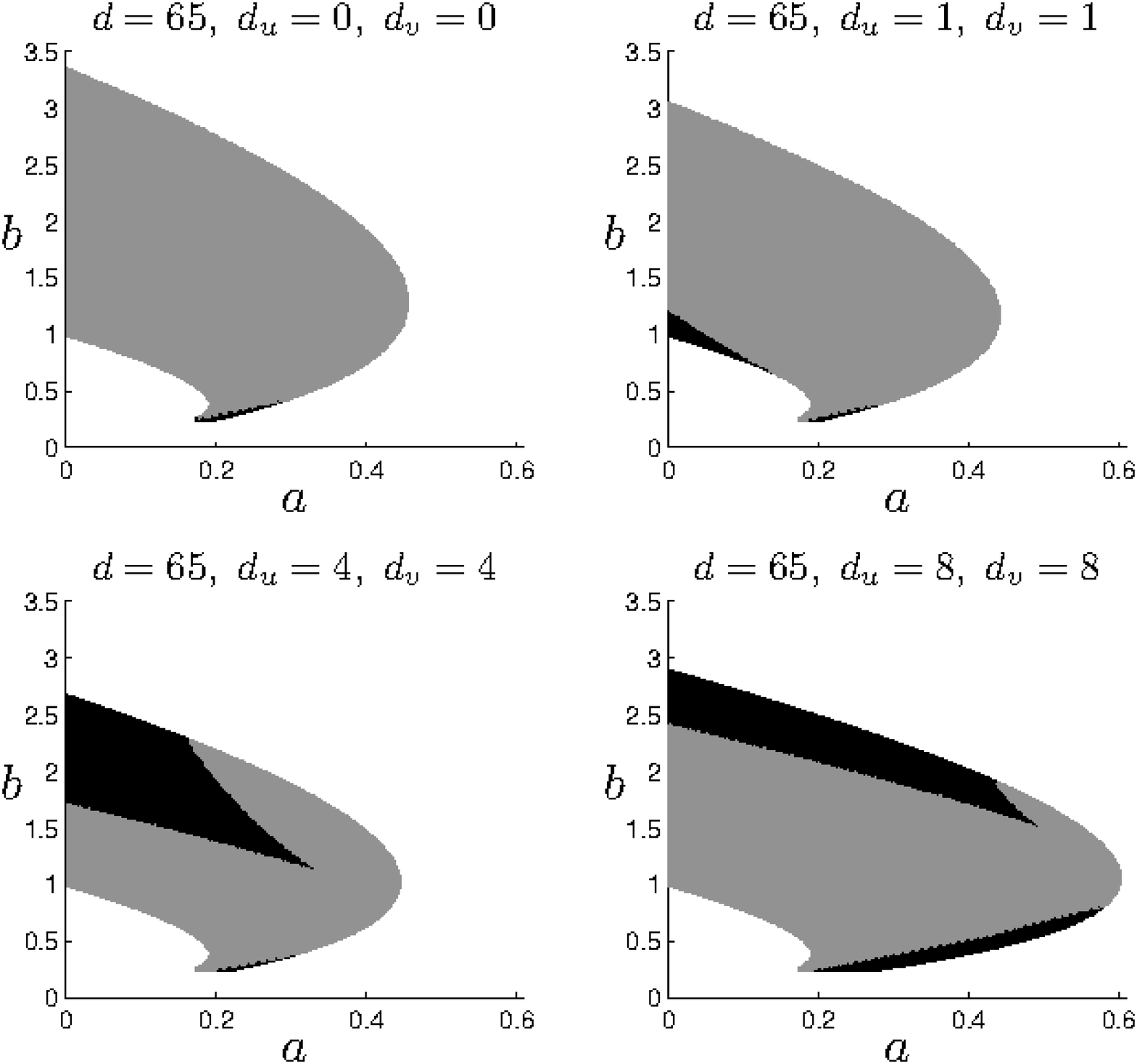}}
\caption{\label{embedding} Turing spaces for $d=15,30,65,90$ in order to demonstrate the different spaces obtained.}
\end{center}
\end{figure}

In order to focus how cross-diffusion terms, $d_u$ and $d_v$, enhances the instability of the equilibrium $P_0$, in Fig. \ref{embedding} we present some comparisons between different cases of the Turing spaces that we can summarize in the following.

\begin{itemize}
\item[-] The parameter space becomes larger and larger as the value of the diffusion coefficient $d$ increases.
\item[-] Cross-diffusion in the inhibitor component only ($d_v=0$) produces the smallest parameter space.
\item[-] Cross-diffusion in the activator component only ($d_u=0$) gives the biggest parameter space and it contains the former one.
\item[-] For some values of the bifurcation parameter $d$, we obtain, in perfect agreement with the results described in \cite{Madzvamuse}: the parameter space corresponding to the reaction-diffusion system without cross-diffusion ($d_u=0$ and $d_v=0$) is a subspace of the Turing space corresponding to the reaction-diffusion system with cross-diffusion in both $u$ and $v$, as is possible to see in Fig. \ref{embedding}(a). As $d$ grows this fact is not more true, in fact, as shown in Fig. \ref{embedding}(b) the two spaces are not included one in the other. If we increase further the value of $d$ we obtain the reverse situation, as shown in Fig. \ref{embedding}(c): the parameter space with cross-diffusion in both the components is a subspace of the parameter space without cross-diffusion.
  \end{itemize}

\begin{figure}[t]
\begin{center}
\subfigure[]{\includegraphics[height=6cm, width=7.5cm]{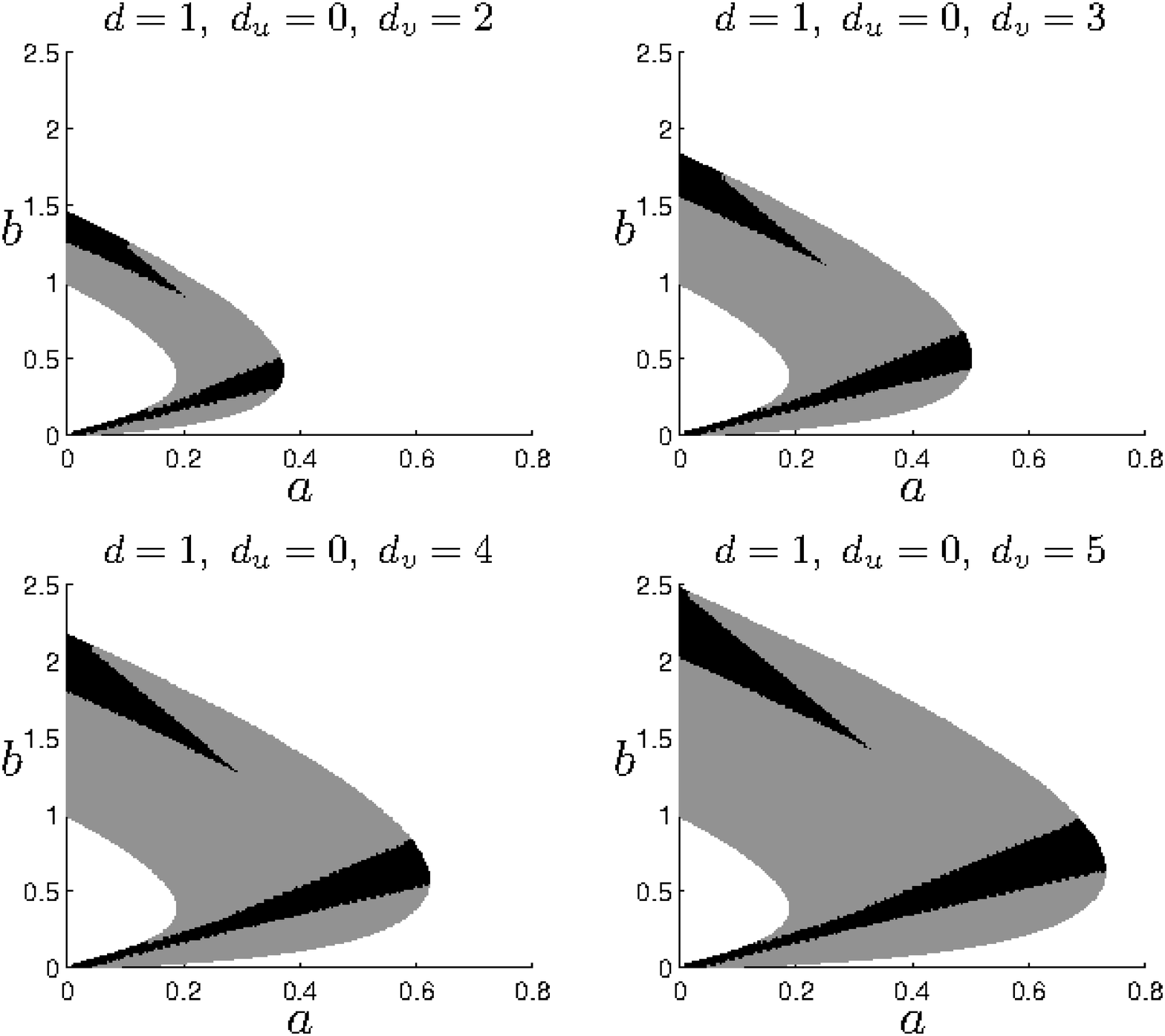}}
\subfigure[]{\includegraphics[height=6cm, width=7.5cm]{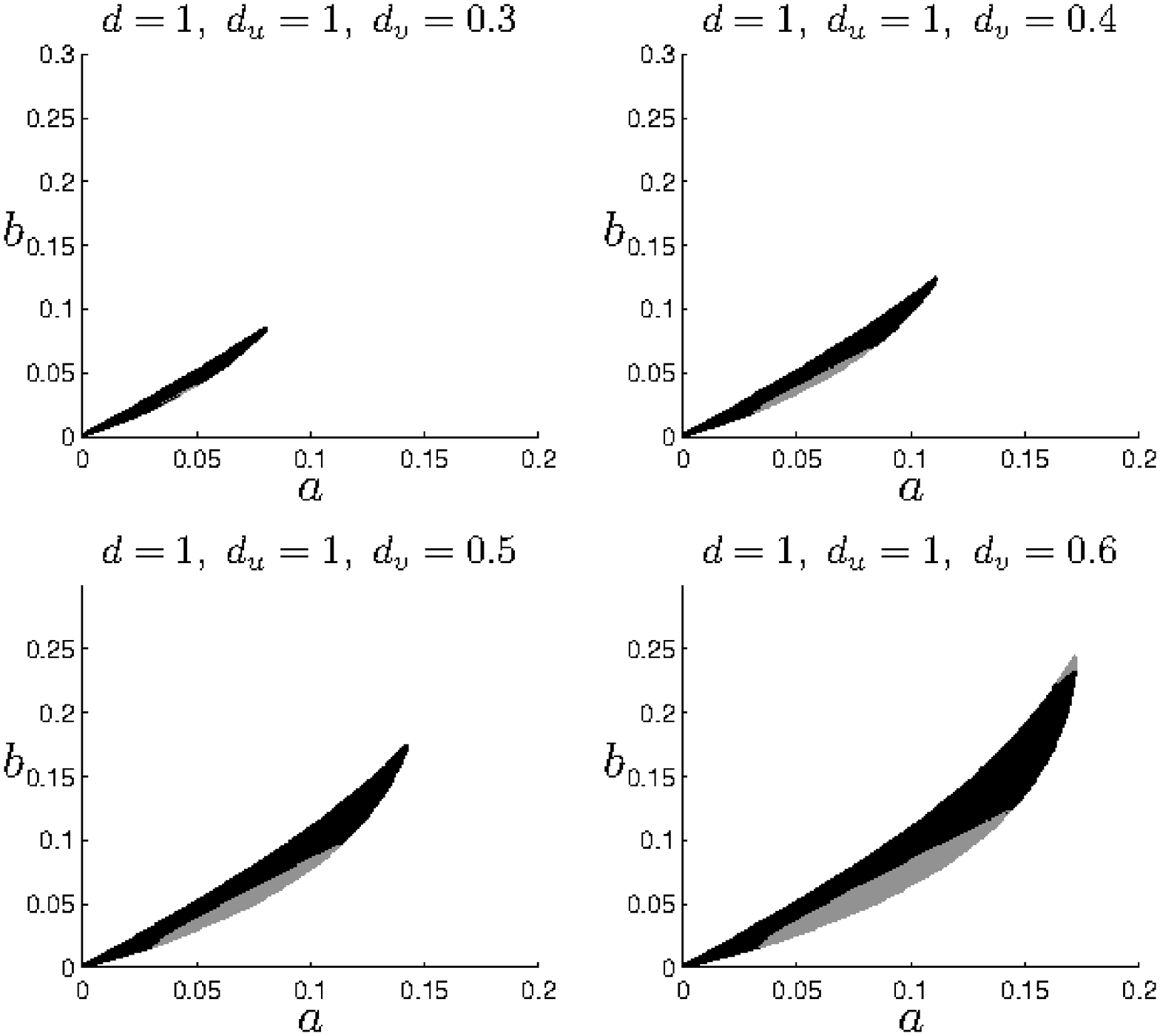}}
\subfigure[]{\includegraphics[height=6cm, width=7.5cm]{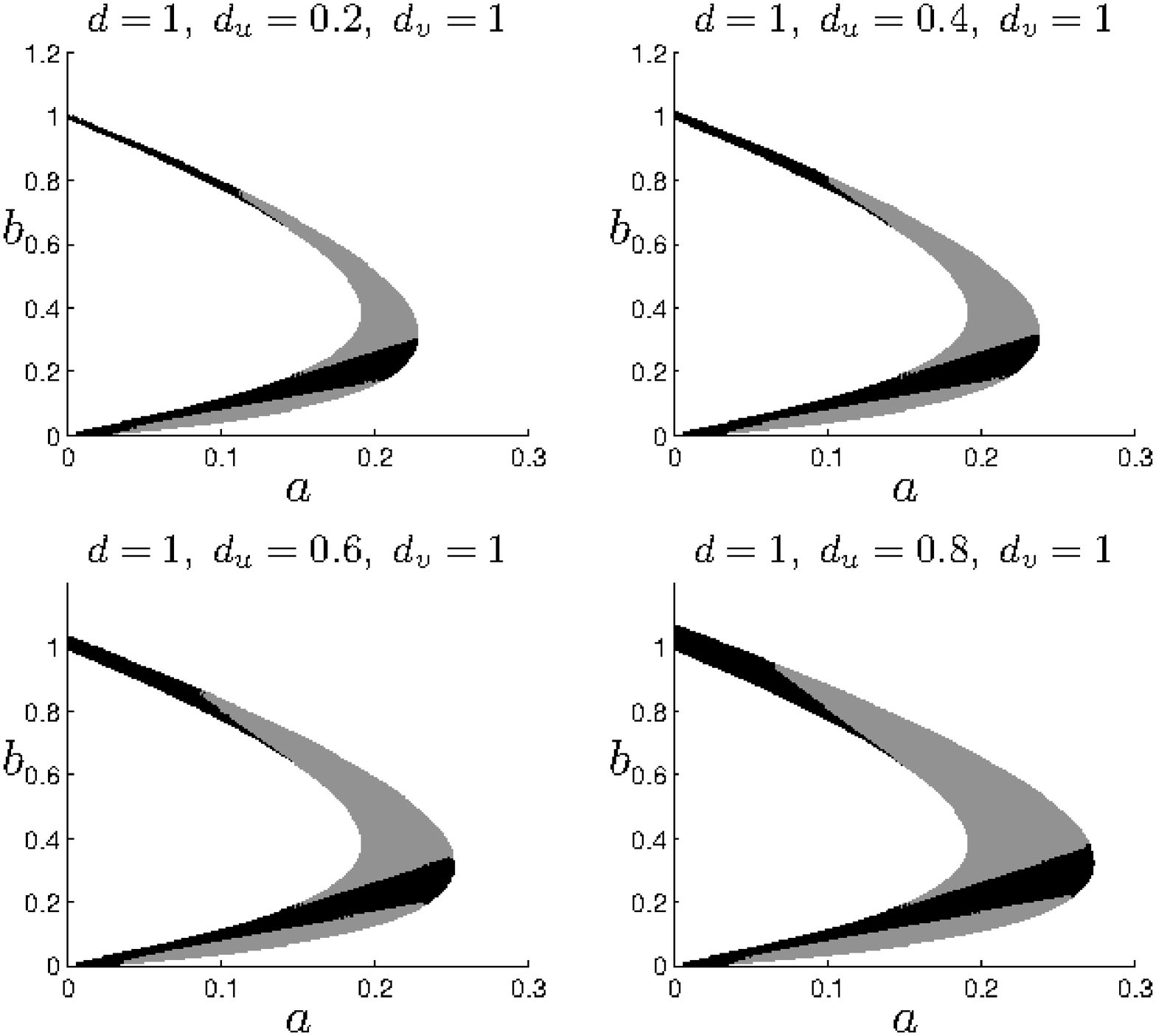}}
\subfigure[]{\includegraphics[height=6cm, width=7.5cm]{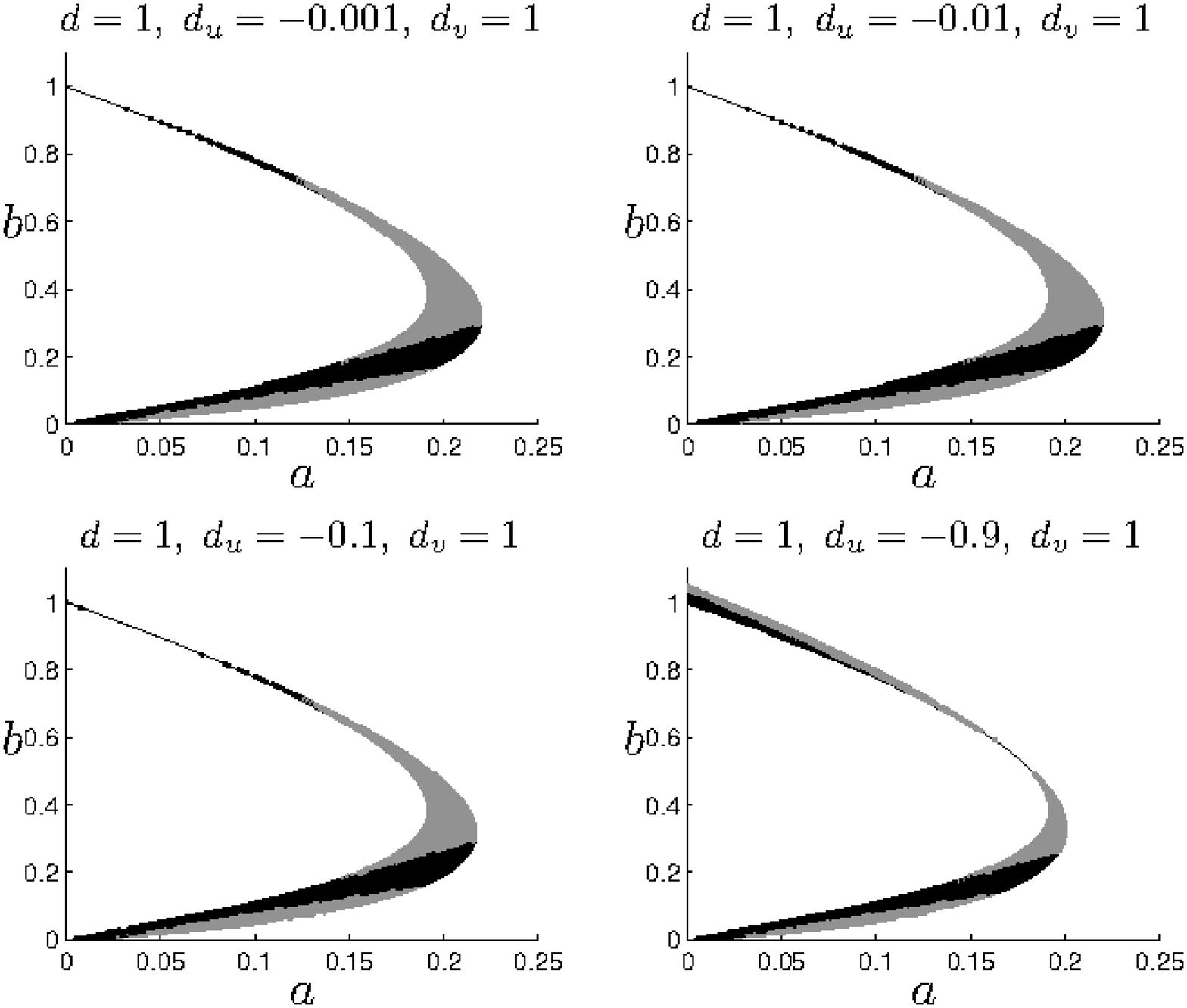}}
\caption{\label{d=1} Turing spaces for $d=1$ in order to demonstrate the effect of cross-diffusion coefficients on diffusion-driven instability.}
\end{center}
\end{figure}
We now perform a weakly nonlinear analysis of the uniform steady state $P_0$ when it is
stable to linear homogeneous perturbations.\\
By expanding the solution $\textbf{w}$, the time $t$ and the bifurcation parameter $d$, and employing the Fredholm solvability condition, we obtain the Stuart-Landau equation for the amplitude $A(T)$:
\begin{equation}
\label{SL3}
\frac{dA}{dT}=\sigma A-L A^3
\end{equation}

where the coefficients are obtained in terms of the parameters of the original system \eqref{model} and their explicit expression together with all the other details are given in Appendix \ref{AppA}.\\
Depending on the sign of the coefficient $L$ the we distinguish the supercritical (when $L>0$) from the subcritical ($L<0$) case.\\
Now we can study how cross-diffusion affects the Turing region.
\begin{itemize}
\item[-] As $d_v$ increases, both the parameter space and the subcritical region increase (see Fig. \ref{embedding}(d)).
\item[-] As $d_u$ increases, the parameter space decreases while the subcritical region increases (see Fig. \ref{embedding}(e)).
\item[-] Putting $d_u=d_v$ and increasing their values, the resulting effect is an increment of the subcritical region. As regards the extension of the Turing space, because of the presence of opposite contributions, it depends, time by time, on the choice of the parameters $d_u$ and $d_v$ (see Fig. \ref{embedding}(f)).
\item[-] $\gamma$ plays no role in the Turing space's representation, as it does not appear in the expressions that give the necessary conditions to destabilize the equilibrium. Moreover, $\gamma$ contributes to the variation of the value of the Landau coefficient, but not of its sign; therefore, as $\gamma$ varies, the region of subcriticality remains unchanged.
\end{itemize}

Observing Fig. \ref{du_dv}(b), in contrast with standard reaction-diffusion systems where Turing instability can occur only if $d>1$, we note that the minimum value of the Turing threshold, attained at the boundaries of $S$, is $d_{min}=0.81$, i.e. a value less than 1.\\
In other words, we do not require short-range activation, long-range inhibition which is the standard mechanism for a diffusively-driven instability and so a Turing instability is possible even if the diffusion coefficient of the activator is larger than that of the inhibitor.\\
To further validate this claim in Fig. \ref{d=1} we plot the parameter space when $d=1$ in the following cases:
\begin{itemize}
\item[(a)] If $d=1$ and $d_u=0$, the parameter space becomes larger and larger as the cross-diffusion $dv$ increases (see Fig. \ref{d=1}(a)). In this case we have not any restrictions in the choice of $d_v$ (provided it is positive), as the inequality $d-d_ud_v>0$ is always satisfied.
\item[(b)] If $d=1$ and $d_u=1$ we obtain the same results of the previous case (see Fig. \ref{d=1}(b)), but this time the inequality \eqref{cond_inst3} is satisfied if $d_v\in (0,1)$.
\item[(c)] If $d=1$ and $d_v=1$ we get parameter spaces that increase as $d_u$ increases in the interval $(0,1)$ (see Fig. \ref{d=1}(c)).
\end{itemize}
(Note that we do not consider the case in which $d_v=0$, because it represents the small parameter space, as seen above, and so it would not give any relevant results, being contained in the standard Turing space.)
Because of none of these spaces exist in the absence of cross-diffusion, these facts prove that a reaction-diffusion system with cross-diffusion can give rise to cross-diffusion-driven instability also putting $d=1$.\\
Finally we investigate the effects of negative cross-diffusion. The introduction of negative cross-diffusion in a reaction-diffusion system has already been studied in previous works. In \cite{lattice}, the author explains that negative cross-diffusion is a factor favoring the probability of spatial instabilities. Nevertheless, this case in quite rare, because it represents an unnatural tendency of a species moves against the concentration gradient of the other species. Recalling that when $d_u>0$ one get the smallest parameter space, in Fig. \ref{d=1}(d) we plot the parameter spaces with $d=1$, $d_v=1$ and $d_u\in(-1,0)$, to conclude that, also in this case, there are the possibility of the formation of spatial structures.
In particular, we note that the parameter spaces decreases as one get values close to $-1$. \\
\begin{figure}[t]
\label{Turing}
\centering
\subfigure[] {\epsfxsize=2.5 in \epsfbox{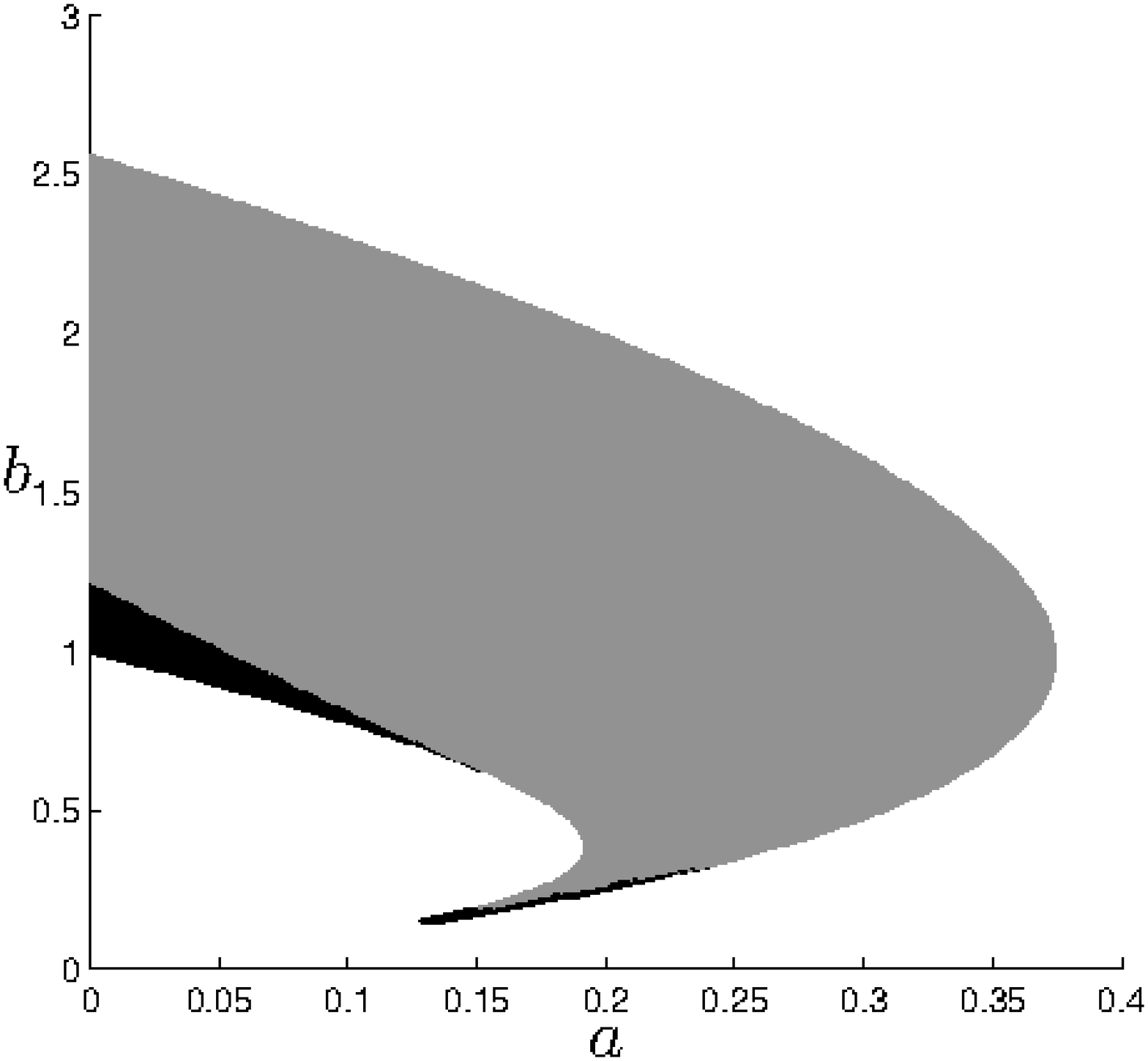}}
\subfigure[] {\epsfxsize=2.5 in \epsfbox{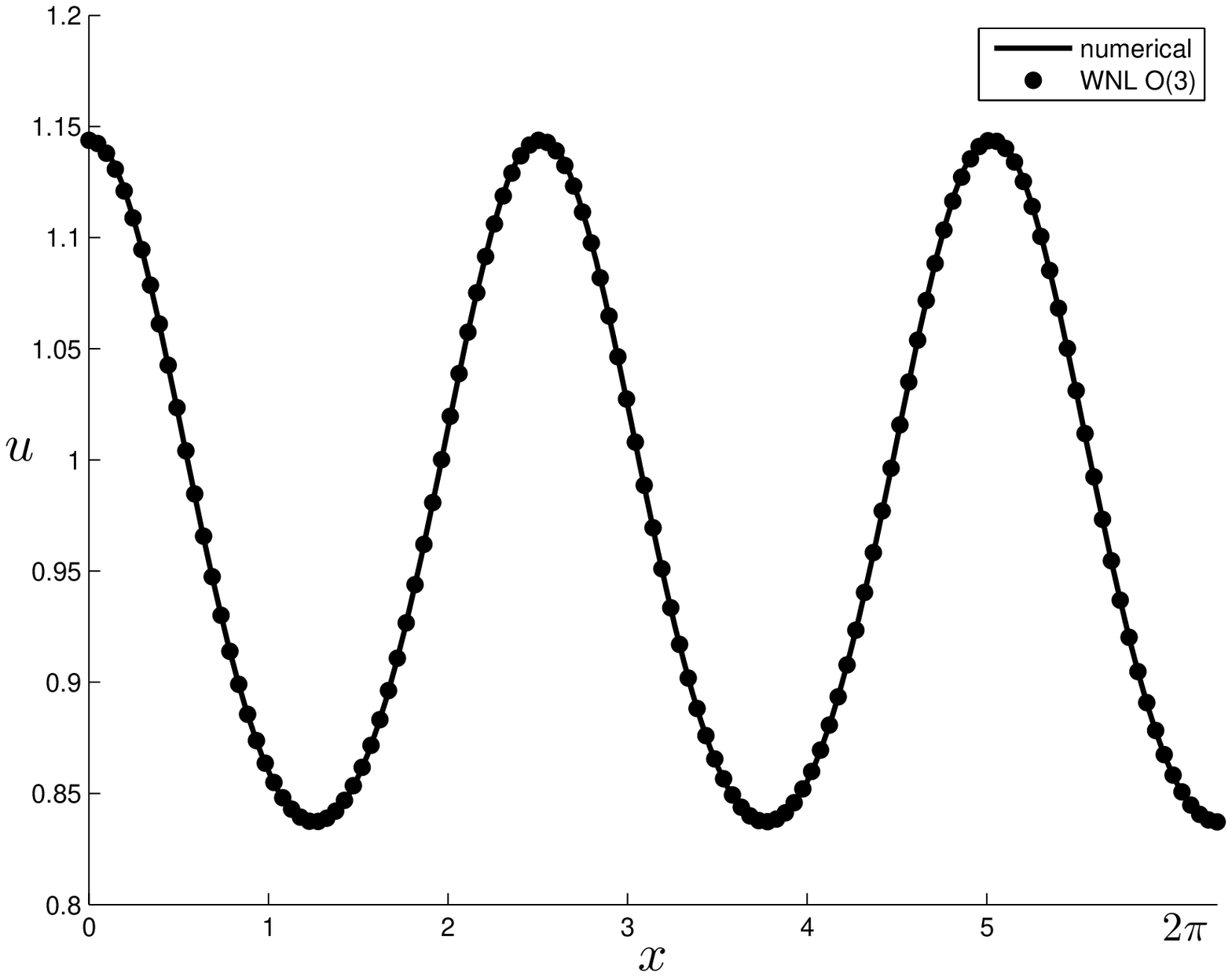}}
\caption{(a) The instability Turing region in the plane $(a,b)$ is shadowed: in gray the supercritical region, in black the subcritical region (see details in Section 3). The parameters are chosen as $a=0.34$, $b=0.64$, $d_u=1$, $dv=1$, $d=45$. (b) Comparison between the WNL approximated solution (dotted line) and the numerical solution of the full system \eqref{model} (solid line).  With the choice of the parameters as above and $d=d_c(1+\varepsilon^2)$, with $\varepsilon=0.2$, one has $d_c=43.9864$, while $\bar{k}_c= 2.5$.}
\end{figure}


\subsection{The Stuart-Landau equation}
According to the sign of $L$ in \eqref{SL3}, it is possible to consider the following two cases:
\vskip.2cm
\begin{enumerate}
\item {\bf The supercritical case.}
If $L>0$, the Stuart-Landau equation admits a stable equilibrium
solution $A_\infty=\sqrt{{\sigma}/{L}}$. Assuming that only the critical wavenumber $\bar{k}$ is admitted in the instability
interval, the long-time behavior of the solution of the reaction diffusion system \eqref{model} is given by $\mathbf{w}=\varepsilon A_\infty\ro \cos(\bar{k}_c x)$, where $\ro$ is defined in \eqref{ro}.
Choosing the system parameters in the supercritical parameter region (see Fig.~\ref{Turing}(a)), we compare in Fig.~\ref{Turing}(b) the asymptotic solution predicted by
the WNL analysis and the numerical solution of the system \eqref{model} computed via spectral methods starting from a random periodic perturbation
of the constant state.

The two solutions show a good agreement; in particular, in all the performed tests we have verified that the distance,
evaluated in the $L^1$ norm, between the WNL approximation
and the numerical solution of the system is $O(\varepsilon^3)$.
\vskip.2cm
\item {\bf The subcritical case}\\ If $L<0$ the Stuart-Landau equation \eqref{SL3} does not admit any stable equilibrium and finite-amplitude effects tend
to enhance infinitesimal disturbances growth.
This is an instance of a subcritical instability and higher order terms must be considered in our WNL analysis to determine the true long-time behavior \cite{BMS09}. In particular, pushing the analysis up to $O(\varepsilon^5)$, we recover the following
quintic Stuart-Landau equation
\begin{equation}
\label{SL5}
\frac{dA}{dT}=\bar{\sigma} A-\bar{L} A^3+\bar{R} A^5
\end{equation}
which mimics the amplitude of the pattern in the subcritical region shadowed in black in Fig.~\ref{Turing}(a).
\begin{figure}[t]
\begin{center}
\subfigure[] {\epsfxsize=3 in \epsfbox{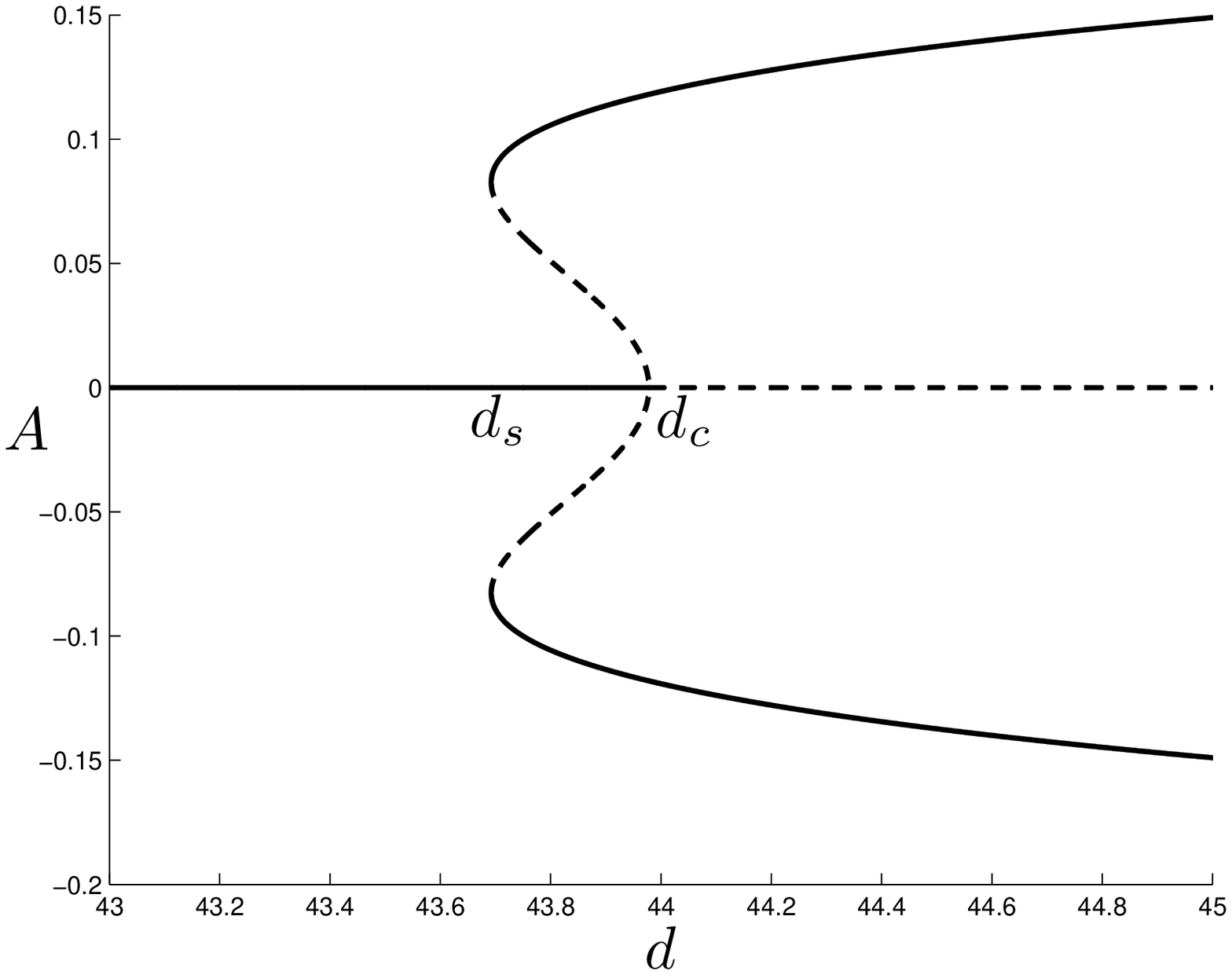}}
\subfigure[] {\epsfxsize=3 in \epsfbox{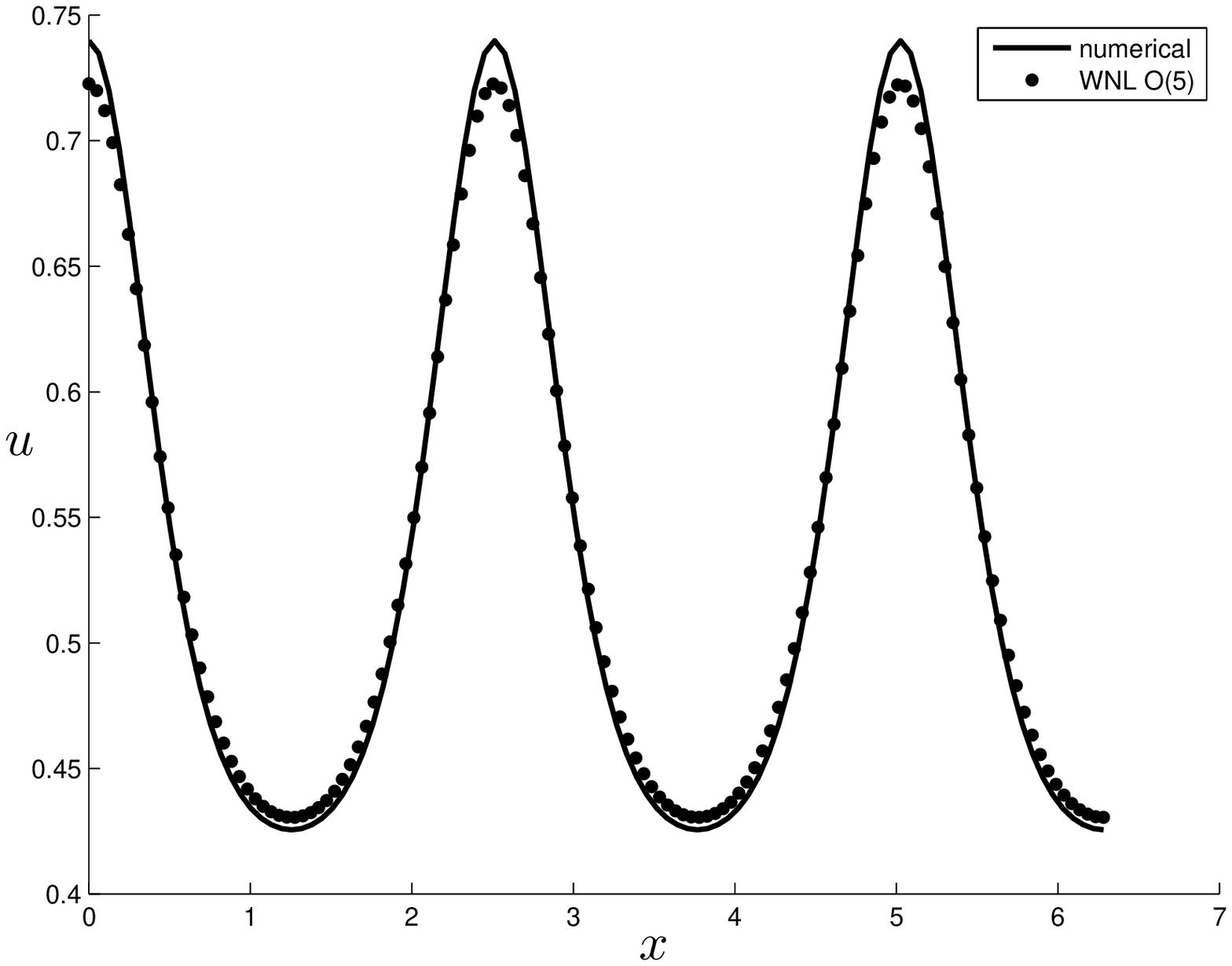}}
\end{center}
\caption{(a) The bifurcation diagram in the subcritical case (the stable branches are drawn with solid line, the unstable ones with dashed line). (b) Comparison between the weakly
nonlinear solution (solid line) and the numerical solution of \eqref{model}. The parameters are chosen as: $a=0.23$, $b=0.31$, $\gamma=76$, $d_u=1$, $d_v=1$, and $\varepsilon=0.02$; $d_c=43.9782$.}
\label{bif_sub}
\end{figure}
The equation \eqref{SL5} predicts the long-time behavior of the amplitude of the pattern
when $\bar{\sigma}>0$, $\bar{L}<0$ and
$\bar{R}<0$, as it admits two symmetric real stable equilibria. In Fig.~\ref{bif_sub}(b) we show the comparison between the numerical and the approximated solutions predicted by the WNL. Notice that the agreement is almost rough as the amplitude is relatively insensitive to the size of the bifurcation parameter.
\end{enumerate}
An interesting phenomenon well described by the equation \eqref{SL5} is the hysteresis, which typically emerges when
qualitatively different stable states coexist (here happens in the range $d_s<d<d_c$, as it is possible to observe in Fig. \ref{bif_sub}). The hysteresis cycle in Fig. \ref{isteresi_1D} shows that starting with a value of the parameter above $d_c$ the solution stabilizes to a pattern with the amplitude corresponding to the stable branch of the bifurcation diagram. Decreasing
$d$ below $d_c$ the pattern does not disappear as the stable amplitude solution persists on the upper
branch. Still decreasing
$d$ below $d_s$ the pattern disappears as the amplitude solution jumps to the constant steady
state. To obtain again the formation of the pattern, we have to increase the
parameter $d$ above $d_c$.
\begin{figure}[h]
\centering
\subfigure{\includegraphics[height=8cm, width=12cm]{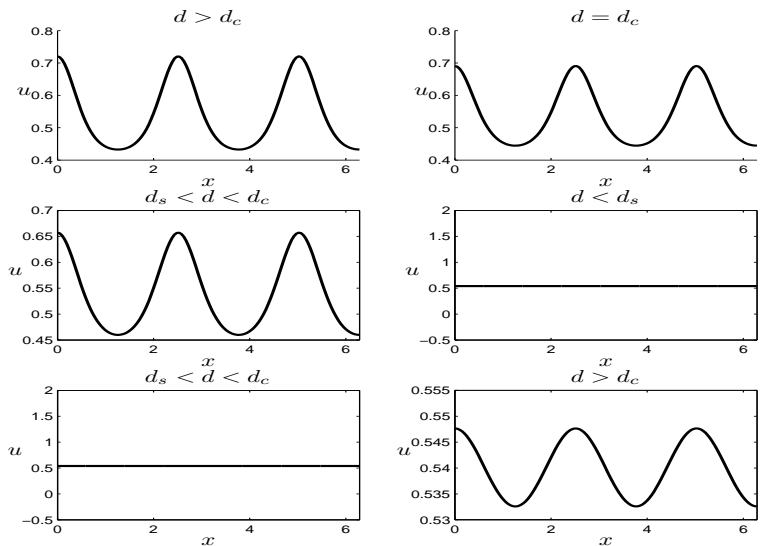}}
\caption{A hysteresis cycle and the corresponding pattern
evolution in the subcritical case.
The values of the other parameters are the same as in Fig.~\ref{bif_sub}}
\label{isteresi_1D}
\end{figure}
\section{Pattern formation in a two-dimensional domain}\label{Sec3}
In this section our analysis will focus on the pattern formation occurrence in a rectangular domain $\Omega=[0,L_x]\times[0,L_y]$.
We stress that the geometry of the domain does not affect the computation of the bifurcation value and the most unstable wavenumber, their values are still given respectively in formulas \eqref{dc} and \eqref{kc_val}.

In what follows we will assume that there exists only
one unstable eigenvalue $\lambda(\bar{k}_c^2)$, admitted in the instability band.
Once imposed the Neumann boundary conditions, the solution to the linear vector system \eqref{sequence_1}, obtained via the WNL analysis at $O(\varepsilon)$, is given by:
\begin{eqnarray}\label{solWNL2d}
&&{\bf w}_1=\sum_{i=1}^m A_i(T_1, T_2)\ro\cos(\phi_i x)\cos(\psi_i
y), \\
&&\bar{k}_{c}^2=\phi_i^2+\psi_i^2, \ {\rm where}\ \ \phi_i\equiv \frac{p_i\pi}{L_x},\ \
\psi_i\equiv  \frac{q_i\pi}{L_y},\label{multip}
\end{eqnarray}
where $A_i$ are the slowly varying amplitudes, $\ro \in {\rm Ker}(J-\bar{k}_c^2D^{d_c})$ and $m$, the multiplicity, reflects the degeneracy phenomenon: in our
analysis $m$ will take the values $1$ or $2$ depending on whether one or two pairs $(p_i,q_i)$  exist such that
$\bar{k}_c^2=\phi_i^2+\psi_i^2$. Once fixed the domain geometry, the types of the supported patterns strictly depend on the multiplicity.
\subsection{Simple eigenvalue, $m=1$}\label{simple}
When $m=1$ in formula \eqref{solWNL2d}, the WNL analysis traces the same steps as in one-dimensional domain and,
employing the Fredholm solvability condition at $O(\varepsilon^3)$, we still recover the Stuart-Landau equation \eqref{SL3}
ruling the evolution of the pattern amplitude.
Under the hypothesis that only one unstable eigenvalue $\lambda(\bar{k}_c^2)$ is admitted in the instability band, here
our investigation on the stability properties of equation \eqref{SL_3_5} is just limited to the supercritical and the subcritical cases.

\begin{figure}[t]
\centering
\subfigure[]{\includegraphics[height=5cm, width=7cm]{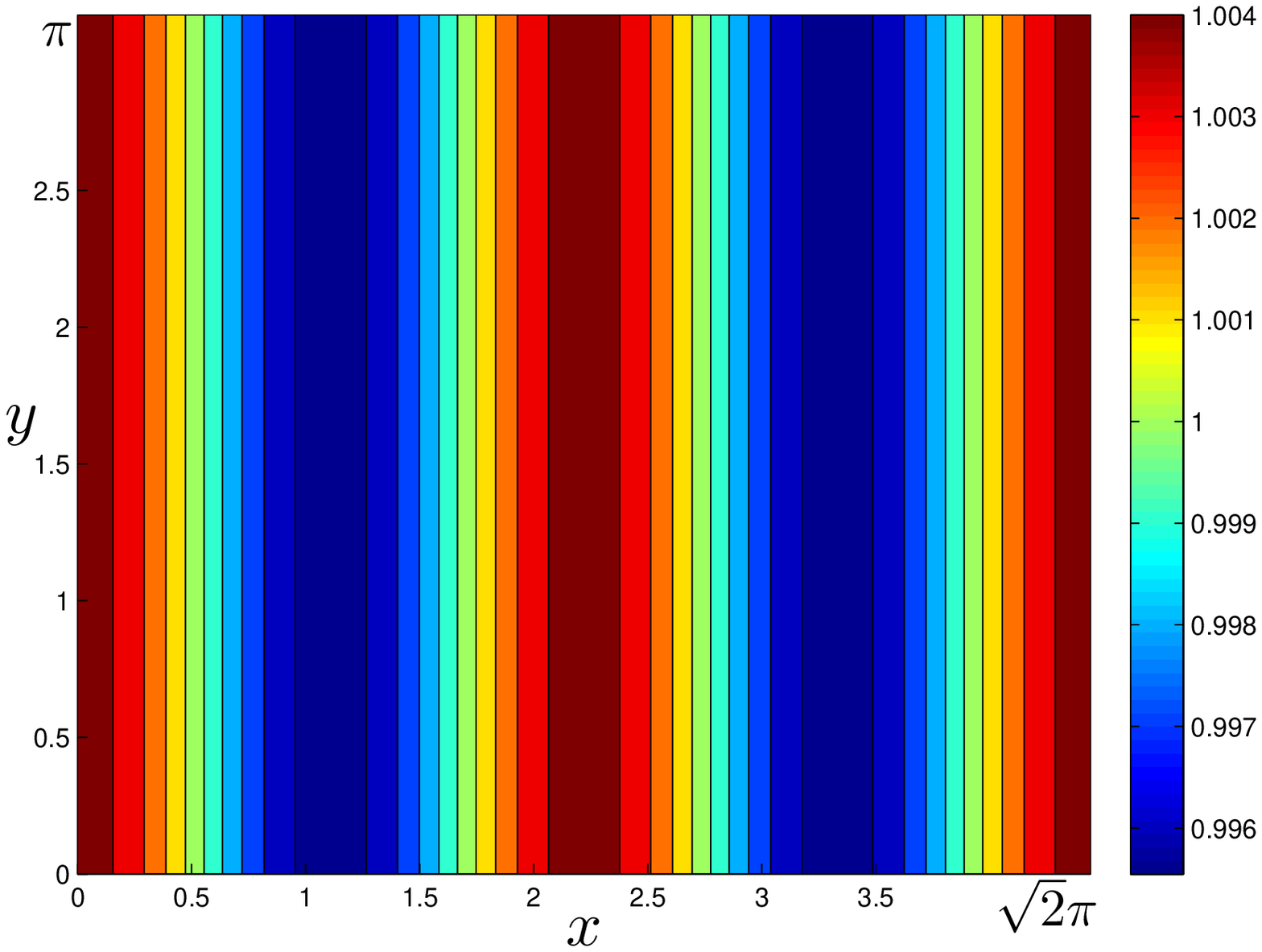}}
\subfigure[]{\includegraphics[height=5cm, width=7cm]{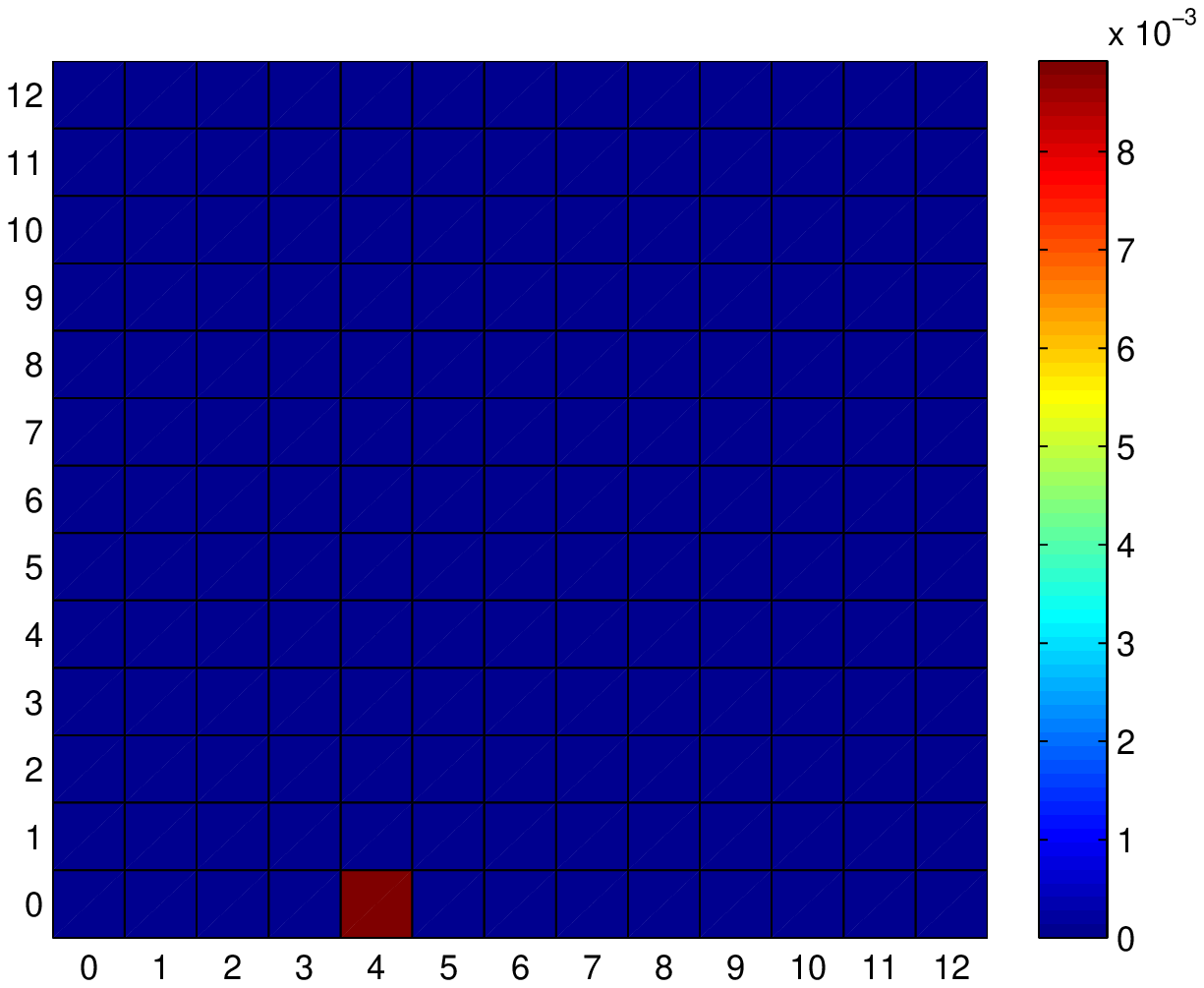}}
\caption{\label{roll} Rolls. (a) The numerical solution $u$ of the full system \eqref{model}. (b) Spectrum of the numerical solution. The parameters are  $a=0.3$, $b=0.7$, $\gamma=41$, $d_u=1$, $d_v=1$, where $d_c=12.2643$ and $\varepsilon= 0.01$.}
\end{figure}

\begin{figure}[t]
\centering
\subfigure[]{\includegraphics[height=5cm, width=7cm]{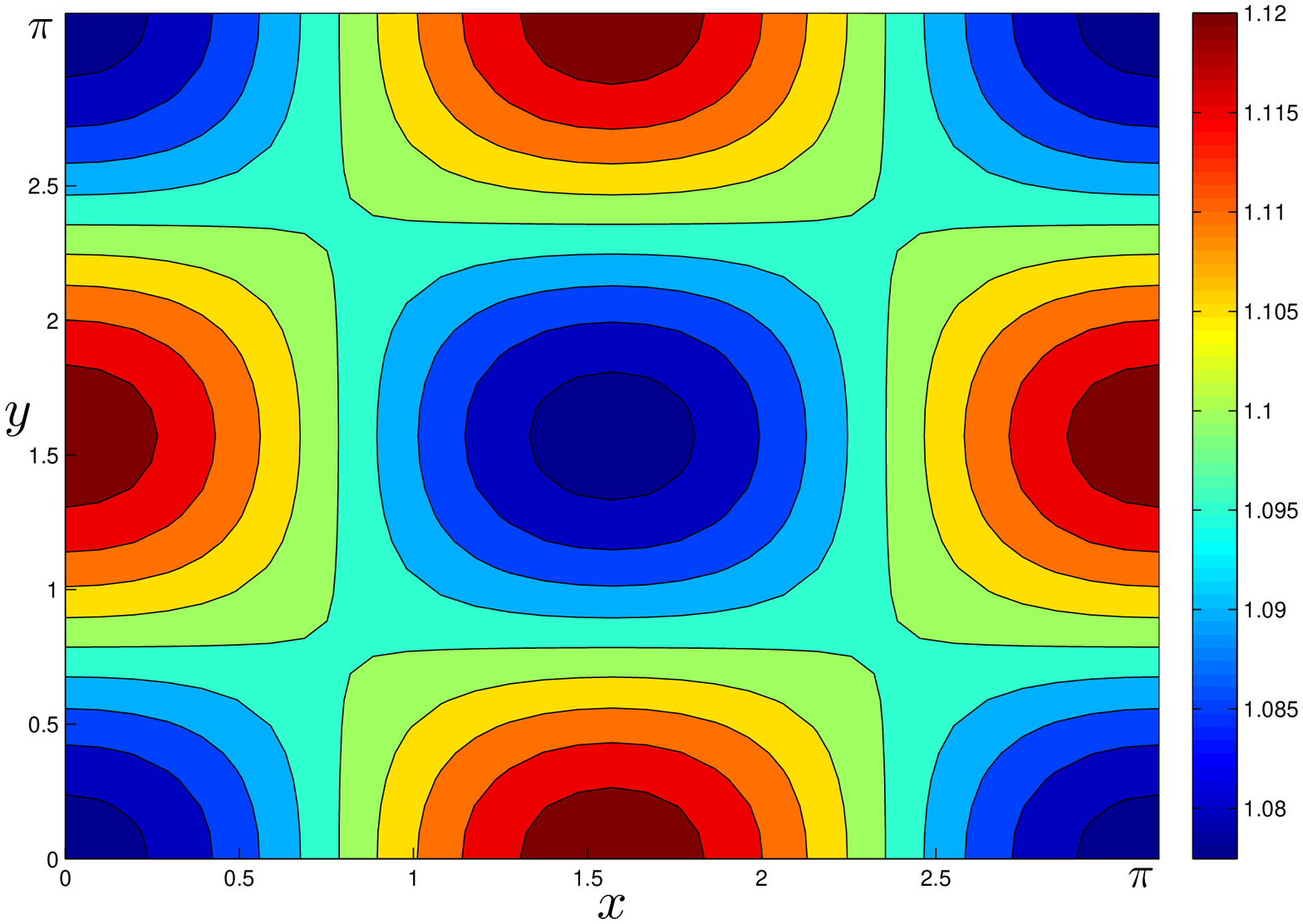}}
\subfigure[]{\includegraphics[height=5cm, width=7cm]{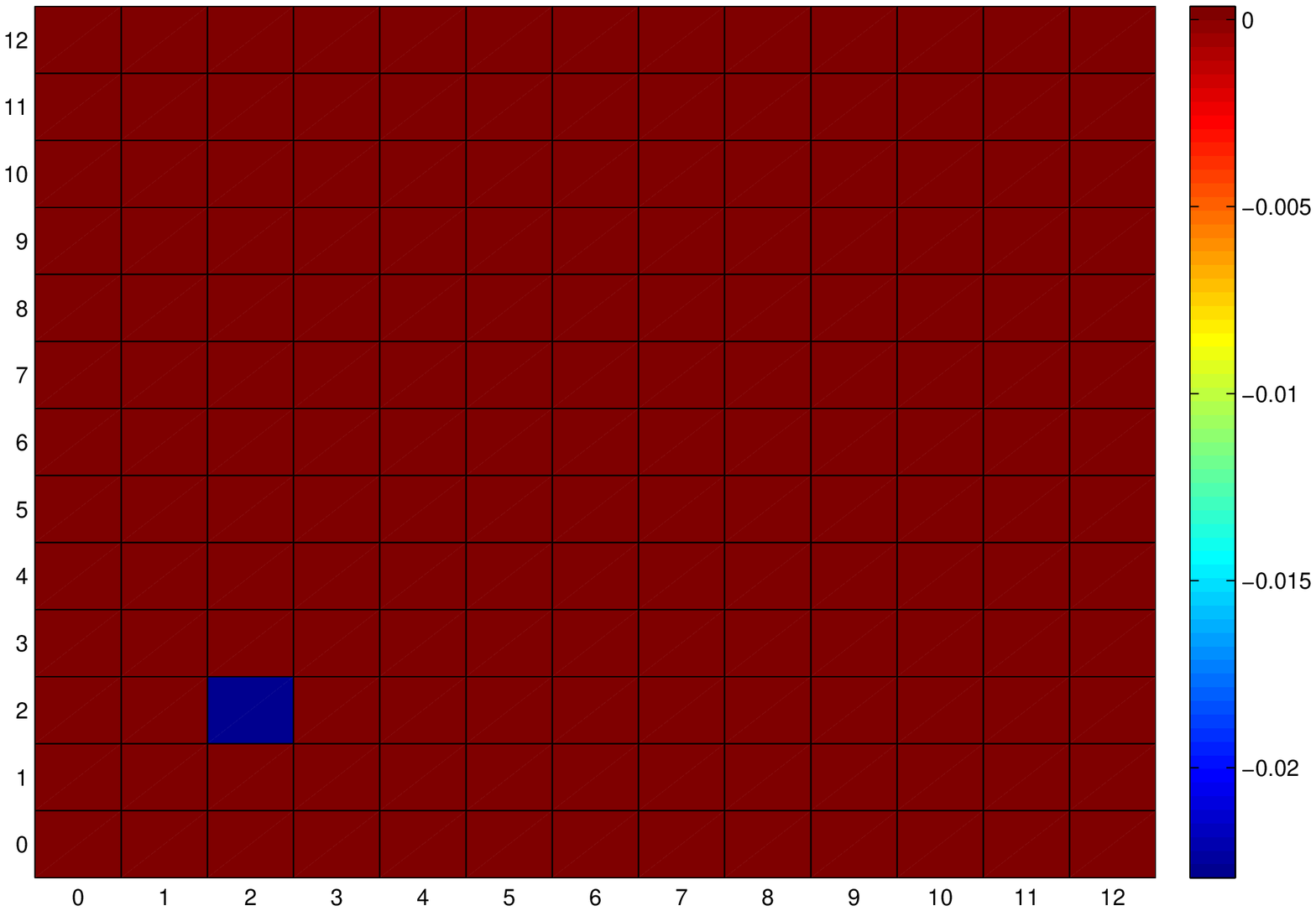}}
\caption{\label{square_sup} Squares. (a) The numerical solution $u$ of the full system \eqref{model}. (b) Spectrum of the numerical solution. The parameters are  $a=0.3$, $b=0.8$, $\gamma=37$, $d_u=1$, $d_v=1$, where $d_c=26.6243$ and $\varepsilon=0.01$.
}
\end{figure}

All the following numerical simulations have been performed via spectral methods employing $32$ modes both in the $x$ and in the $y$ axis. However the use
of a higher number of modes in the scheme (we tested the method up to $128$ modes) does not
appreciably affect the results. The initial conditions are random periodic perturbation about $P_0$.
Notice that, for a better presentation of the results, the amplitude of the zero mode (corresponding to the equilibrium solution) has been set equal to zero into the figures representing the spectrum of the solution.
\vskip.2cm
\textbf{Case (1): $L>0$.}
In the supercritical case the Stuart-Landau equation admits the stable equilibrium
solution $A_\infty=\sqrt{{\sigma}/{L}}$ and the asymptotic solution predicted by
the WNL analysis is given by:
\begin{equation}\label{sol_m1}
\mathbf{w}=\varepsilon \ro A_{\infty} \cos(\phi x)\cos(\psi y)+O(\varepsilon^2),
\end{equation}
where $(\phi, \psi)$ is the only pair such that $\bar{k}_{c}^2=\phi^2+\psi^2$.
In a rectangular domain the expected types of patterns corresponding to the solution \eqref{sol_m1} are rhombic pattern (see \cite{CMM97,CWKC14}), whose special cases are the rolls (when $\phi$ or $\psi$ is zero) or the squares (when $\phi=\psi$).
\vskip.2cm
\textbf{Rolls.} Choosing the system parameters as in Fig.~\ref{roll} and fixing the deviation from the bifurcation value $\varepsilon=0.01$, in the rectangular domain $L_x=\sqrt{2}\pi$, $L_y=\pi$ the only unstable critical wavenumber is $\bar{k}_c^2=8$ and the condition \eqref{multip} is satisfied by the unique mode pair $(p,q)=(4,0)$.
In this case the expected solution \eqref{sol_m1} represents a roll pattern, in good agreement with the numerical solution of the full system \eqref{model} shown in Fig.~\ref{roll}(a).
In particular, the values of the most excited modes respectively of the numerical solution (shown in Fig.~\ref{roll}(b)) and the WNL approximated solution truncated at $O(\varepsilon)$ are $0.0089$ and $0.0093$. \\
\vskip.2cm
\textbf{Squares.} Let us choose the square domain $L_x=L_y=\pi$, the deviation from the bifurcation value $\varepsilon=0.1$, the system parameters values
as in Fig.~\ref{square_sup}. In this case the unique discrete unstable mode is $\bar{k}_c^2=8$ and the conditions in \eqref{multip} are satisfied
only by the mode pair $(p,q)=(2,2)$. The predicted solution via the WNL analysis \eqref{sol_m1} is a square pattern, whose values of the amplitudes of the most excited mode $(2,2)$ (computed at $O(\varepsilon)$ of
the WNL analysis) and of the subharmonics $(4,0)$, $(0,4)$ and $(4,4)$ (computed at $O(\varepsilon^2)$ of
the WNL analysis) are in good agreement with the corresponding amplitude modes of the numerical solution of the full system \eqref{model}, as summarized in Table \ref{tab_square}. The numerical solution, together
with its spectrum, are given in Fig.~\ref{square_sup}.

\begin{figure}[h!]
\centering
\subfigure{\includegraphics[height=6cm, width=8cm]{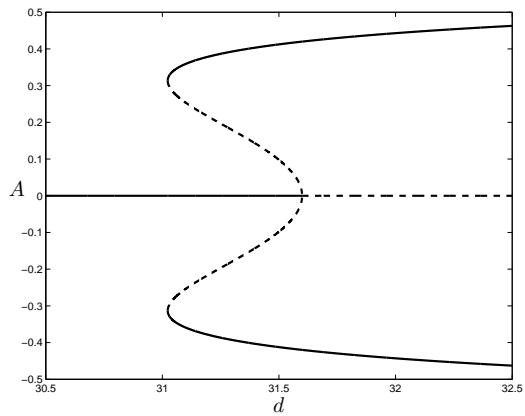}}
\caption{\label{diagr_bif_sub_2D}The bifurcation diagram (the stable branches are drawn with solid line and the unstable ones with dashed line). The parameters are  $a=0.32$, $b=0.76$, $\gamma=41$, $d_u=1$, $d_v=1$, $d=d_c(1+\varepsilon^2)$, where $\varepsilon=0.01$ and $d_c=31.5993$.
}
\end{figure}

\begin{figure}[t]
\begin{center}
{\includegraphics[height=3.8cm, width=6cm]{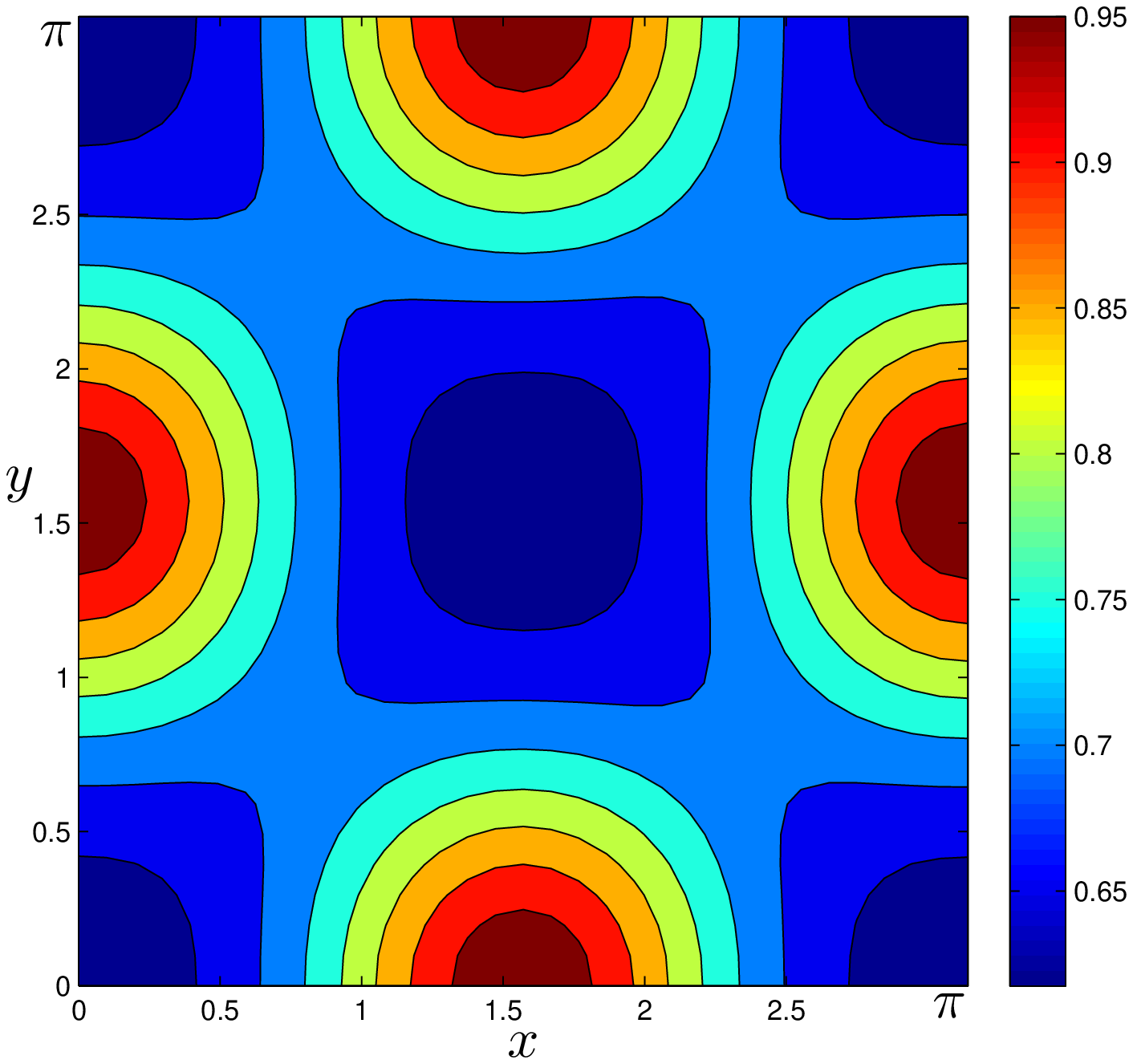}}
{\includegraphics[height=3.8cm, width=6cm]{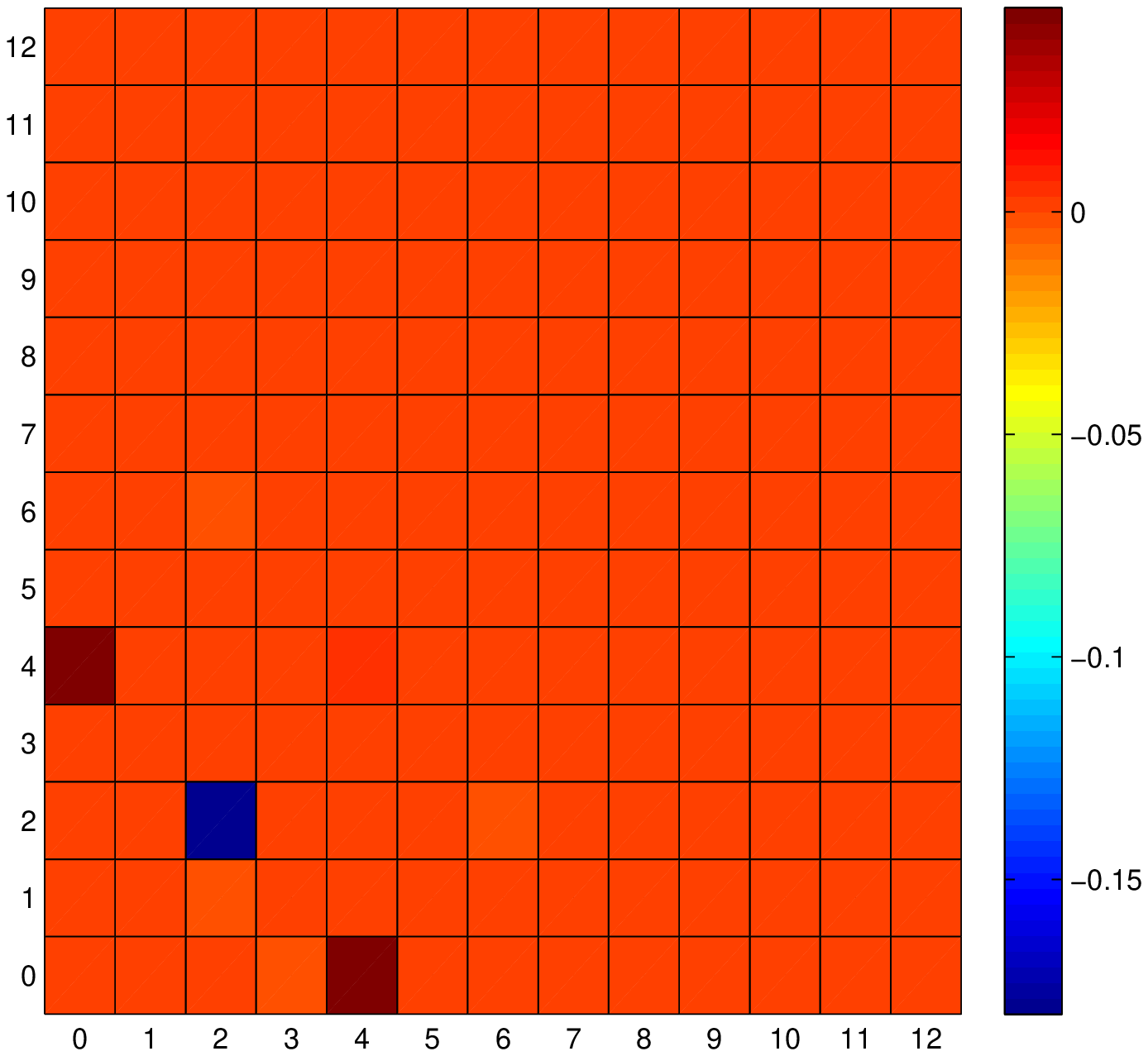}}
{\includegraphics[height=3.8cm, width=6cm]{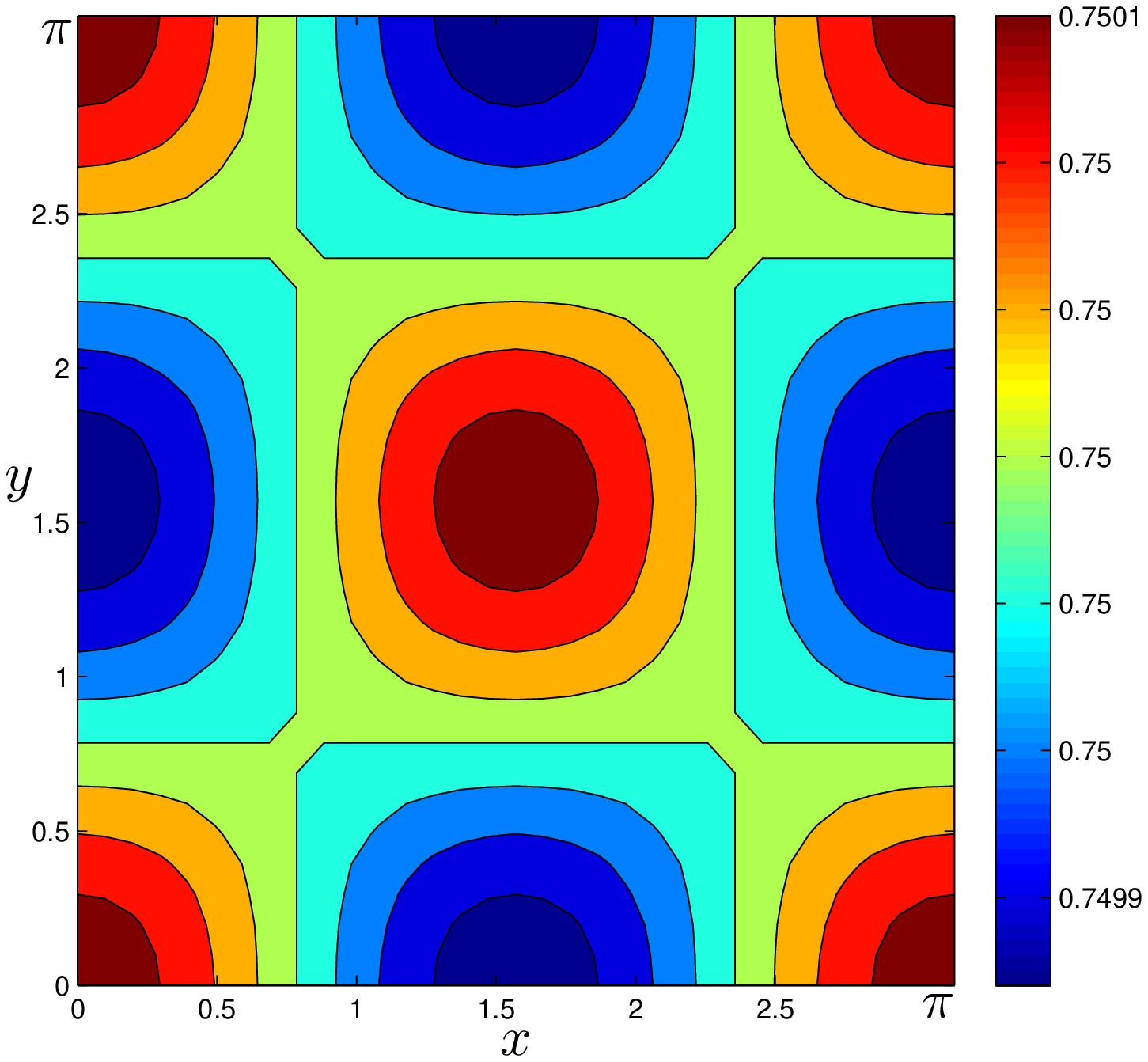}}
{\includegraphics[height=3.8cm, width=6cm]{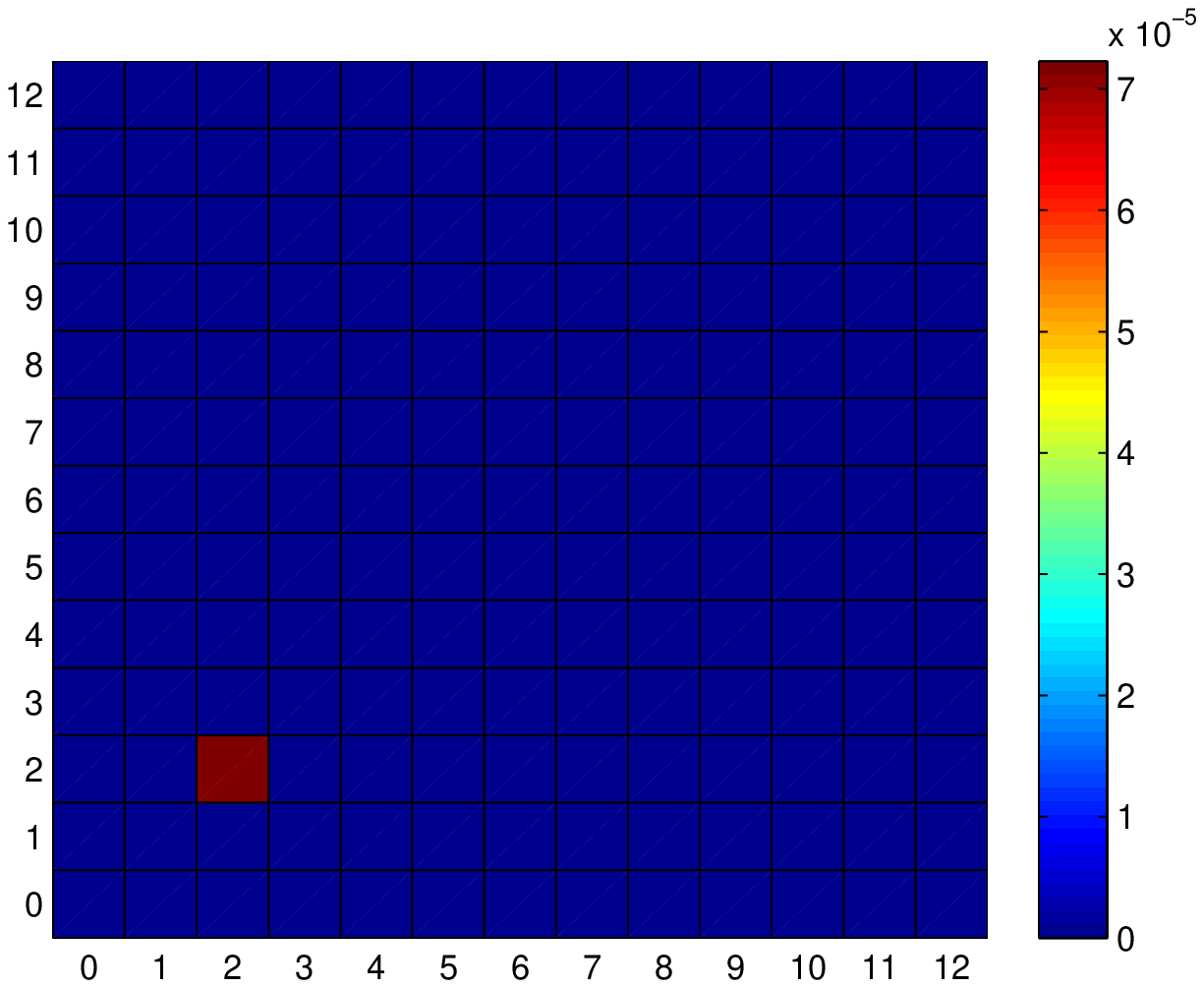}}
{\includegraphics[height=3.8cm, width=6cm]{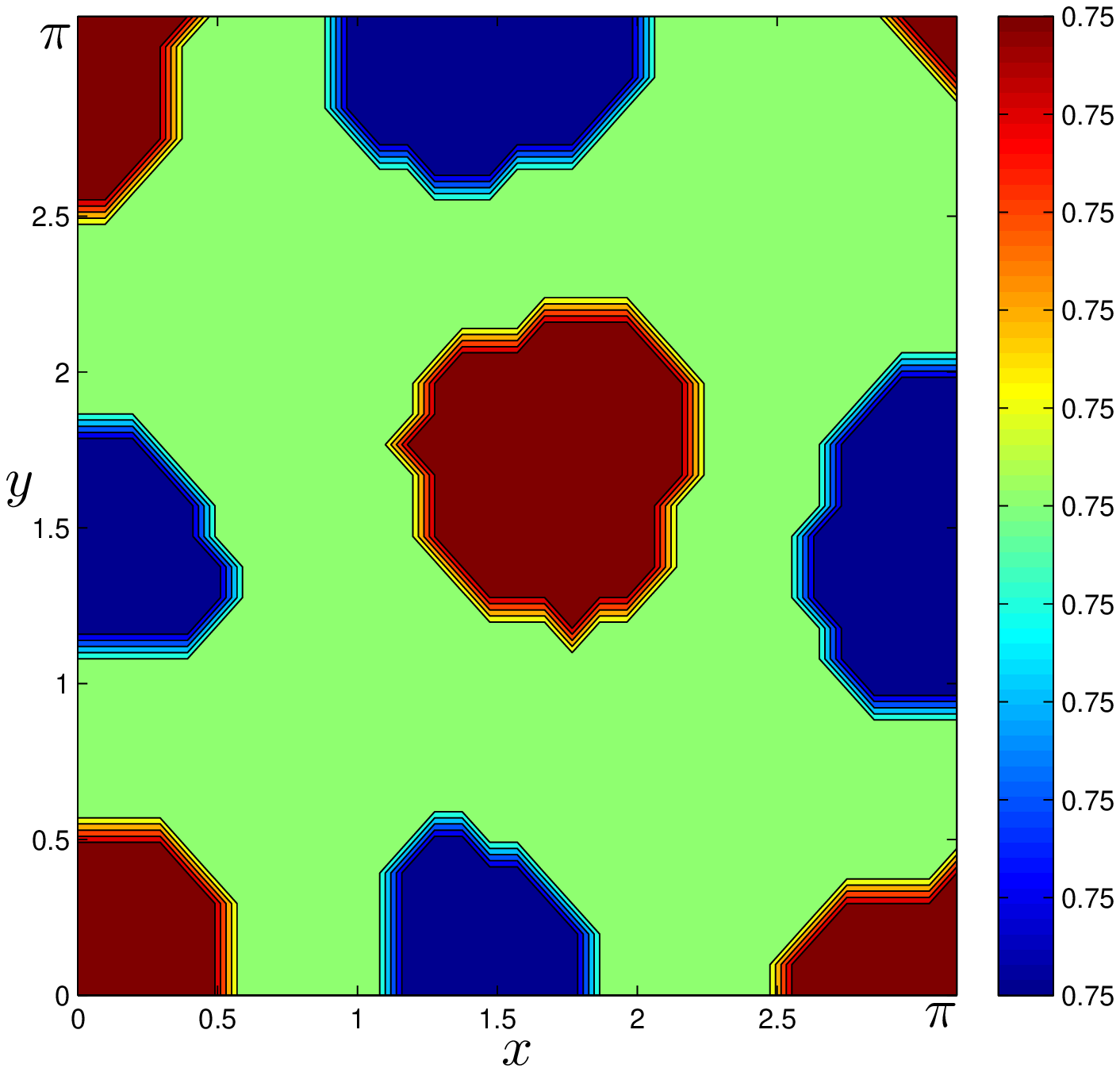}}
{\includegraphics[height=3.8cm, width=6cm]{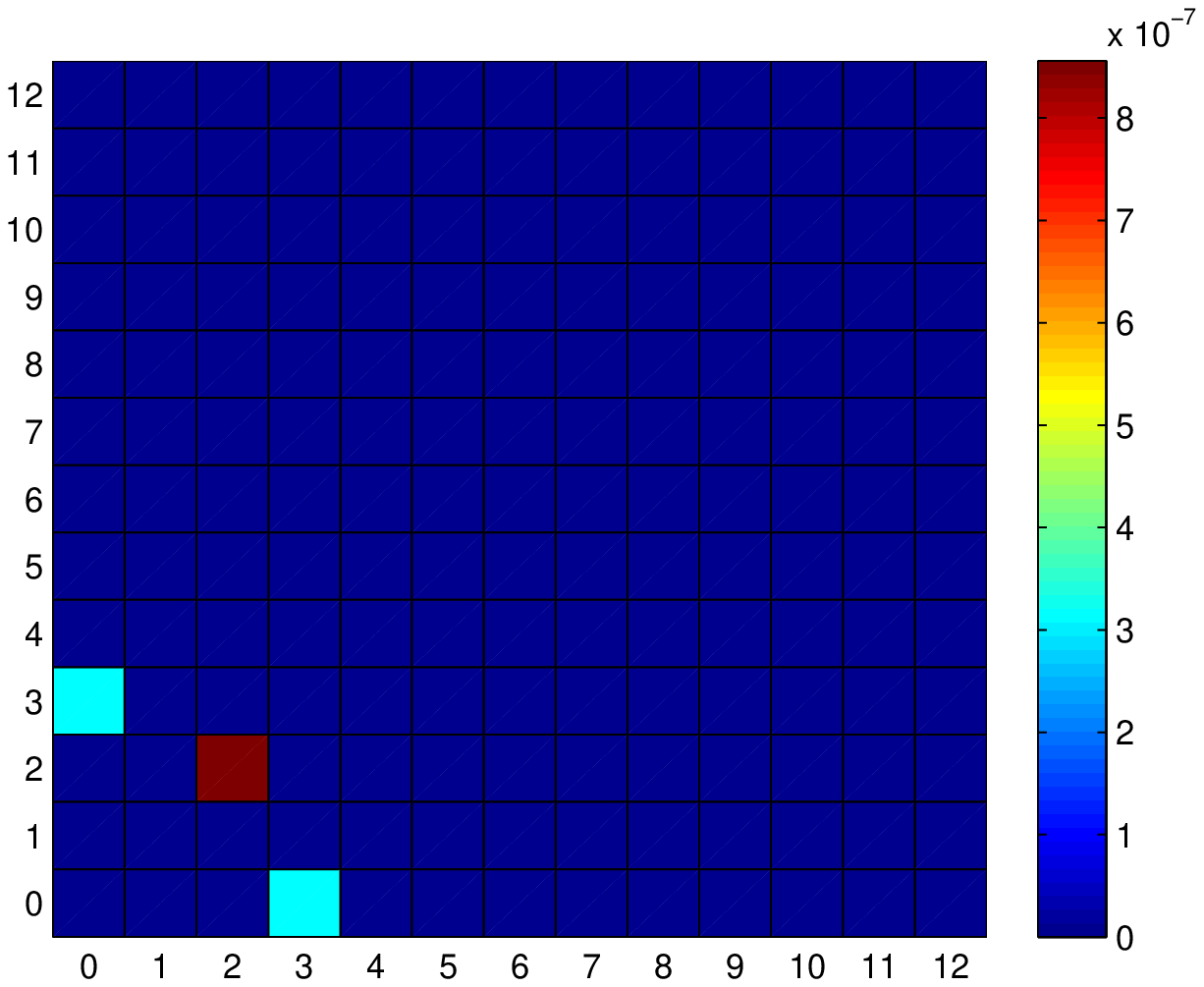}}
{\includegraphics[height=3.8cm, width=6cm]{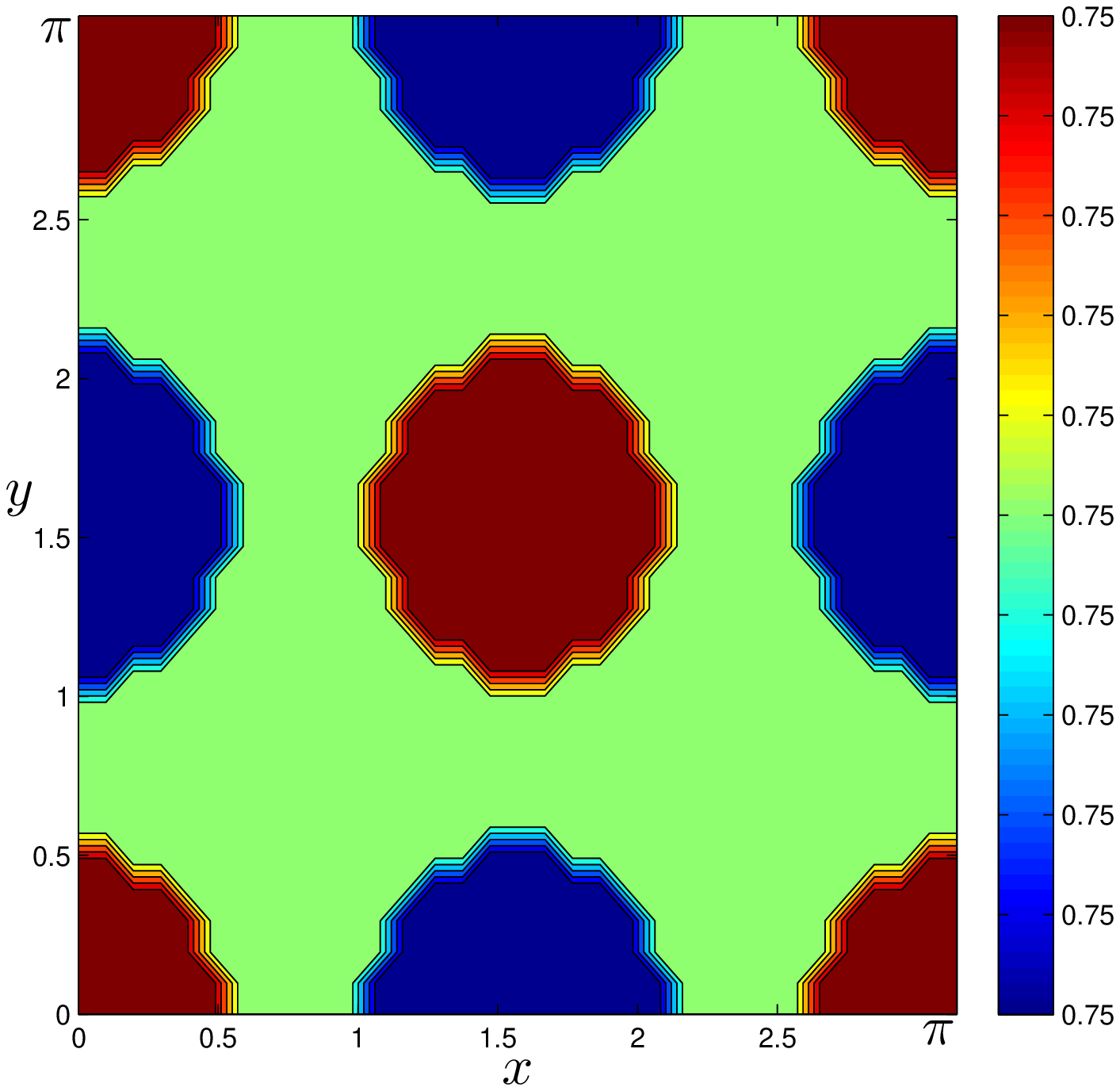}}
{\includegraphics[height=3.8cm, width=6cm]{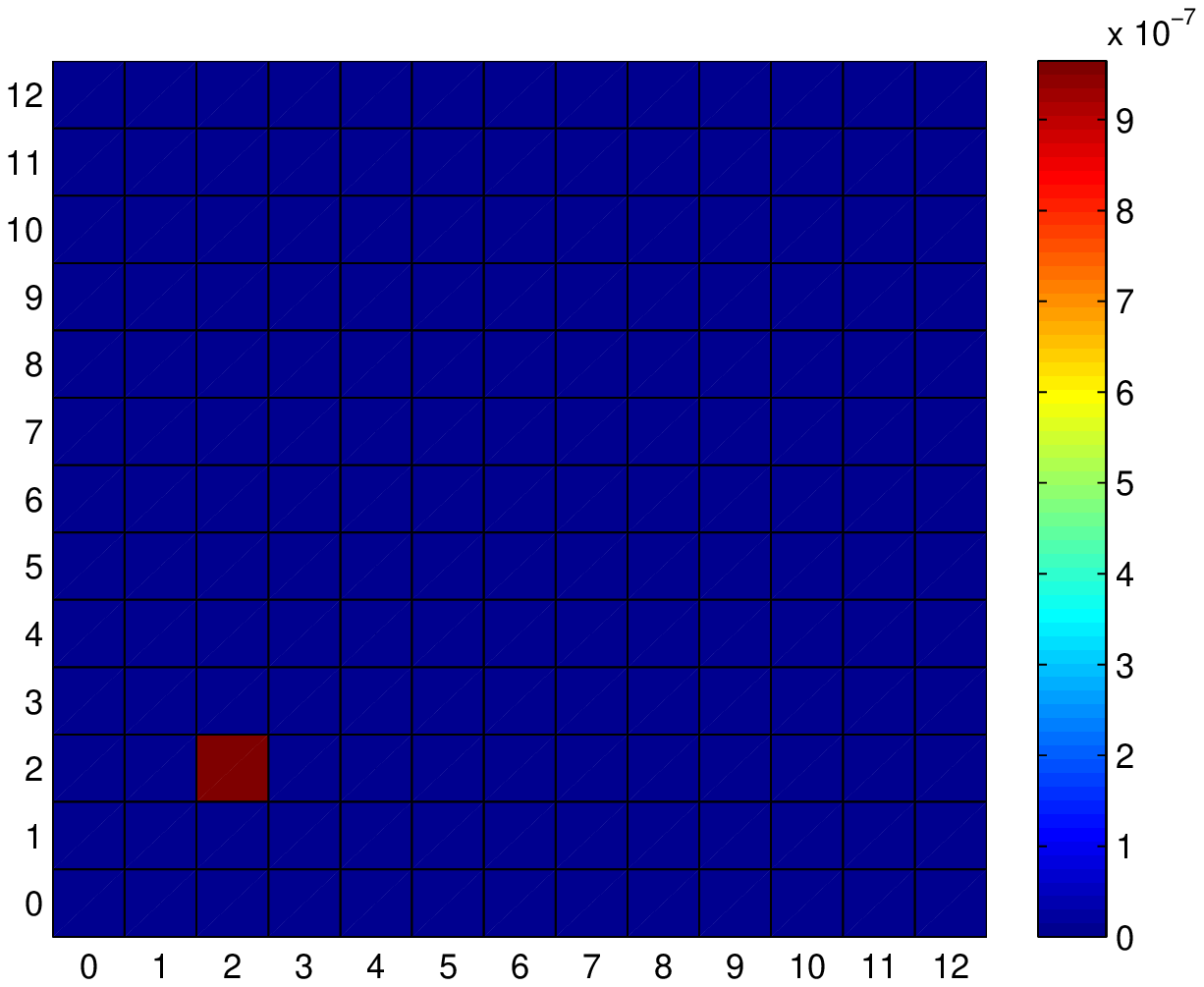}}
\subfigure[]{\includegraphics[height=3.8cm, width=6cm]{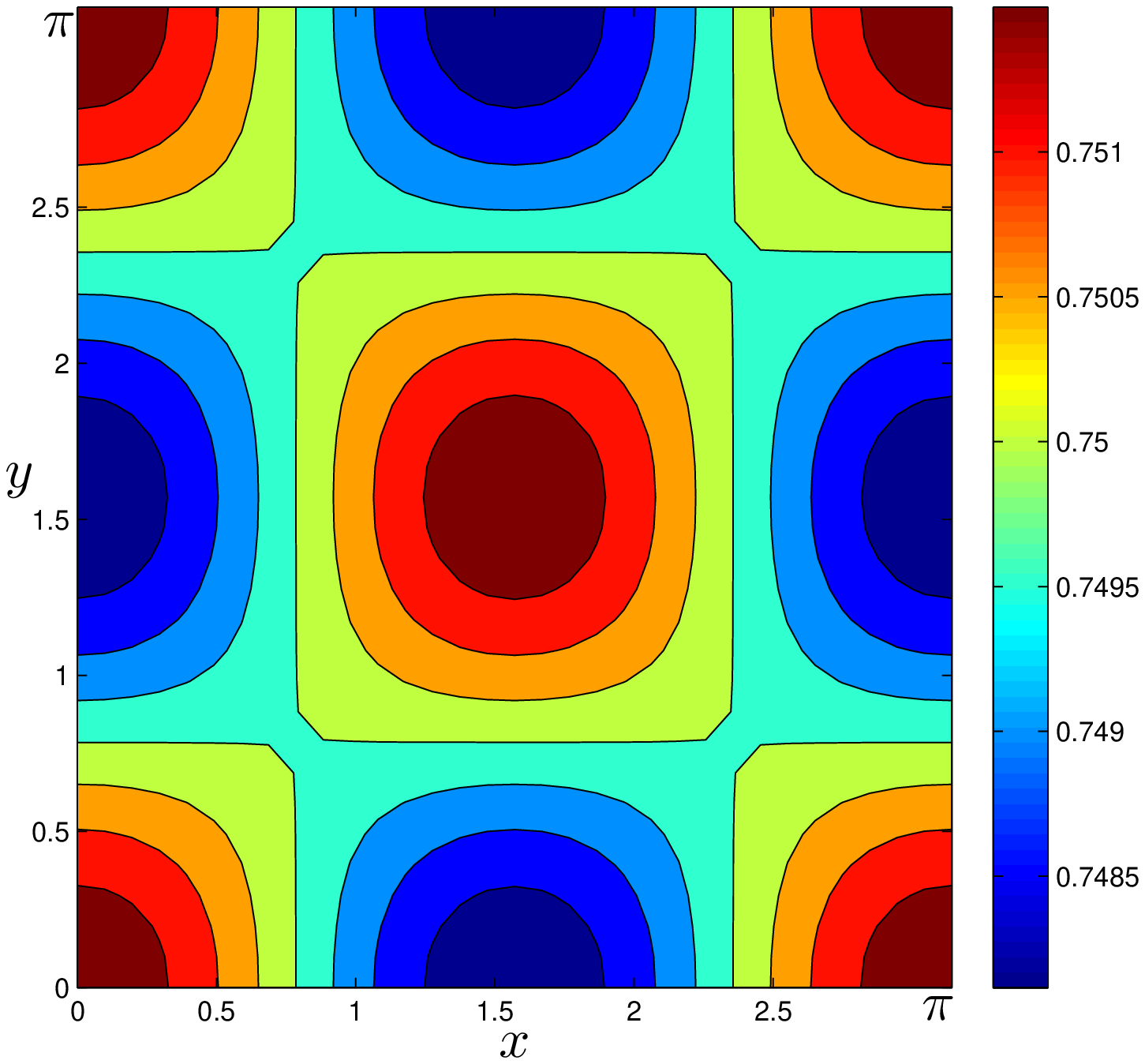}}
\subfigure[]{\includegraphics[height=3.8cm, width=6cm]{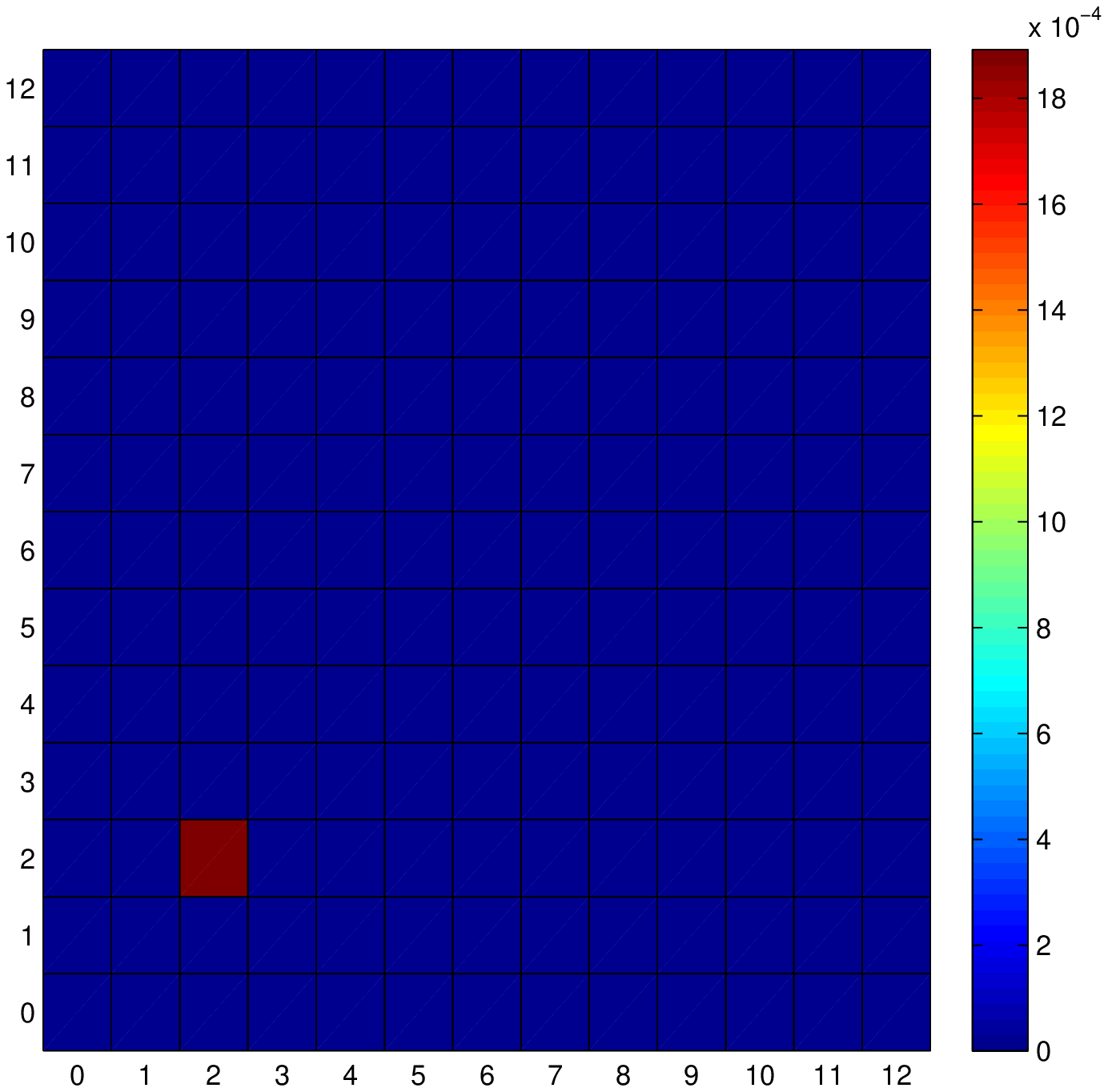}}
\caption{\label{isteresi_sub} Hysteresis cycle in the subcritical case. (a) The numerical solution $u$ of the full system \eqref{model}. (b) Spectrum of the numerical solution. (The parameters are  chosen  in Fig.~\ref{diagr_bif_sub_2D}).
}
\end{center}
\end{figure}

\vskip.2cm
\textbf{Case (2): $L<0$.} In the subcritical case the Stuart-Landau equation \eqref{SL3} does not admit any stable equilibrium and, pushing the analysis up to $O(\varepsilon^5)$ as in a one-dimensional domain,
we recover the quintic Stuart-Landau equation \eqref{SL5}. The stability properties of equation \eqref{SL5} have been already discussed in Case (2) of Section \ref{Sec2}.
The corresponding expected solution via WNL analysis is of the form \eqref{sol_m1}, where $A_{\infty}$ is the stable amplitude branch predicted by the quintic Stuart-Landau \eqref{SL5}.
Therefore, in this case, the system still supports rhombic patterns.
Here we just report a numerical experiment showing the emergence of the hysteresis phenomenon for a square pattern. We choose the domain size $L_x=L_y=\pi$ and the system parameters as given in Fig.~\ref{diagr_bif_sub_2D}.
The bifurcation diagram in Fig.~\ref{diagr_bif_sub_2D} shows that two qualitatively
different stable states coexist, therefore we expect the system supports a hysteresis cycle.
Choosing the initial deviation from the bifurcation value $\varepsilon=0.1$, in such a way that the unique discrete unstable mode is $\bar{k}_c^2=8$ and only the mode pair $(p,q)=(2,2)$ satisfies the condition \eqref{multip}, the numerical simulation in the first plot of Fig.\ref{isteresi_sub} shows the formation of a square pattern (as predicted via WNL analysis). Following the order and the direction of the arrows in Fig.~\ref{diagr_bif_sub_2D},
the corresponding numerical simulations of the full system shows the hysteresis cycle in Fig.~\ref{isteresi_sub}.
\begin{table}[h]
\centering
\begin{tabular}{c c c }
\hline
Modes & Numerical solution & Approximated solution\\
\hline
$\cos(2x)\cos(2y)$ & $0.02295$ & $0.02608$\\
$\cos(4x)$ & $0.00023$ & $0.00035$\\
$\cos(4y)$ & $0.00023$ & $0.00035$\\
$\cos(4x)\cos(4y)$ & $0.00005$ & $0.00004$\\
\hline
\end{tabular}
\caption{Squares in the supercritical case. The parameters are chosen as in Fig.~\ref{square_sup}}
\label{tab_square}
\end{table}
\clearpage

\subsection{Multiple eigenvalue, $m=2$, and the no-resonance condition holds}\label{double_nores}
In this section we assume that the multiplicity of the eigenvalue $\lambda(\bar{k}_c^2)$ is $m=2$ and the following no-resonance condition
holds:
\begin{eqnarray}
 \phi_i+\phi_j\neq \phi_j   \quad  &\mbox{or}&
\quad\psi_i-\psi_j\neq \psi_j \nonumber \\
 & \mbox{and}&  \label{4.13} \\
\phi_i-\phi_j\neq\phi_j \quad  &\mbox{or}& \quad
\psi_i+\psi_j\neq\psi_j   \nonumber
\end{eqnarray}
with $i,j=1,2$ and $i\neq j$. Performing the weakly nonlinear analysis and employing the Fredholm solvability condition at
$O(\varepsilon^3)$ to the vector system \eqref{sequence_3} leads to
the following two coupled Landau equations for the amplitudes $A_1$ and $A_2$:\\
\begin{subequations}\label{4.19}
\begin{eqnarray}\label{4.19a}
\frac{dA_1}{dT}&=&\sigma A_1-L_1 A_1^3+R_1 A_1 A_2^2,\\
\frac{dA_2}{dT}&=&\sigma A_2-L_2 A_2^3+R_2 A_1^2\
A_2.\label{4.19b}
\end{eqnarray}
\label{no_risonance_system}
\end{subequations}
All the details on the parameters of the system \eqref{4.19} are given in Appendix \ref{AppB}.
The stationary states of system \eqref{4.19} are the trivial equilibrium and the points $P_1^\pm\equiv\left(\pm\sqrt{\frac{\sigma}{L_1}},0\right)$, $P_2^\pm\equiv\left(0,\pm\sqrt{\frac{\sigma}{L_2}}\right)$ and
$P_3^{(\pm,\pm)}\equiv\left(\pm\sqrt{\frac{\sigma(L_2+R_1)}{L_1L2-R_1R_2}},\pm\sqrt{\frac{\sigma(L_1+R_2)}{L_1L2-R_1R_2}}\right)$ and their stability properties are summarized in Table \ref{stab_2_no_ris}.
\vskip.5cm
\begin{center}
\begin{tabular}{ccc}
  \hline
  $\ $ & Existence & Stability \\
  \hline\\
  $P_1^{\pm}$ & $L_1>0$ & $L_1+R_{2}<0$ \\
  \ \\
  \hline\\
  $P_2^{\pm}$ & $L_2>0$ & $L_2+R_{1}<0$ \\
   \ \\
  \hline\\
  $\ $ &
  $
  \left\{
  \begin{array}{lll}
  L_1L_2-R_1R_2<0,\\
  L_1+R_2<0\\
  L_2+R_1<0
  \end{array}
  \right.
  $
  & always unstable \\
  $P_3^{(\pm, \pm)}$ & or & $\ $ \\
  $\ $ &
  $
  \left\{
  \begin{array}{lll}
  L_1L_2-R_1R_2>0,\\
  L_1+R_2>0\\
  L_2+R_1>0
  \end{array}
  \right.
  $
  & $L_1R_1+L_2R_2+2L_1L_2<0$ \\
  \ \\
  \hline
\end{tabular}\label{stab_2_no_ris}
\end{center}
\vskip.5cm
When the system \eqref{4.19} admits at least one stable equilibrium $(A_{1\infty}, A_{2\infty})$, the long-time behavior of the solution of the reaction diffusion system \eqref{model} is given by:
\begin{equation}\label{m2nores}
\mathbf{w}=\varepsilon \ro \sum_{i=1}^2 A_{i\infty}\cos(\phi_i x)\cos(\psi_i y)+O(\varepsilon^2).
\end{equation}
Depending on the values of $(A_{1\infty}, A_{2\infty})$ and $(\phi_i,\psi_i)$, the solution \eqref{m2nores} describes the following types of patterns:
\begin{itemize}
\item[(i)]\noindent if $P_1^\pm$ or $P_2^\pm$ is stable, the solutions in \eqref{m2nores} are the rhombic spatial patterns
described in Section \ref{simple};
\item[(ii)]\noindent if $P_3^{(\pm, \pm)}$ is stable, more complex structures arise due to the interaction
of different modes $\phi_i$, $\psi_i$, the so-called mixed-mode patterns.
\end{itemize}
\begin{figure}[t]
\begin{center}
\subfigure[]{\includegraphics[height=6cm, width=7cm]{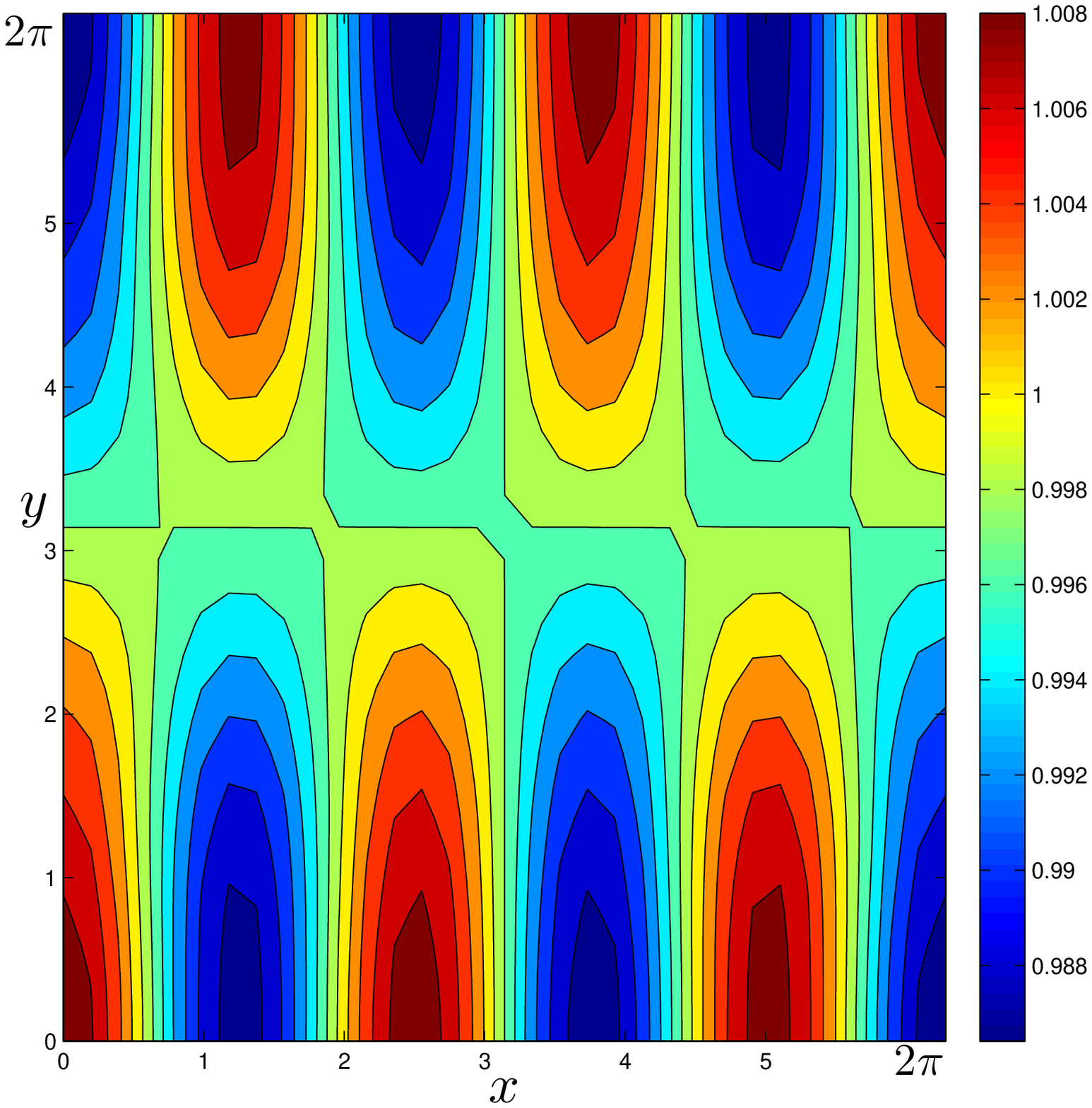}}
\hspace{1cm}
\subfigure[]{\includegraphics[height=6cm, width=7cm]{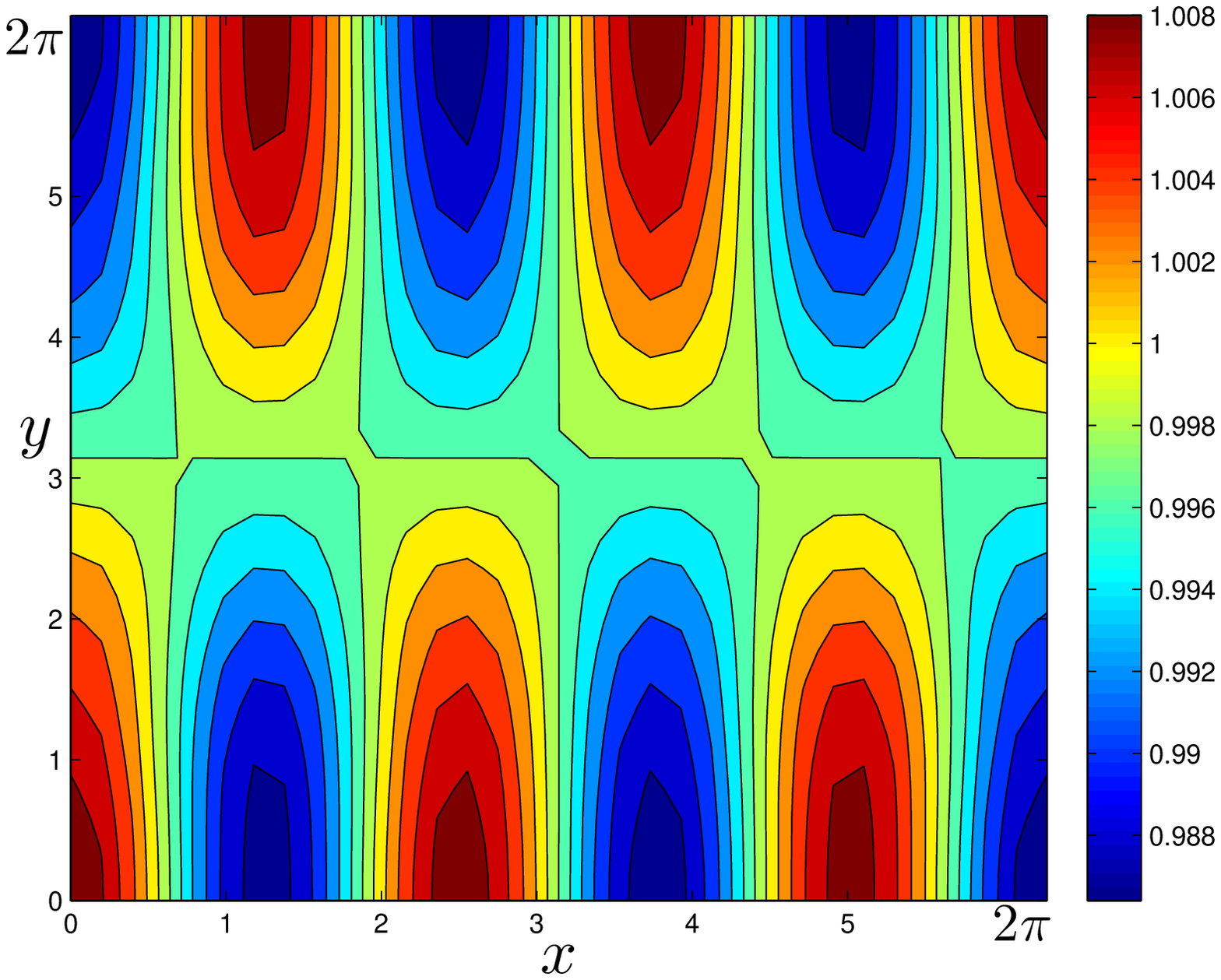}}
\end{center}
\caption{\label{rectangle} Rhombic pattern. Comparison between the numerical solution (a) and the weakly nonlinear first order approximation of the solution (b). The
system parameters are chosen as follows: $a=0.2$, $b=0.798$, $\gamma=21$, $d_u=1$, $d_v=1$, where $d_c=11.3960$ and $\varepsilon= 0.1$.}
\end{figure}
In the following numerical experiments, we will show how the WNL analysis predicts different types of patterns as described in (i)-(ii) and in the simulations we fix the square domain $L_x=L_y=2\pi$.
\vskip.2cm
\textbf{Case (i): Rhombic patterns.} We choose the system parameters as in Fig.~\ref{rectangle} and the deviation from the bifurcation values $\varepsilon= 0.1$.
Picking up these values the only admitted unstable mode is $\bar{k}_c^2=6.5$ and the conditions in \eqref{multip} are satisfied by the two couples $(1,5)$ and $(5,1)$ (for which the no-resonance conditions \eqref{4.13} hold).
The only stable states of the amplitude system \eqref{no_risonance_system} are $P_1^{\pm}$ and $P_2^{\pm}$ and the predicted solution via WNL analysis (depending on the randomic initial data), respectively, the following:
\begin{equation}
\label{rettangoli}
\begin{split}
&\mathbf{w}=\varepsilon \ro  A_{1\infty}\cos(2.5 x)\cos(0.5 y)+O(\varepsilon^2),\\
&\mbox{or}\\
&\mathbf{w}=\varepsilon \ro  A_{2\infty}\cos(0.5 x)\cos(2.5 y)+O(\varepsilon^2),
\end{split}
\end{equation}

where $A_{1\infty}$ is the nonzero coordinate of the points $P_1^\pm$, while $A_{2\infty}$ is the nonzero coordinate of the points $P_2^\pm$.
Our numerical tests starting from a random periodic perturbation of the equilibrium show that the solution evolves to the rectangular pattern predicted in \eqref{rettangoli}. In Fig. \ref{rectangle} we show the agreement (with $\varepsilon=0.1$) between the numerical solution and
the solution expected on the basis of the weakly nonlinear analysis.

\begin{figure}[t]
\begin{center}
\subfigure[]{\includegraphics[height=6.5cm, width=7cm]{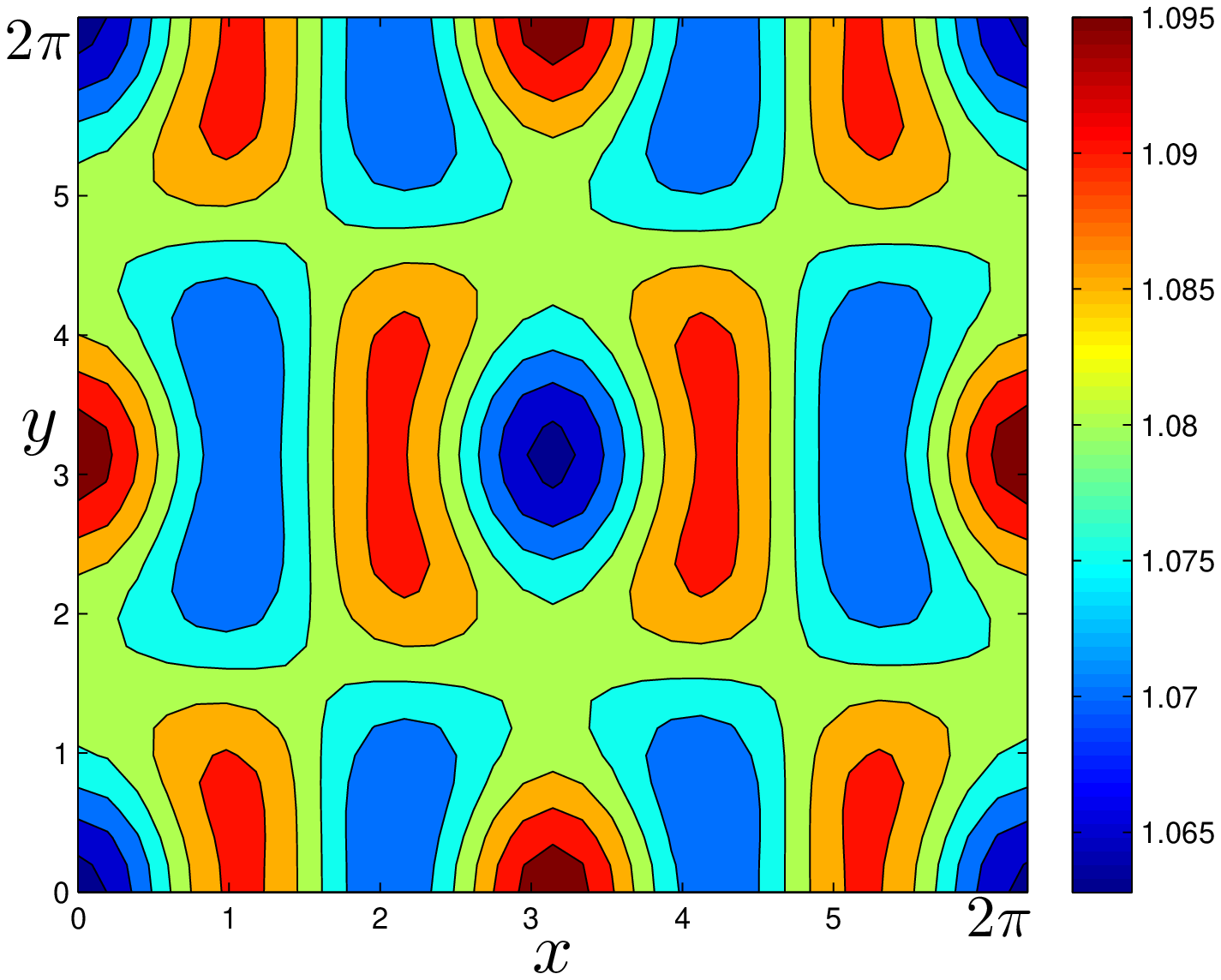}}
\hspace{1cm}
\subfigure[]{\includegraphics[height=6.5cm, width=7cm]{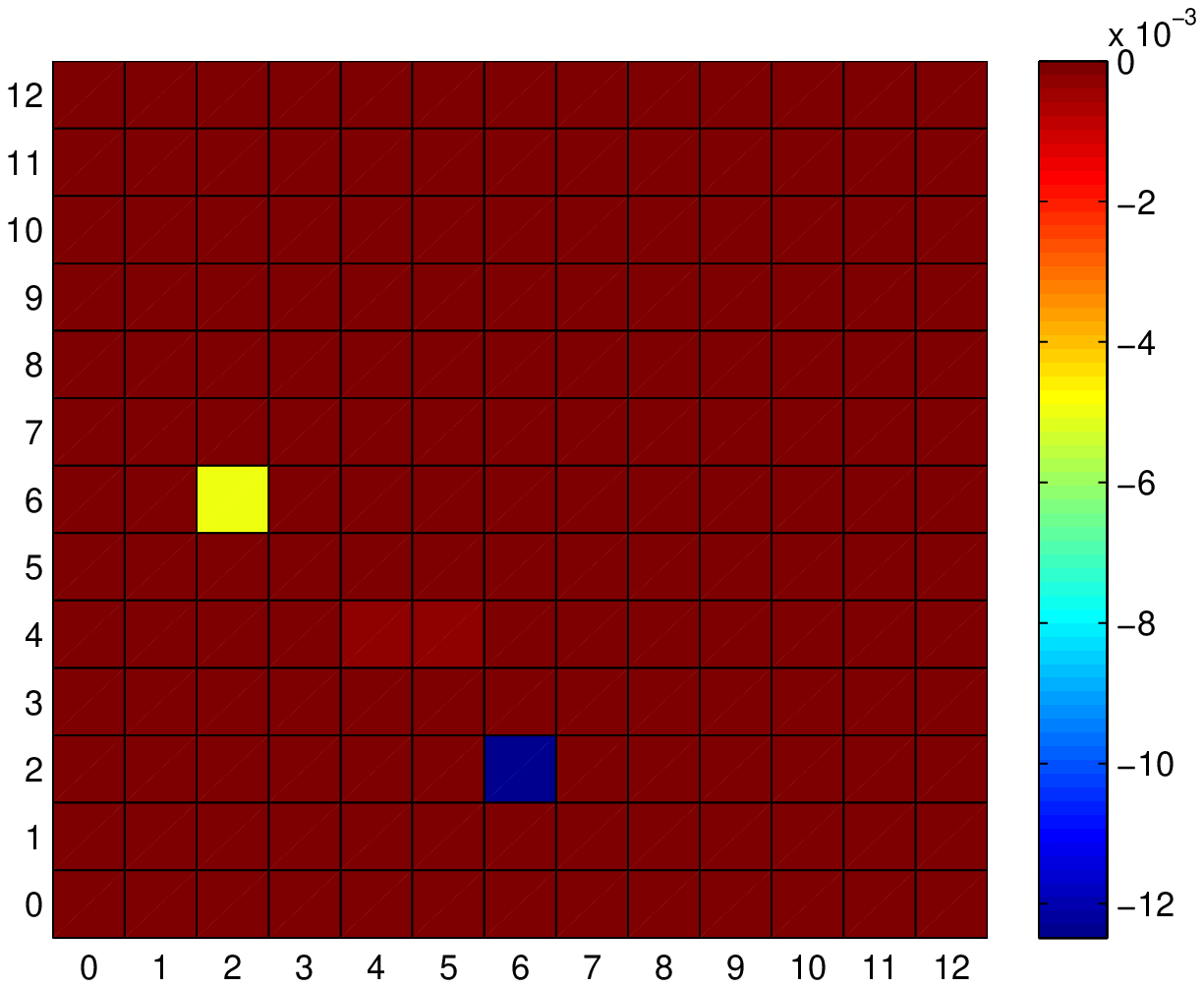}}
\end{center}
\caption{Mixed-modes pattern. (a) The numerical solution $u$ of the full system \eqref{model}. (b) Spectrum of the numerical solution.
$a=0.18$, $b=0.901$, $\gamma=30$, $d_u=1$, $d_v=1$, where $d_c=11.5168$ and $\varepsilon= 0.01$.}
\label{mixed}
\end{figure}

\vskip.2cm
\textbf{Case (ii): Mixed-modes patterns.}
Let us choose the parameter values as in Fig.~\ref{mixed} in such a way that only the most unstable discrete
mode $\bar{k}^2_c=10$ falls within the band of unstable modes allowed
by the boundary conditions. The two mode pairs
$(0,5)$ and $(2,3)$ satisfy the conditions \eqref{multip} and \eqref{4.13}. With this choice of the parameters the only stable equilibria of the system \eqref{4.19} are
$P_3^{(\pm, \pm)}$, therefore the predicted long-time solution is the following mixed-mode pattern:
\begin{equation}\label{m2nores_mix}
\textbf{w}=\varepsilon \ro\left(A_{1\infty}
\cos\left(2x\right)\cos\left(6y\right)+A_{2\infty}
\cos\left(6x\right)\cos\left(2y\right)\right)+O(\varepsilon^2),
\end{equation}
where $A_{1\infty}, A_{2\infty}$ are the coordinates of the point $P_3^{(\pm,\pm)}$.

Choosing the parameter as in Fig.~\ref{mixed}, we note a good agreement between the amplitude of the most excited modes ($\cos(x)\cos(3y)$ and $\cos(3x)\cos(y)$ have the same amplitude due to symmetry):  $0.0125$ and $0.0114$ for the numerical and approximated solution, respectively.
\vskip.2cm

\begin{figure}[t]
\begin{center}
\subfigure[]{\includegraphics[height=6.5cm, width=7cm]{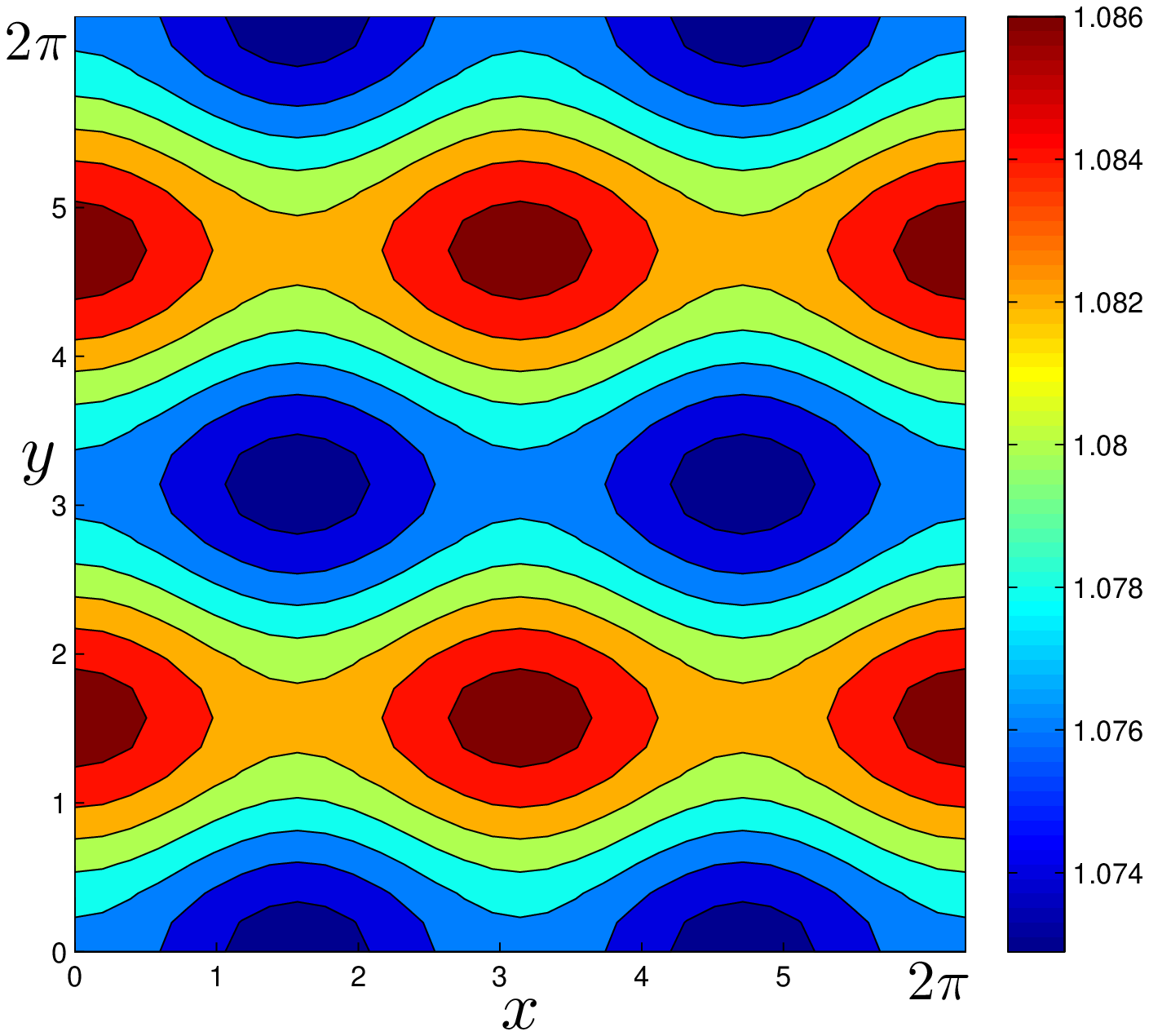}}
\hspace{1cm}
\subfigure[]{\includegraphics[height=6.5cm, width=7cm]{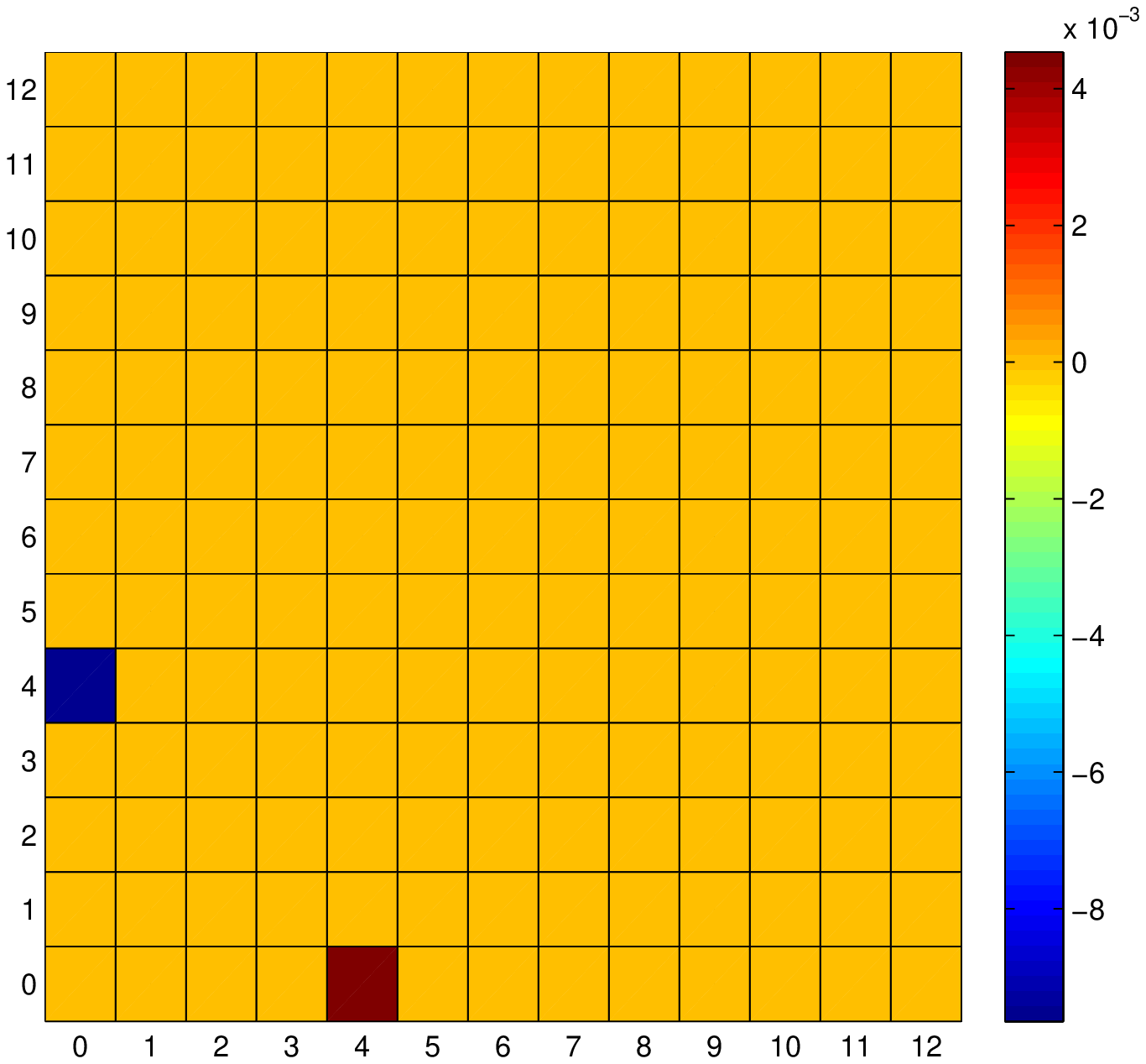}}
\end{center}
\caption{Mixed-modes pattern. (a) The numerical solution $u$ of the full system \eqref{model}. (b) Spectrum of the numerical solution.
$a=0.18$, $b=0.901$, $\gamma=30$, $d_u=1$, $d_v=1$, where $d_c=11.5168$ and $\varepsilon= 0.01$.}
\label{flux}
\end{figure}
In another numerical test, with the choice of the parameters as in Fig. \ref{flux}, the most unstable mode is $\bar{k}^2_c=4$ and the conditions in \eqref{multip} are satisfied by the two couples $(4,0)$ and $(0,4)$. The only stable equilibria are $P_3^{(\pm,\pm)}$ and the asymptotic solution is the following:
\begin{equation}
\textbf{w}=\varepsilon \ro\left(A_{1\infty}
\cos(2x)+A_{2\infty}\cos(2y)\right)+O(\varepsilon^2),
\end{equation}
In Fig.~\ref{flux} the numerical solution of the full system \eqref{model} is shown, together with its spectrum. The numerical solution is very close to the predicted WNL approximated solution, in particular the amplitude of the most unstable modes ($\cos(2x)$ and $\cos(2y)$ have the same amplitude due to symmetry) are $0.0071$ (for the numerical solution) and $0.0078$  (for the approximated solution).

\begin{figure}[htb]
\begin{center}
\subfigure[]{\includegraphics[height=6cm, width=7cm]{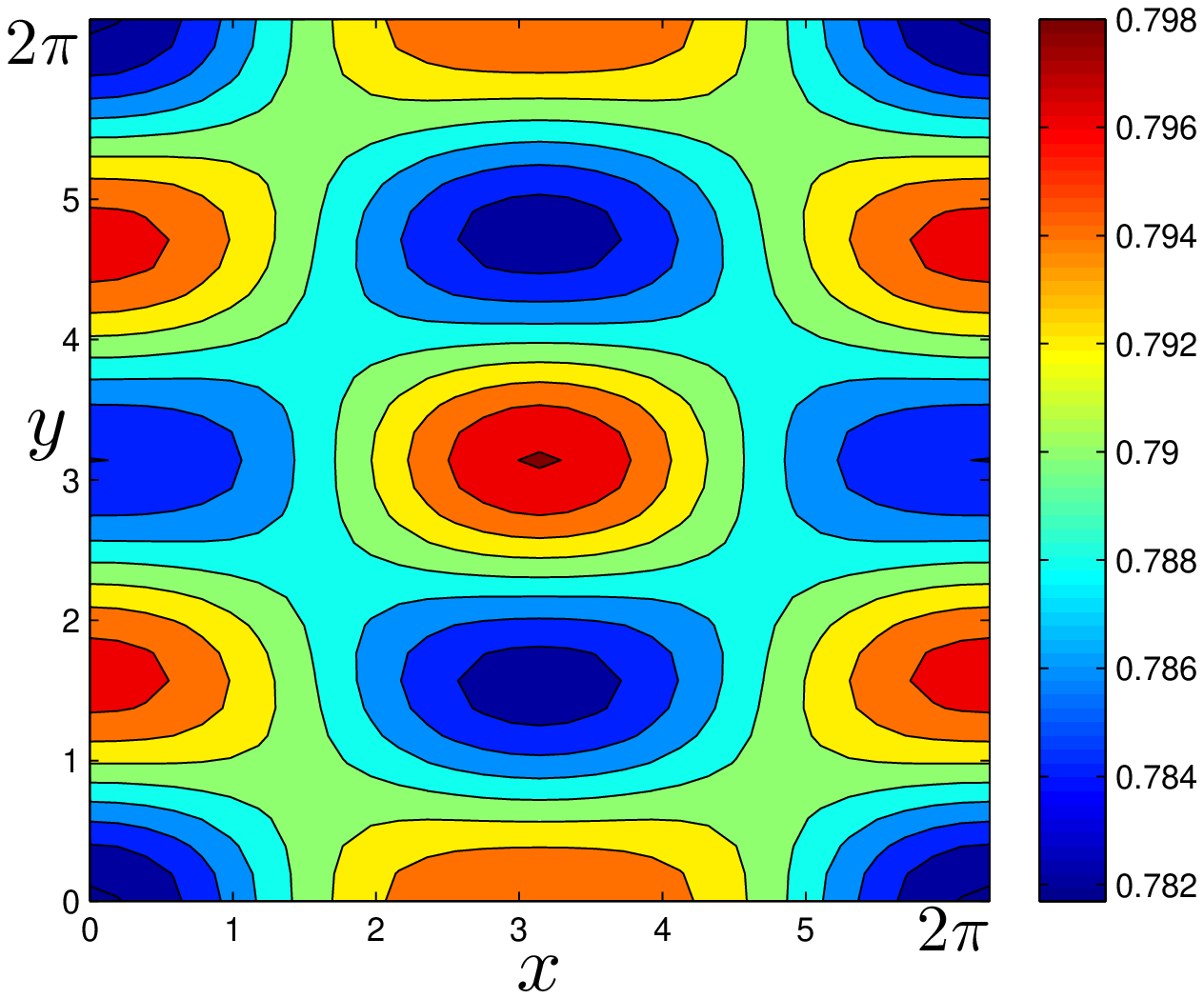}}
\hspace{1cm}
\subfigure[]{\includegraphics[height=6cm, width=7cm]{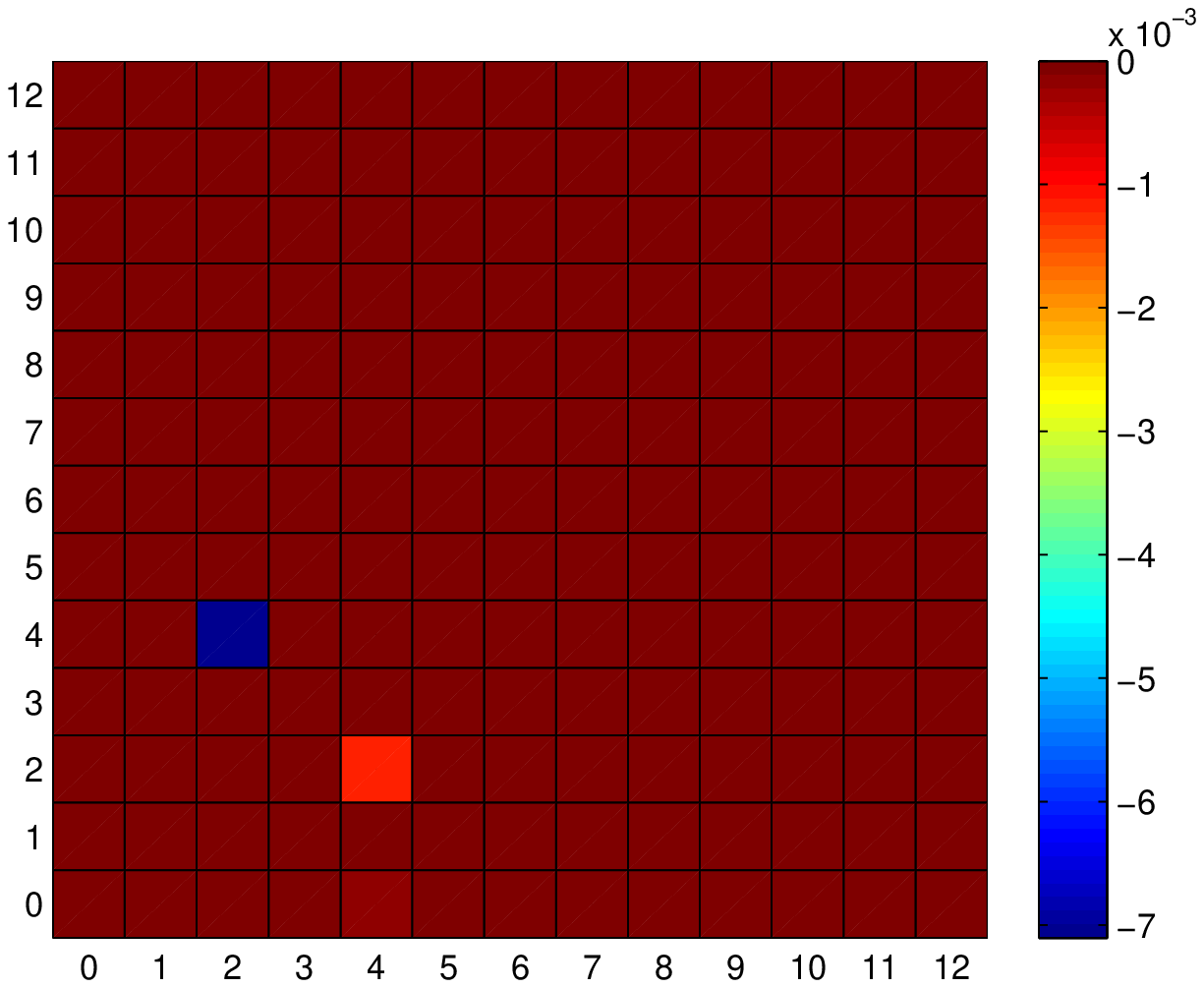}}
\end{center}
\caption{\label{supersquare} Super-squares pattern.  (a) The numerical solution $u$ of the full system \eqref{model}. (b) Spectrum of the numerical solution.
$a=0.19$, $b=0.6$, $\gamma=15.3$, $d_u=1$, $d_v=1$, where $d_c=6.8274$ and $\varepsilon= 0.1$.}
\end{figure}

\vskip.2cm
In a further numerical test picking the parameter values as in Fig.~\ref{supersquare}, the most unstable discrete mode is exactly $\bar{k}^2_c=5$ and the condition \eqref{multip} is satisfied by the two mode pairs $(2,4)$ and $(4,2)$. The WNL analysis predicts that the only stable equilibria are $P_3^{(\pm,\pm)}$ and that the asymptotic solution is the following:
\begin{equation}
\textbf{w}=\varepsilon \ro\left(A_{1\infty}
\cos\left(x\right)\cos\left(2y\right)+A_{2\infty}
\cos\left(2x\right)\cos\left(y\right)\right)+O(\varepsilon^2),
\end{equation}
where $A_{1\infty}, A_{2\infty}$ are the coordinates of the point $P_3^{(\pm,\pm)}$.
In Fig.~\ref{supersquare} the numerical solution of the full system \eqref{model} is shown, together with its spectrum. The numerical solution is very close to the predicted WNL approximated solution, in particular the amplitude of the most unstable modes ($\cos(x)\cos(2y)$ and $\cos(2x)\cos(y)$ have the same amplitude due to symmetry) are $0.0071$ (for the numerical solution) and $0.0072$  (for the approximated solution).

\subsection{Multiple eigenvalue, $m=2$, and the resonance condition holds}\label{double_res}
Let us assume that the multiplicity of the eigenvalue $\lambda(\bar{k}_c^2)$ is $m=2$ and the following resonance
conditions are satisfied:
\begin{eqnarray}
 \phi_i+\phi_j = \phi_j   \quad  &\mbox{and}&
\quad\psi_i-\psi_j = \psi_j \nonumber \\
 & \mbox{or}&  \label{res_si} \\
\phi_i-\phi_j  =\phi_j \quad  &\mbox{and}& \quad
\psi_i+\psi_j = \psi_j   \nonumber
\end{eqnarray}
with $i, j=1, 2$  and $i\neq j$.

\noindent In what follows we hypothesize, without loss of generality, that the second condition in \eqref{res_si} holds with $i=2$ and $j=1$, therefore
taking into account the relation in \eqref{multip}, we obtain $\phi_2=2\phi_1$, $\psi_2=0$,
$\psi_1= \sqrt{3}\phi_1$, $\phi_1=  \kcb/2$ and also $L_y=\sqrt{3}L_x$.
The Fredholm solvability condition on the system \eqref{sequence_2}  at $O(\varepsilon^2)$ allows to derive a ODEs system for the amplitudes which
does not admit stable equilibrium in any parameter regimes and the WNL analysis has to be pushed to
higher order, see \cite{GLS13}.
At $O(\varepsilon^3)$ the following system
for the amplitudes $A_1$ and $A_2$ is found:
\begin{equation}\label{4.32}
\begin{split}
\frac{dA_1}{dT}=&\,\sigma_1 A_1-L_1 A_1A_2+R_1 A_1^3+S_1A_1 A_2^2,\\
\frac{dA_2}{dT}=&\,\sigma_2 A_2-L_2A_1^2+R_2 A_2^3+S_2A_1^2\, A_2,
\end{split}
\end{equation}
where ${\sigma_i}$ and ${L_i}$ are $O(1)$ perturbation of the coefficients of the amplitude equations found
at $O(\varepsilon^2)$, while ${R_i}$ and $S_i$ are $O(\varepsilon)$.\\
In Appendix \ref{AppC} one can find the expression of the parameters of system \eqref{4.32}.\\
The emerging long-time solution of the full system \eqref{model} is approximated by:
\begin{equation}
\label{4.15}
\mathbf{w}=\varepsilon \ro (A_{1\infty}\cos(\phi_1 x)\cos(\psi_1 y)
+A_{2\infty}\cos(\phi_2 x)\cos(\psi_2 y))+O(\varepsilon^2),
\end{equation}
where $(A_{1\infty}, A_{2\infty})$ is a stable state of the system \eqref{4.32}.\\
The stability analysis of system \eqref{4.32} is almost rich and provides as stationary states the equilibria $R^{\pm}\equiv(0, \pm\sqrt{-{{\sigma}_2}/{{S}_2}})$ and the six roots $H^{\pm}_i\equiv(A_{1i}^\pm, A_{2i}),\ i=1,2,3,$ of the following system:
$$
\begin{cases}
&A_2^3(S_1S_2-R_1R_2)+A_2^2({L}_1R_2+{L}_2R_1)+A_2\left(S_1{\sigma}_2-{L}_1{L}_2-R_2{\sigma}_1\right)+{L}_2{\sigma}_1=0,\\
&A_1^2=\displaystyle\frac{1}{S_1}\left(-R_1A_2^2\right.+\left.{L}_1A_2-{\sigma}_1 \right).
\end{cases}
$$

\begin{figure}[t]
\begin{center}
\subfigure[]{\includegraphics[height=5cm, width=6.5cm]{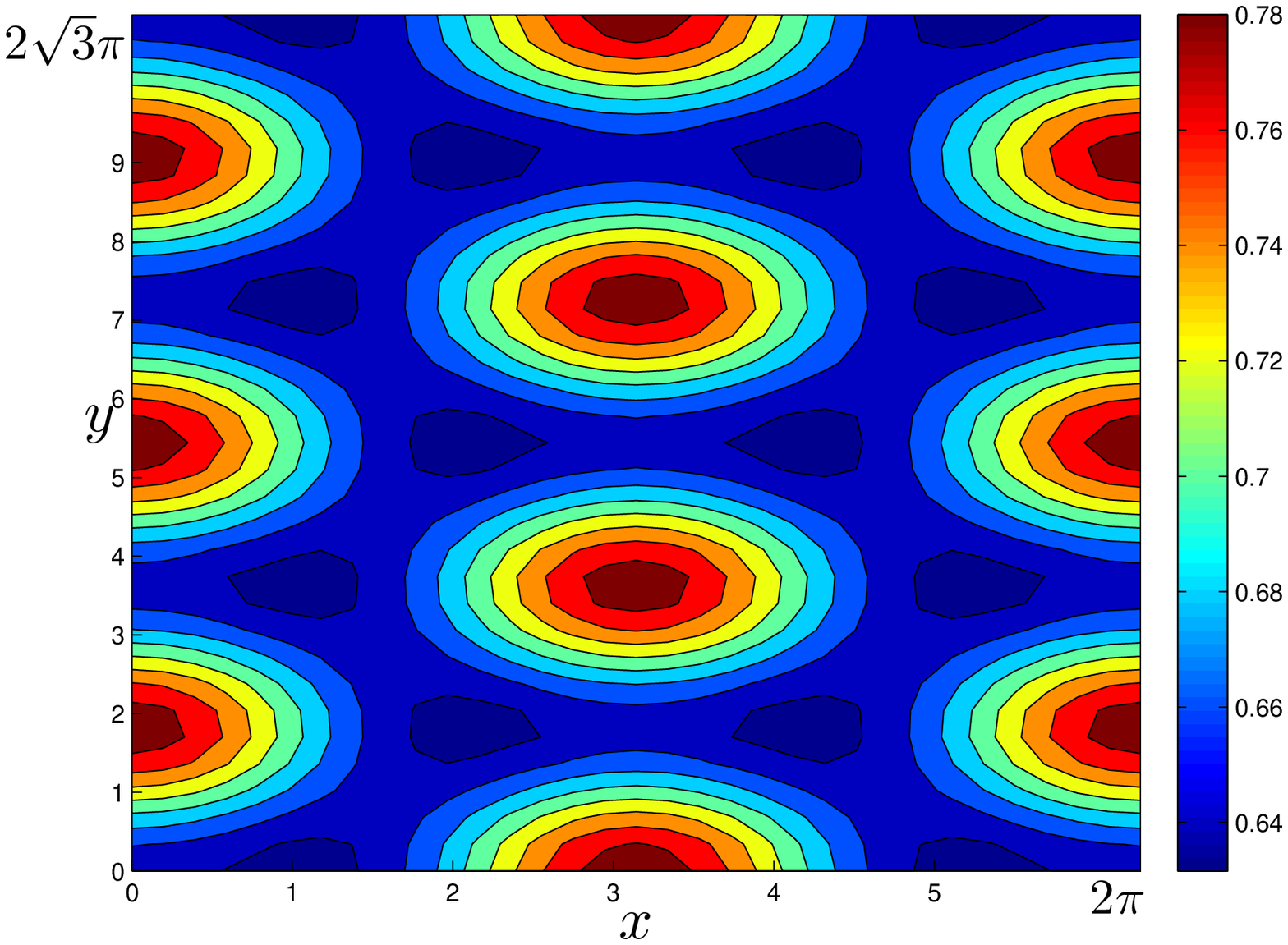} }
\subfigure[]{\includegraphics[height=5cm, width=6.5cm]{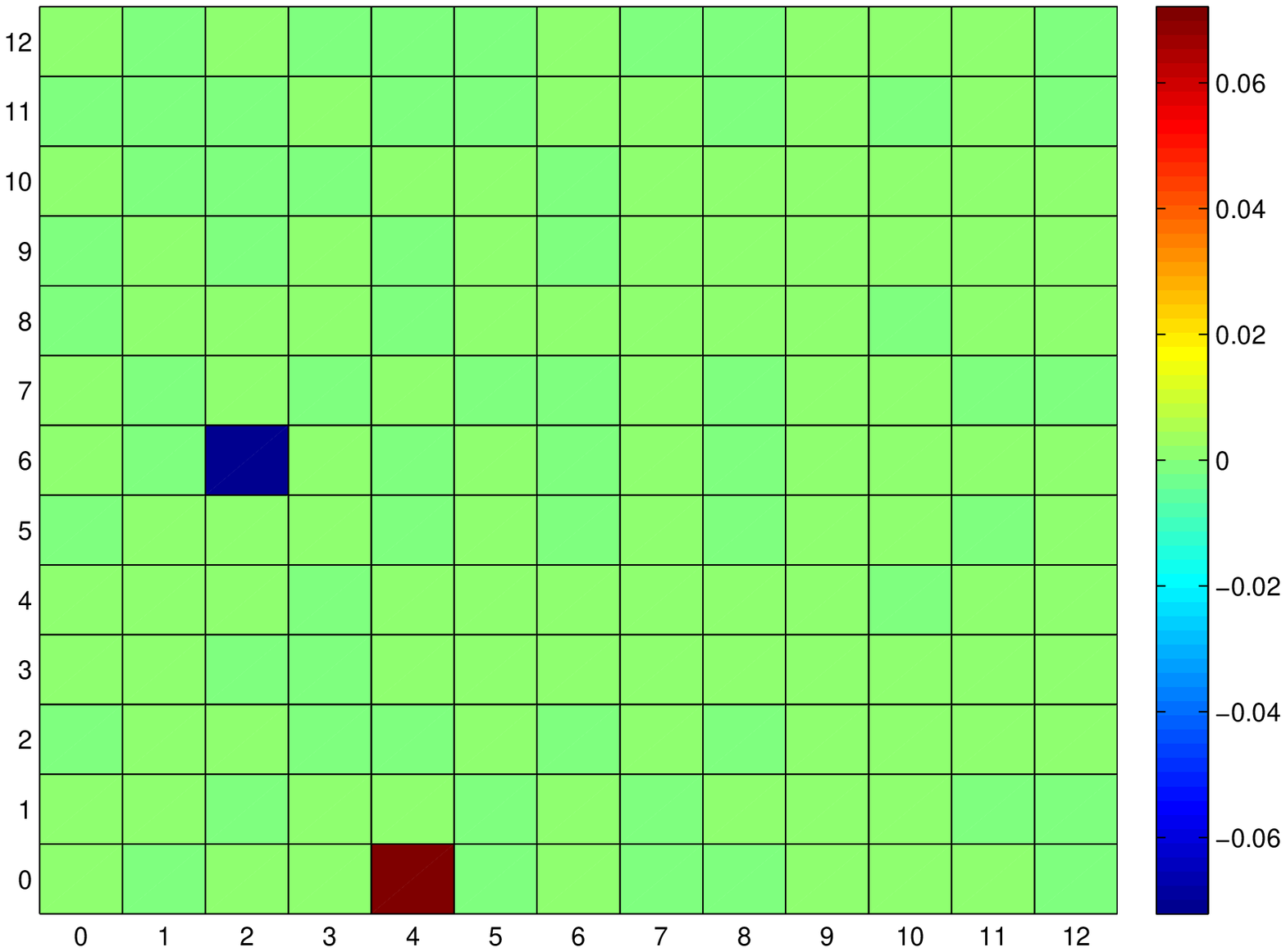}}
\end{center}
\caption{\label{hexa} Hexagonal pattern. (a) The numerical solution $u$ of the full system \eqref{model}. (b) Spectrum of the numerical solution. The parameters are chosen as: $a=0.2$, $b=0.485$, $\gamma=14$, $d_u=1$, $d_v=1$, where $d_c=6.7434$ and $\varepsilon= 0.1$.}
\end{figure}
\begin{figure}[h!]
\begin{center}
\subfigure[]{\includegraphics[height=5cm, width=6.5cm]{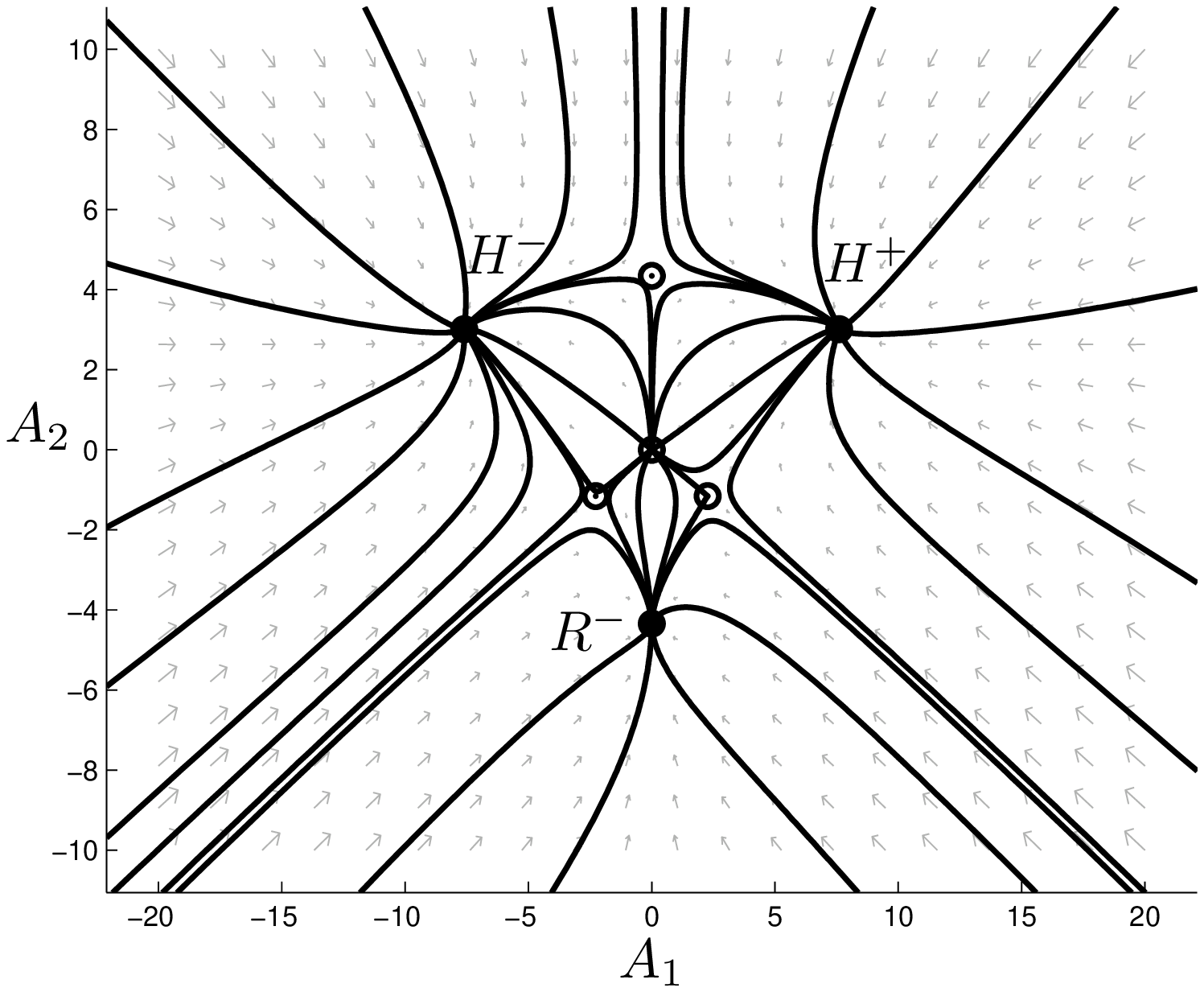}}
\subfigure[]{\includegraphics[height=5cm, width=6.5cm]{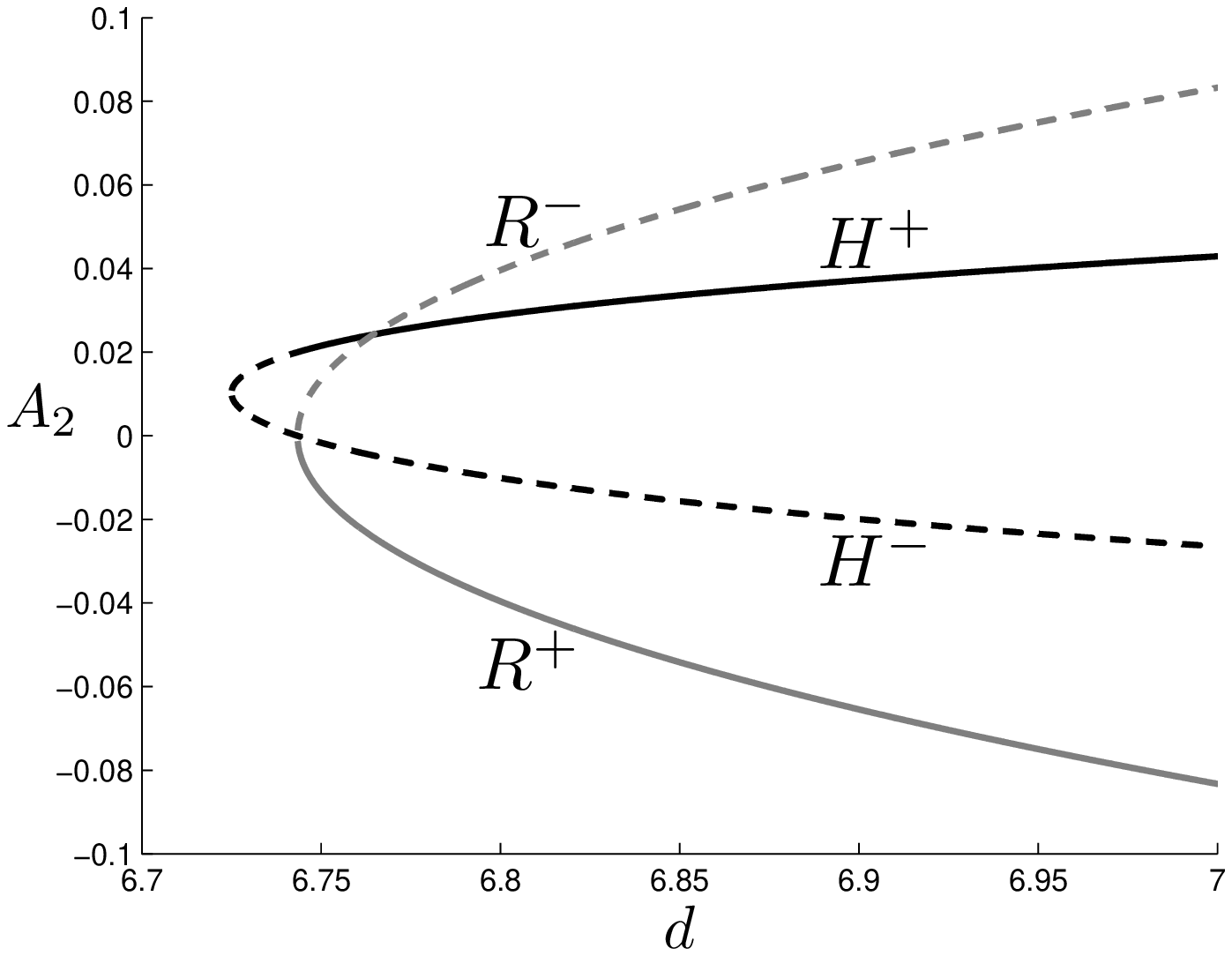}}
\end{center}
\caption{(a) The basin of attraction of system \eqref{4.32} and (b) the bifurcation diagram. The parameters are chosen as in Fig.\ref{hexa}.}
\label{bif_hexa}
\end{figure}

Whether $R^{\pm}$ or $H^{\pm}_i$ exist real and stable, the corresponding asymptotic solution \eqref{4.15} predicts respectively a roll pattern or a hexagonal pattern.
In the numerical experiment shown in Fig.~\ref{hexa}, we pick the parameters in such a way that, in the rectangular domain with $L_x=2\pi$ and $L_y=2\sqrt{3}\pi$, the only admitted discrete unstable mode is $\bar{k}_c^2=4$.
\begin{figure}[t]
\begin{center}
{\includegraphics[height=3.1cm, width=6.5cm]{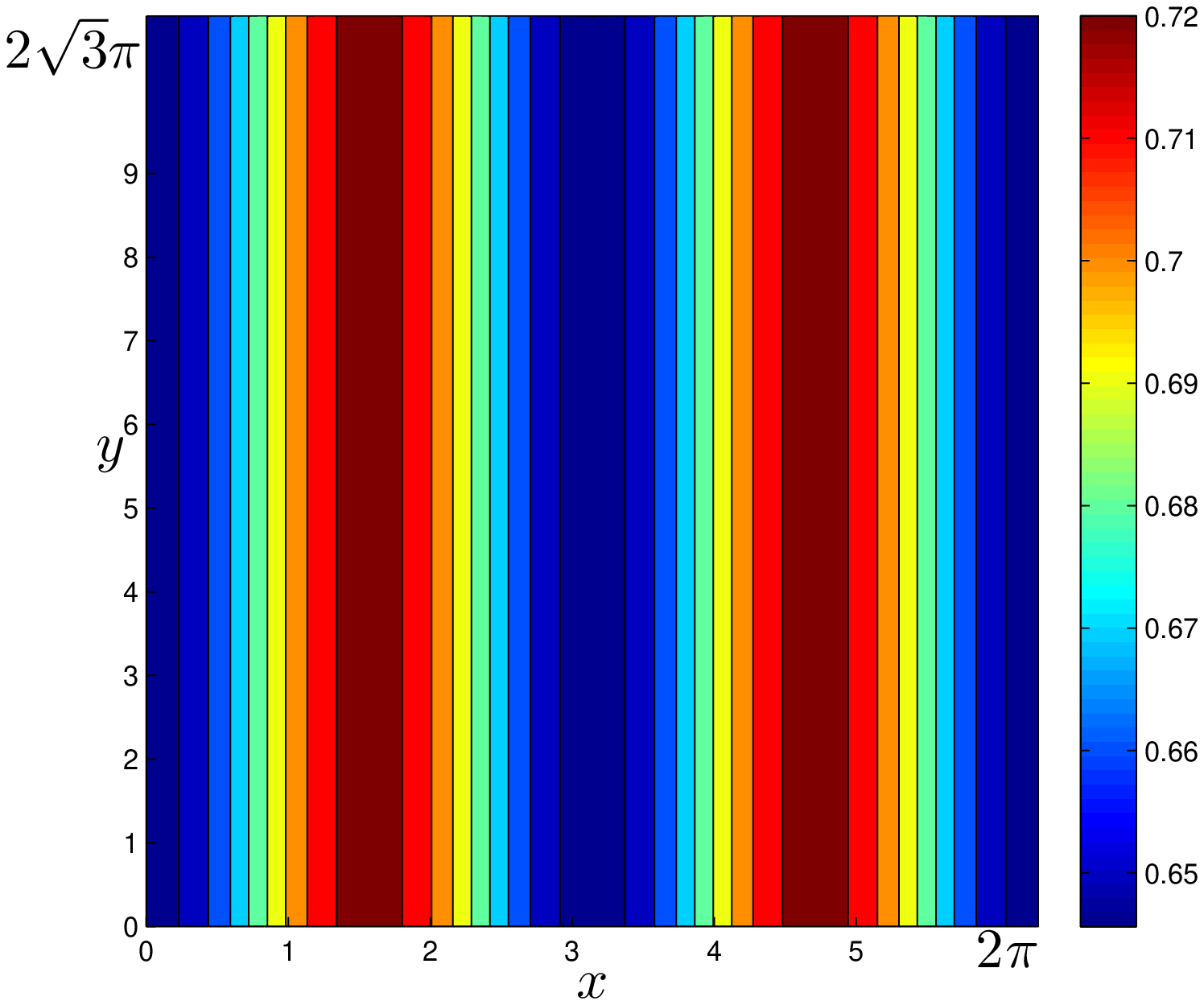}}
{\includegraphics[height=3.1cm, width=6.5cm]{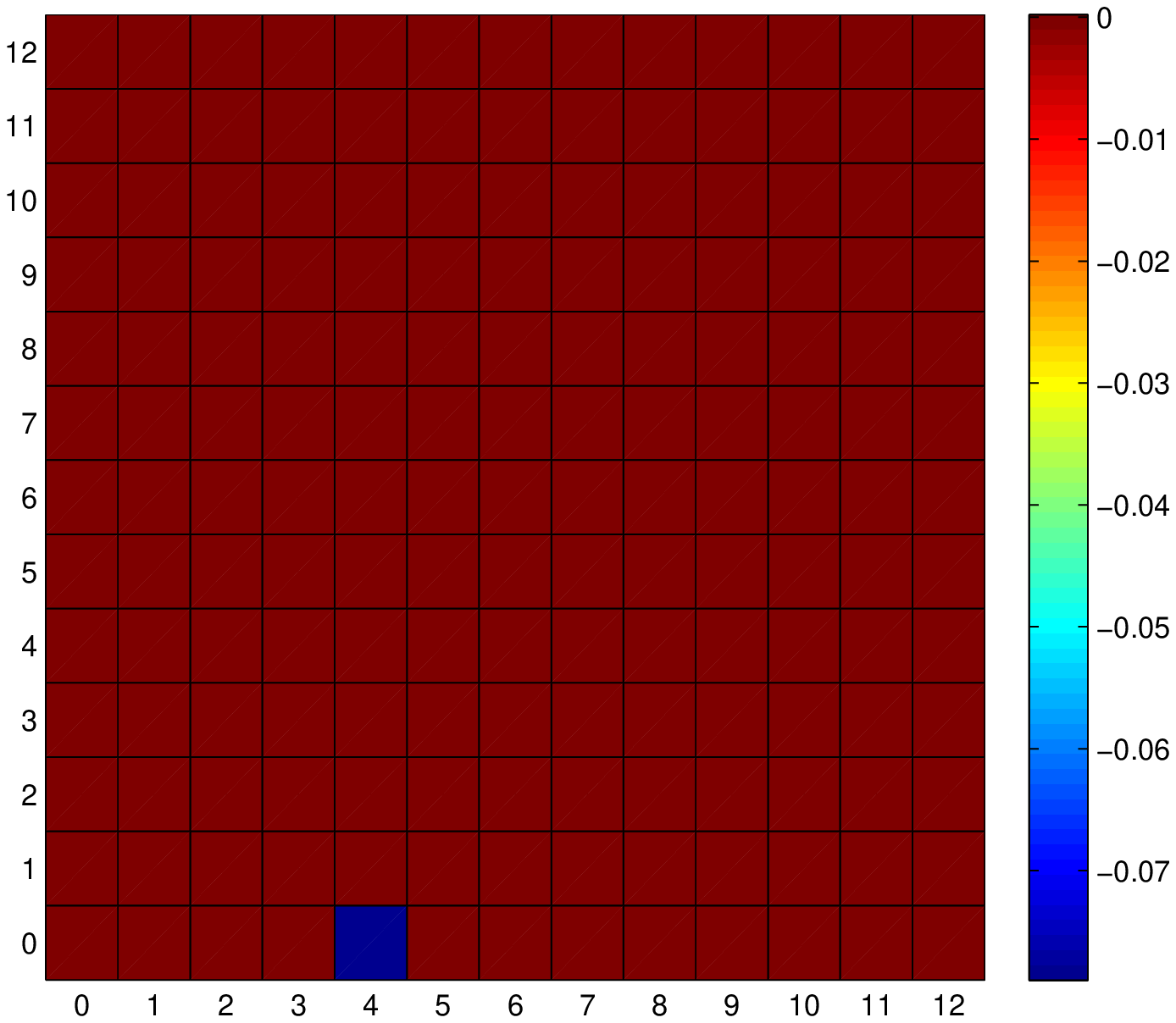}}
{\includegraphics[height=3.1cm, width=6.5cm]{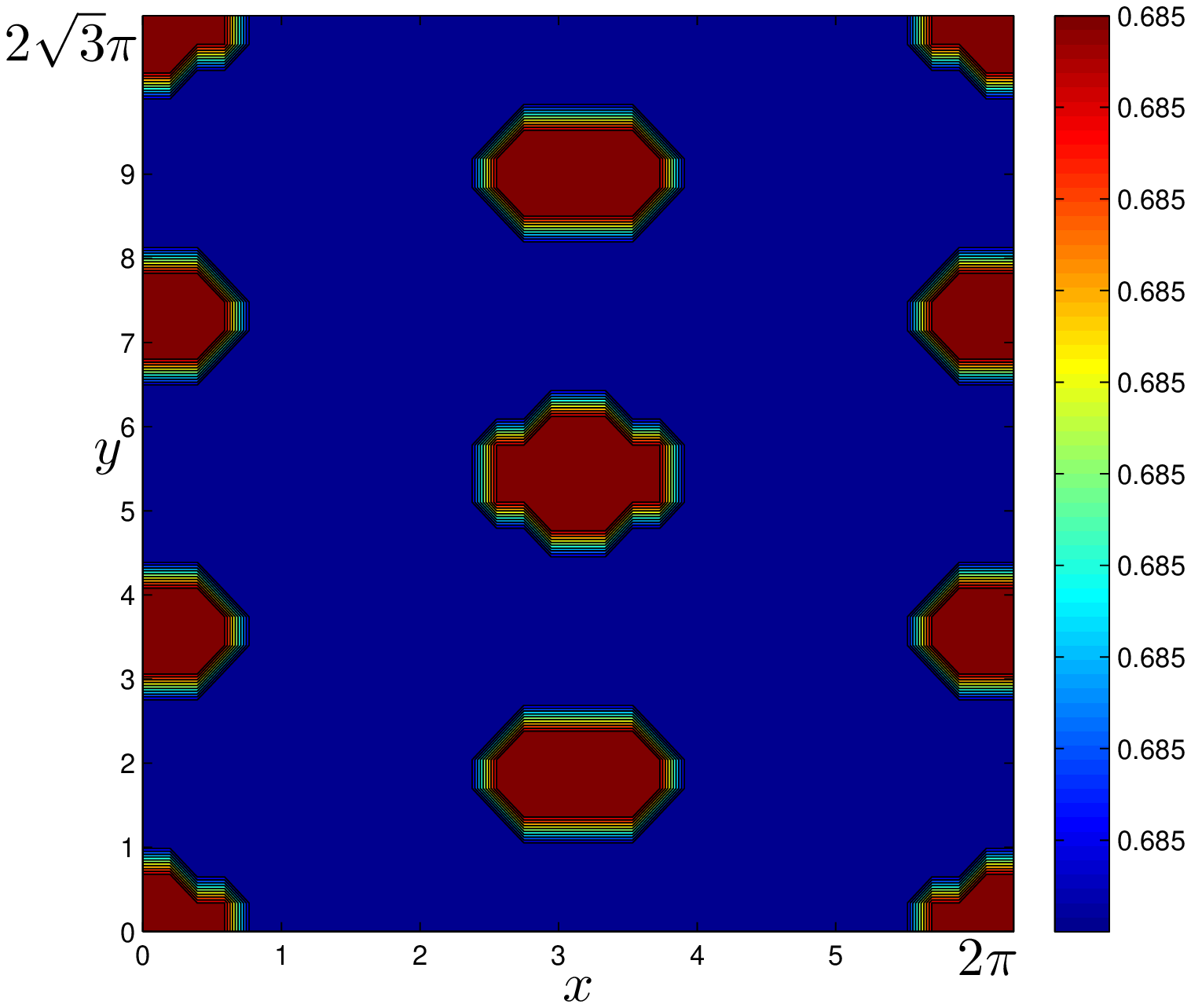}}
{\includegraphics[height=3.1cm, width=6.5cm]{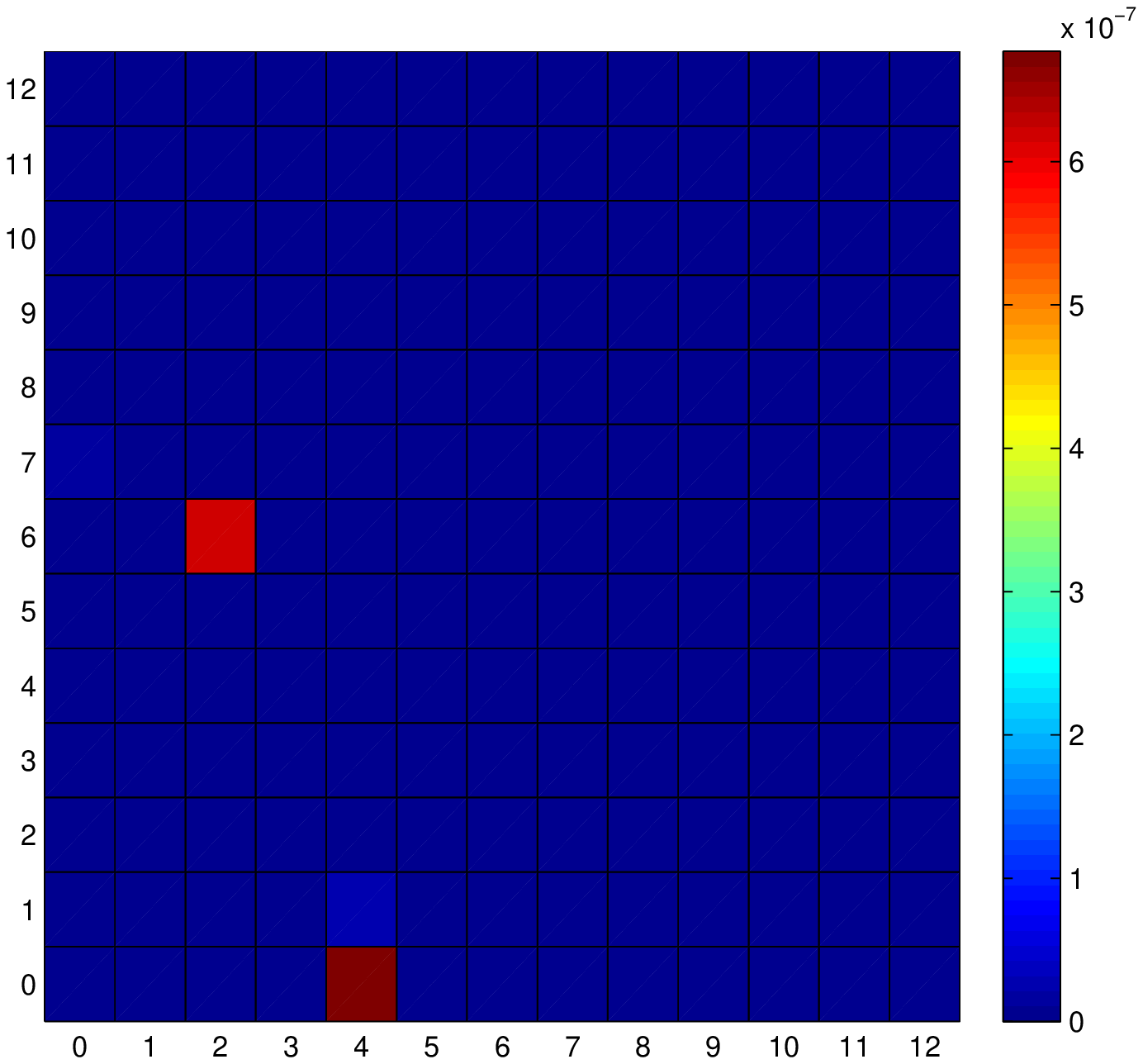}}
{\includegraphics[height=3.1cm, width=6.5cm]{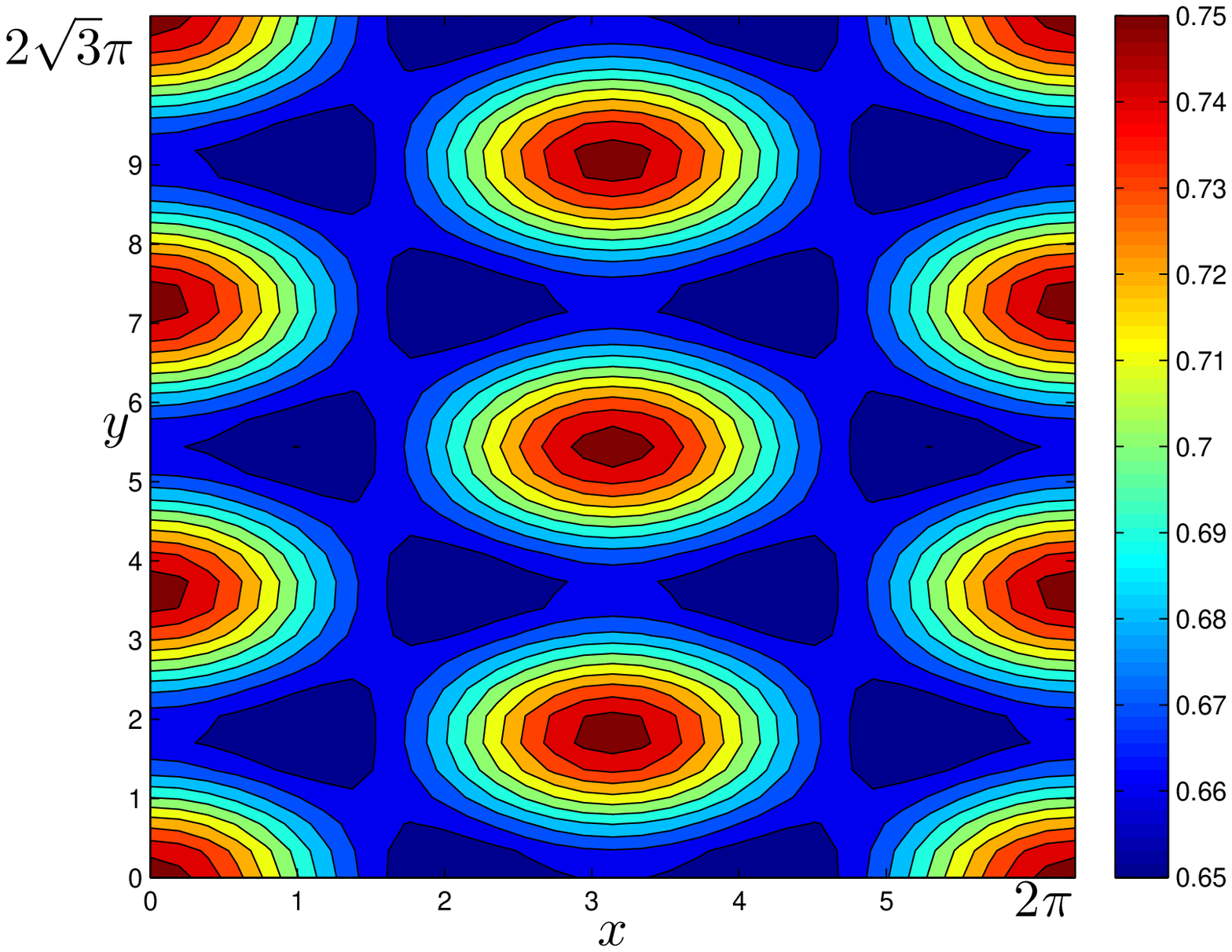}}
{\includegraphics[height=3.1cm, width=6.5cm]{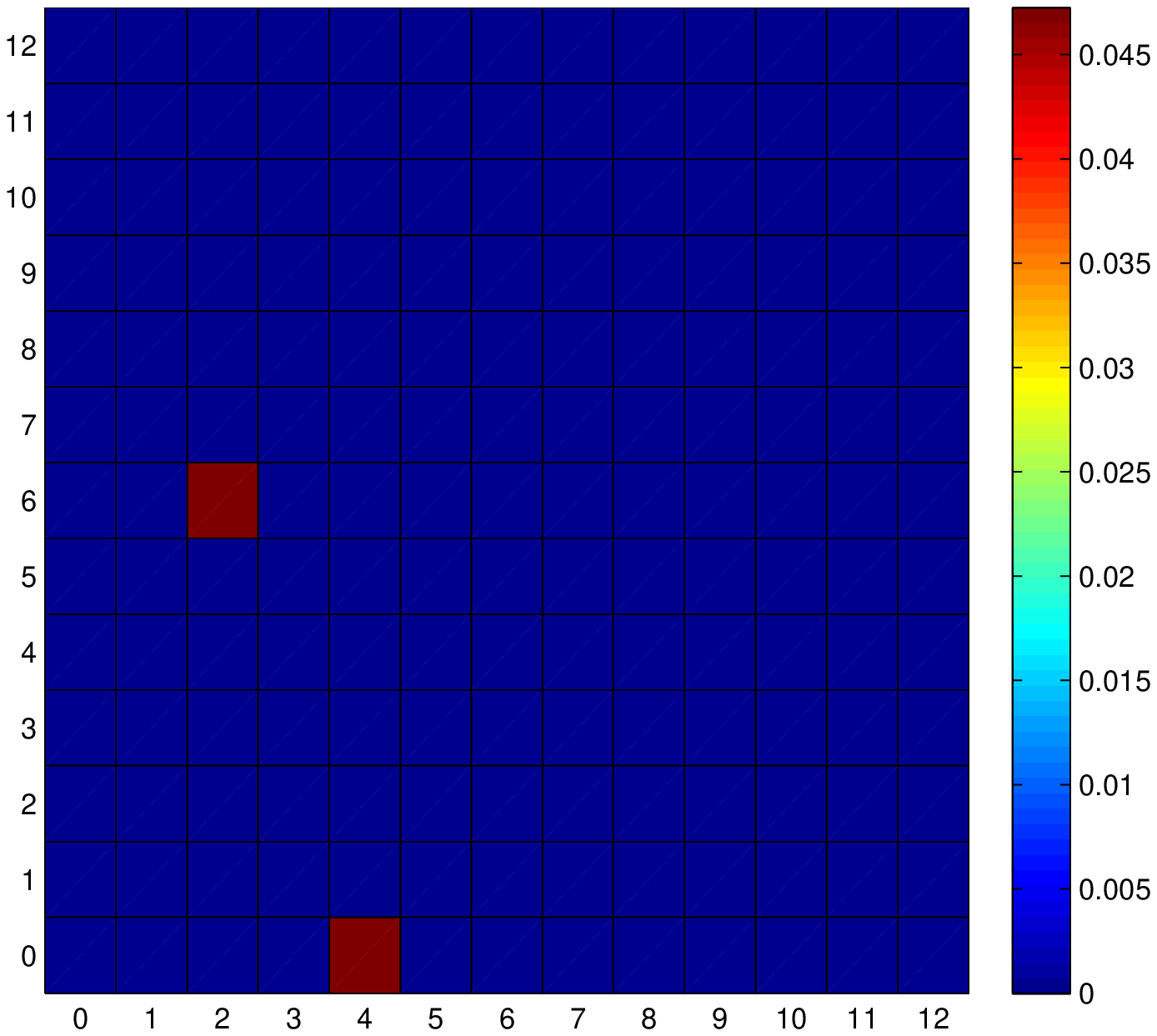}}
\subfigure[]{\includegraphics[height=3.1cm, width=6.5cm]{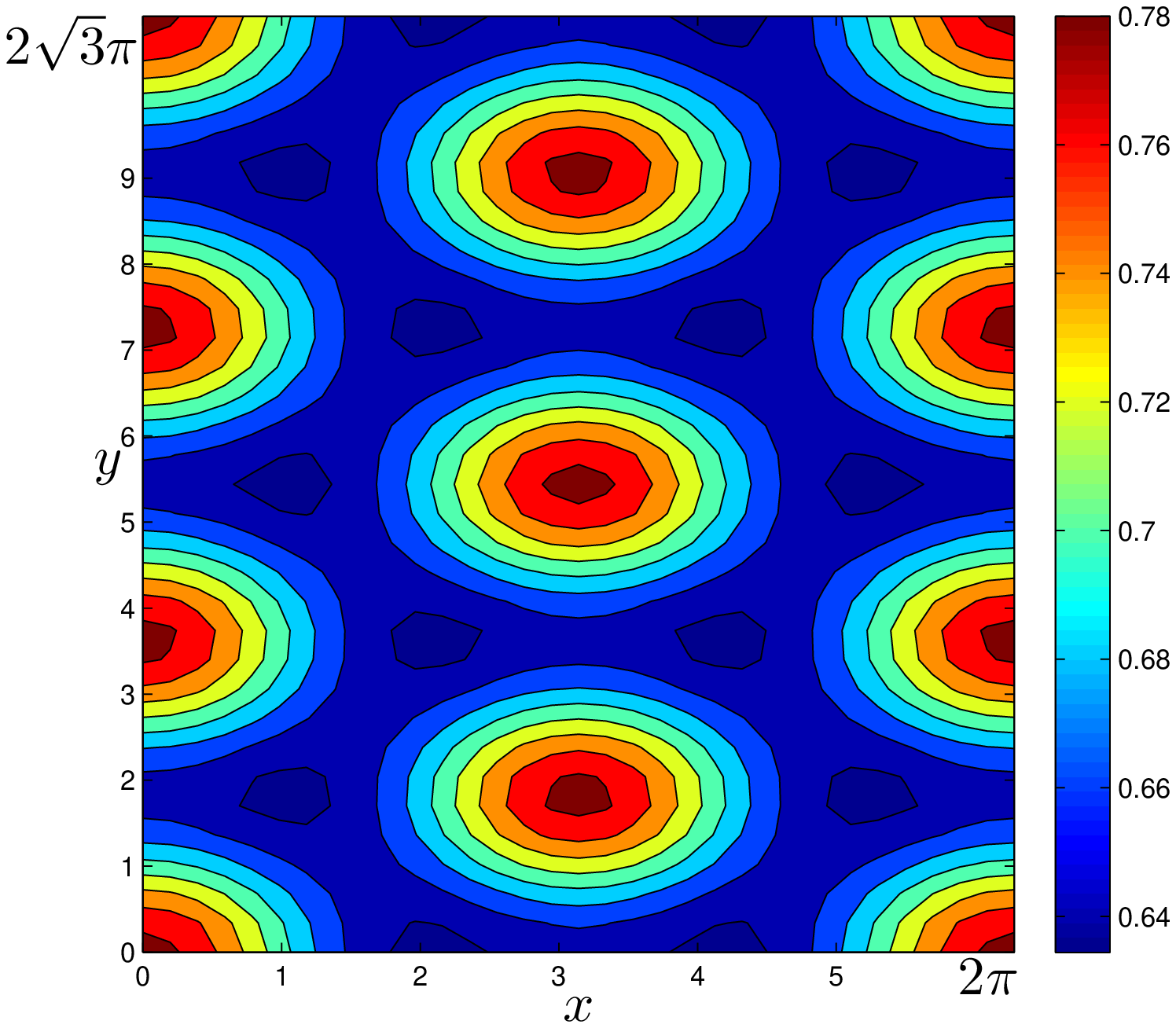}}
\subfigure[]{\includegraphics[height=3.1cm, width=6.5cm]{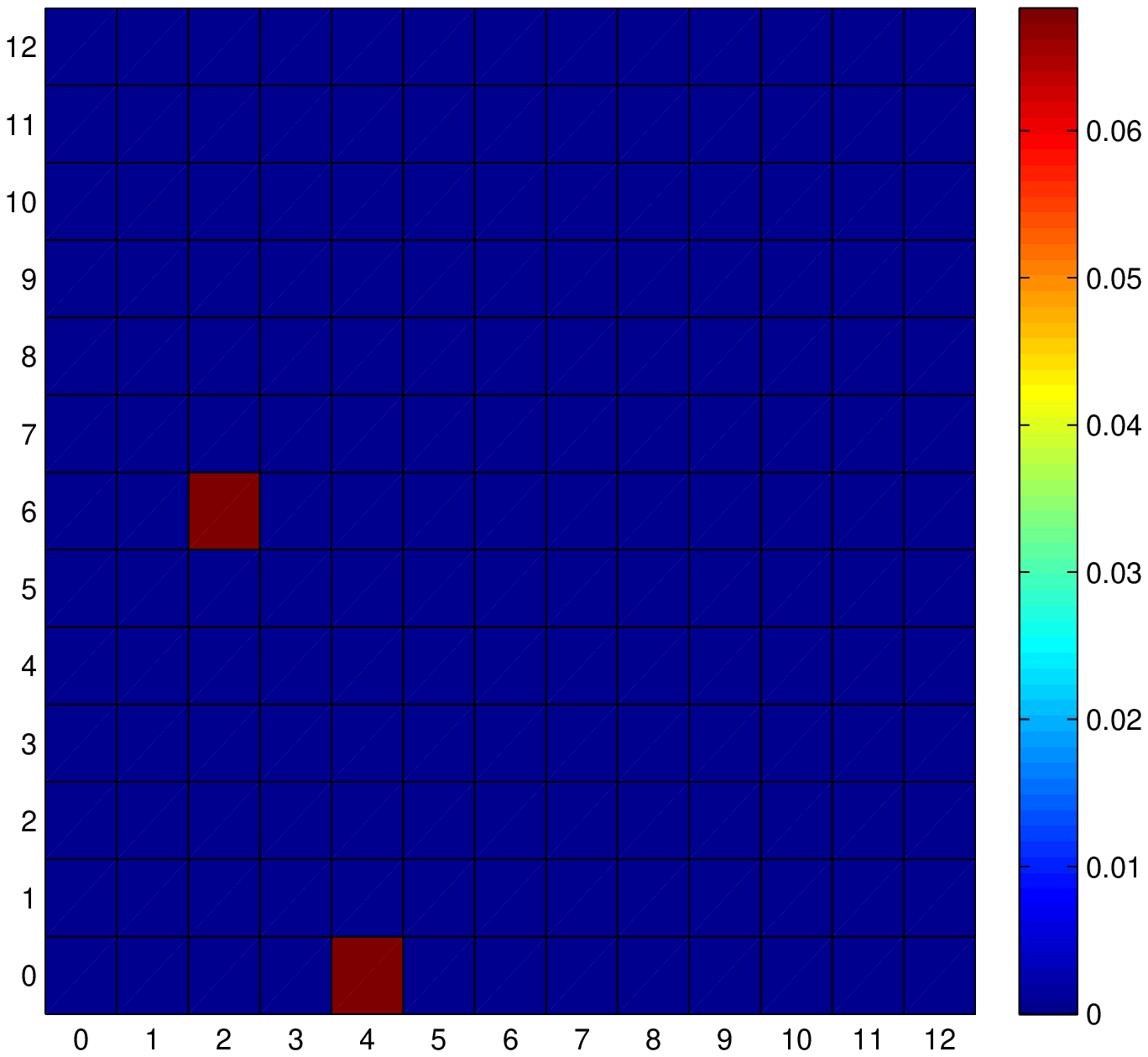}}
\end{center}
\caption {Hysteresis cycle when the resonance conditions \ref{res_si} hold. The parameters are chosen as in Fig. \ref{hexa}(a) The numerical solution $u$ of the full system \eqref{model}. (b) Spectrum of the numerical solution.}
\label{isteresi_ris}
\end{figure}
The conditions \eqref{multip} are satisfied by the two mode pairs $(4,0)$ and $(2,6)$, which also satisfy the resonance conditions \eqref{res_si}.
For the chosen set of parameters, the bifurcation diagram of system \eqref{4.32} in Fig. \ref{bif_hexa}(b) shows a bi-stability regime of hexagonal and roll patterns, which one forms strictly depend on the initial datum.

Performing a thousand simulations, starting from different randomly chosen initial conditions, we have found that the ``preferred'' shape of the pattern is the hexagon shown in Fig.~\ref{hexa}, which agrees almost well with the predicted asymptotic solution in formula \eqref{4.15}.
The existence of multiple stable steady states, once again, gives rise to the phenomenon of hysteresis, as shown in Fig. \ref{isteresi_ris} . To obtain the roll we perform our numerical simulation of the full system \eqref{model} starting from an initial condition chosen in the basin of attraction of the equilibrium $R^+$, as shown in the phase portrait of the system \eqref{4.32} in Fig. \ref{bif_hexa}(a). For $d<d_c$ the pattern disappears because there are no stable branches. If now we increase the parameter $d$ above $d_c$, the system jumps from the rolls to the hexagons. If, instead, we come back under $d_c$ the pattern remains always on the stable branch $H^+$.

\appendix
\section {Appendix}
\label{linear}
\begin{figure}
\begin{center}
\subfigure[]{\includegraphics[height=5cm, width=7.5cm]{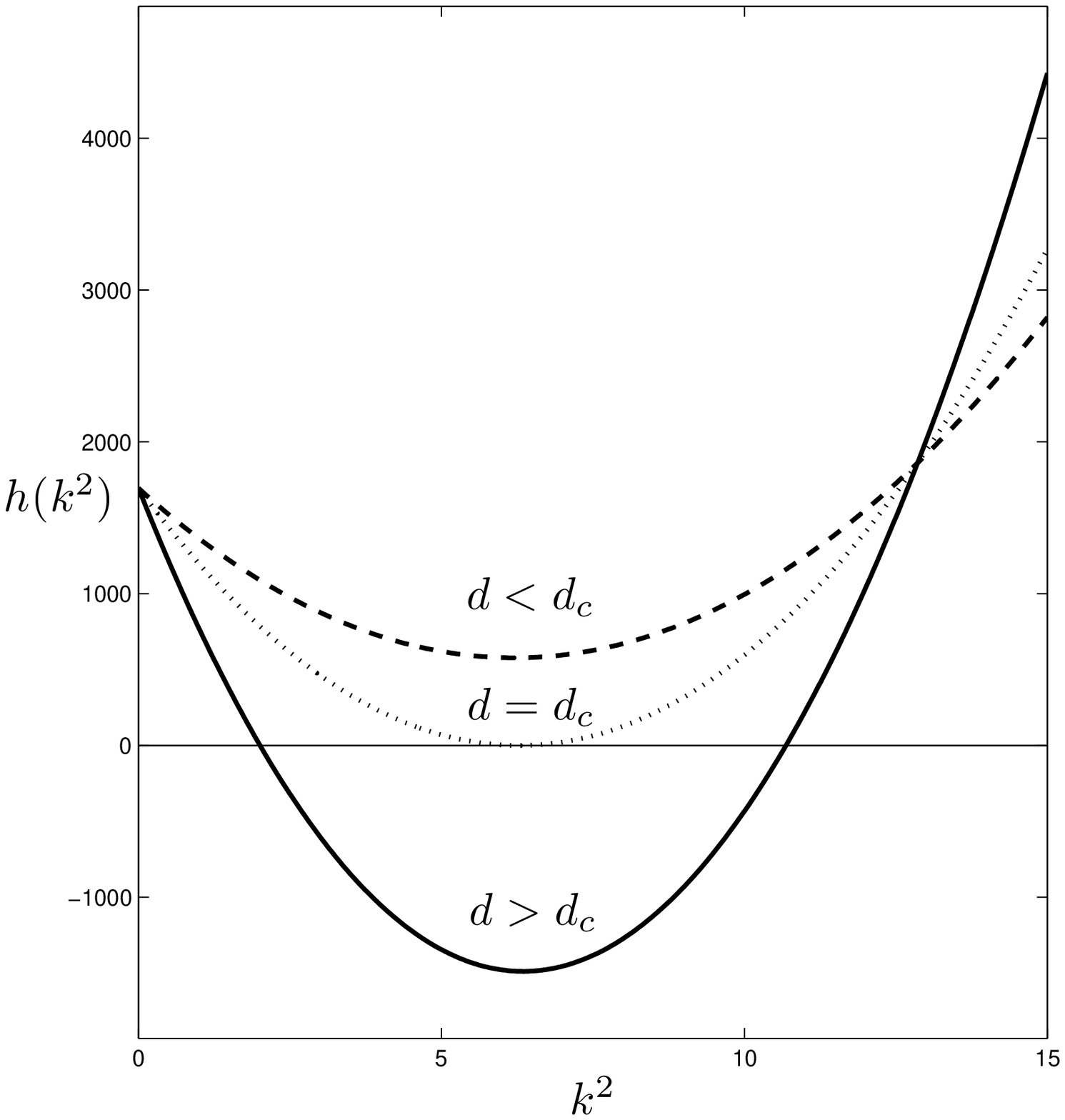}}
\subfigure[]{\includegraphics[height=5cm, width=7.5cm]{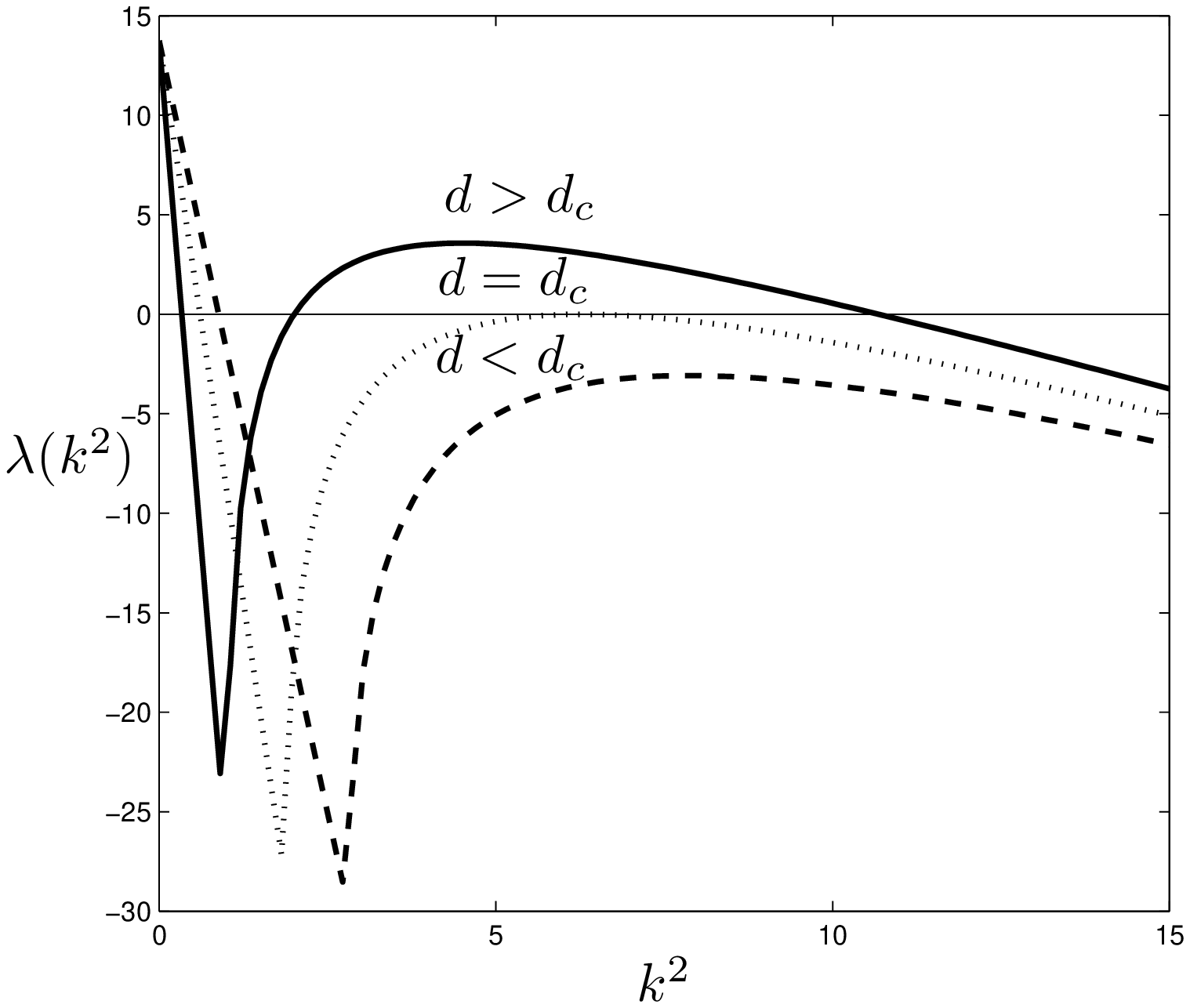}}
\end{center}
\caption{\label{parabola} (a) Plot of $h(k^2)$ with fixed reaction parameter values $a=0.34$, $b=0.64$, $\gamma=42$, $d_u=1$, $d_v=1$. With this choice of the parameters one has $d_c=43.9864$, while $\bar{k}_c=2.5$. (a) Growth rate of the $k$-th mode.}
\end{figure}

From here on we just indicate $J(P_e)\equiv J$.
Looking for solutions of system \eqref{sistema_rd} of the form
$e^{ikx+ \lambda t}$ leads to the following dispersion relation,
which gives the eigenvalue $\lambda$ as a function of the wavenumber
$k$:
\begin{equation}
\lambda^2+t(k^2)\lambda+h(k^2)=0,
\label{disp}
\end{equation}
where
\begin{equation*}
\begin{split}
t(k^2)&=k^2(1+d)-\gamma\tr J,\\
h(k^2)&=(d-d_ud_v)k^4d-\gamma k^2(dJ_{11}+J_{22}-d_uJ_{12}-d_vJ_{21})+\gamma^2\det J.
\end{split}
\end{equation*}
Requiring that $P_0$ is stable to the spatially homogeneous mode $k = 0$ entails ${\rm tr}(J)<0$ and ${\rm det}(J)>0$.
In order to have diffusion driven instability, we require $Re(\lambda)>0$ for some $k\neq 0$, which is
equivalent to impose $h(k^2)<0$ for some nonzero $k$ (see Fig. \ref{parabola}(a)).
In order to have a upward opening parabola we require the following condition on the relationship between diffusion and cross-diffusion coefficients to hold:
\begin{equation}
\label{cond3}
\det(D)=d-d_ud_v>0.
\end{equation}
Now, in order for $h(k^2)<0$ for some $k^2$ non-zero, we require that
\begin{equation}
\label{cond4}
dJ_{11}+J_{22}-d_uJ_{12}-d_vJ_{21}>0.
\end{equation}
Finally, for diffusively-driven instability to occur, we also require that there exists real $k^2_{1,2}$
such that $h(k^2_{1,2}) = 0$ and it is easily shown this yields to impose the third and last condition for cross-diffusion-driven-instability:
\begin{equation}
\label{cond5}
(dJ_{11}+J_{22}-d_uJ_{12}-d_vJ_{21})^2-4(d-d_vd_u)\det J>0.
\end{equation}

In Figure \ref{Turing} (a), the region in the parameter space $(a,b)$ where the diffusion driven instability arises,
and spatial patterns can develop, is given for a particular choice of the other system parameters.
Notice that the minimum of $h(k^2)$ is attained when:
\begin{equation}\label{kc_val}
k_c^2=\gamma\sqrt{\frac{\det J}{d_c-d_ud_v}},
\end{equation}
and by imposing $|J|=\frac{(dJ_{11}+J_{22}-d_uJ_{12}-d_vJ_{21})^2}{4(d-d_vd_u)}$ at the bifurcation, we find the bifurcation value for the diffusion parameter $d_c$, referred in \eqref{dc}

%
%
For $d>d_c$ the eigenvalue $\lambda(k^2)$ is positive at some $k\neq 0$ (see Fig.~\ref{parabola}(b)) and the system has a finite $k$ pattern-forming stationary
instability.

\section{Appendix}
\label{AppA}
\begin{small}
 First we expand $\textbf{w}$, the time $t$ and the bifurcation parameter $d$ as:
\begin{equation}
\begin{split}
d&=d_c+\varepsilon d^{(1)}+\varepsilon^2d^{(2)}+\varepsilon^3d^{(3)}+\varepsilon^4d^{(4)}+\varepsilon^5d^{(5)}+O(\varepsilon^6),\\
\ww&=\varepsilon\wuno+\varepsilon^2\wdue+\varepsilon^3\wtre+\varepsilon^4\wqua+\varepsilon^5\wcin+O(\varepsilon^6),\\
t&=\frac{T_1}{\varepsilon}+\frac{T_2}{\varepsilon^2}+\frac{T_3}{\varepsilon^3}+\frac{T_4}{\varepsilon^4}+\frac{T_5}{\varepsilon^5}+O(\varepsilon^6).
\end{split}
\label{expansions}
\end{equation}
After substituting the above expansions into \eqref{sistema_rd} and collecting the terms at each order in $\varepsilon$, we obtain a sequence of vector systems for the expansion coefficients $\textbf{w}_i$:\\
$\ \,O(\varepsilon):$
\begin{equation}\label{sequence_1}
\mathcal{L}^{d_c} {\bf w}_1=\mathbf{0},
\end{equation}
$\ \,O(\varepsilon^2):$
\begin{equation}\label{sequence_2}
\mathcal{L}^{d_c} {\bf w}_2=\mathbf{F},
\end{equation}
$\ \,O(\varepsilon^3):$
\begin{equation}\label{sequence_3}
\mathcal{L}^{d_c} {\bf w}_3=\mathbf{G},
\end{equation}
$\ \,O(\varepsilon^4):$
\begin{equation}\label{sequence_4}
\mathcal{L}^{d_c} {\bf w}_3=\mathbf{H},
\end{equation}
$\ \,O(\varepsilon^5):$
\begin{equation}\label{sequence_5}
\mathcal{L}^{d_c} {\bf w}_3=\mathbf{P},
\end{equation}
where $\mathcal{L}^{d_c}=J + D^{d_c} \nabla^2$ and:
\begin{eqnarray*}
\mathbf{F}&=&\frac{\partial\wuno}{\partial T_1}-\begin{pmatrix}0&0\\0&d^{(1)}\end{pmatrix}\nabla^2\wuno-\gamma\begin{pmatrix}2(a+b)u_1v_1+\frac{b}{(a+b)^2}u_1^2\\-2(a+b)u_1v_1-\frac{b}{(a+b)^2}u_1^2\end{pmatrix},\\
\mathbf{G}&=&\frac{\partial\wuno}{\partial T_2}+\frac{\partial\wdue}{\partial T_1}-\begin{pmatrix}0&0\\0&d^{(1)}\end{pmatrix}\nabla^2\wdue-\begin{pmatrix}0&0\\0&d^{(2)}\end{pmatrix}\nabla^2\wuno
-\gamma\begin{pmatrix}u_1^2v_1 \\ -u_1^2v_1\end{pmatrix}\\
&\ &-\gamma\begin{pmatrix}\frac{2b}{(a+b)^2}u_1+2(a+b)v_1 & 2(a+b)u_1 \\ -\frac{2b}{(a+b)^2}u_1-2(a+b)v_1 & -2(a+b)u_1\end{pmatrix}\wdue, \\
\mathbf{H}&=&\frac{\partial\wuno}{\partial T_3}+\frac{\partial\wdue}{\partial T_2}+\frac{\partial\wtre}{\partial T_1}-\begin{pmatrix}0&0\\0&d^{(1)}\end{pmatrix}\nabla^2\wtre-\begin{pmatrix}0&0\\0&d^{(2)}\end{pmatrix}\nabla^2\wdue-\begin{pmatrix}0&0\\0&d^{(3)}\end{pmatrix}\nabla^2\wuno\\
&\ &-\gamma\begin{pmatrix}\frac{2b}{(a+b)^2}u_1+2(a+b)v_1 & 2(a+b)u_1 \\ -\frac{2b}{(a+b)^2}u_1-2(a+b)v_1 & -2(a+b)u_1\end{pmatrix}\wtre\\
&\ &-\gamma\begin{pmatrix}\frac{b}{(a+b)^2}u_2+2(a+b)v_2+2u_1v_1 & u_1^2\\ -\frac{b}{(a+b)^2}u_2-2(a+b)v_2-2u_1v_1 & -u_1^2\end{pmatrix}\wdue, \\
\end{eqnarray*}
\begin{eqnarray*}
\mathbf{P}&=&\frac{\partial\wuno}{\partial T_4}+\frac{\partial\wdue}{\partial T_3}+\frac{\partial\wtre}{\partial T_2}+\frac{\partial\wqua}{\partial T_1}-\begin{pmatrix}0&0\\0&d^{(1)}\end{pmatrix}\nabla^2\wqua-\begin{pmatrix}0&0\\0&d^{(2)}\end{pmatrix}\nabla^2\wtre\\
&\ &-\begin{pmatrix}0&0\\0&d^{(3)}\end{pmatrix}\nabla^2\wdue-\begin{pmatrix}0&0\\0&d^{(4)}\end{pmatrix}\nabla^2\wuno-\gamma\begin{pmatrix}\frac{2b}{(a+b)^2}u_1+2(a+b)v_1 & 2(a+b)u_1 \\ -\frac{2b}{(a+b)^2}u_1-2(a+b)v_1 & -2(a+b)u_1\end{pmatrix}\wqua\\
&\ &-\gamma\begin{pmatrix}\frac{2b}{(a+b)^2}u_2+2(a+b)v_2+2u_1v_1 & 2(a+b)u_2+u_1^2\\ -\frac{2b}{(a+b)^2}u_2-2(a+b)v_2-2u_1v_1 & -2(a+b)u_2-u_1^2\end{pmatrix}\wtre\\
&\ &-\gamma\begin{pmatrix}u_2v_1 & 2u_1u_2 \\ -u_2v_1 & -2u_1u_2\end{pmatrix}\wdue. \\
\end{eqnarray*}
The solution to the linear problem
\eqref{sequence_1}, satisfying the Neumann boundary conditions, is of the form:
\begin{equation}
\wuno=A(T)\ro \cos(\bar{k}_c x), \quad \mbox{with} \quad \ro\in (J-\bar{k}_c^2D^{d_c}),\label{sol1}
\end{equation}
where the amplitude of the pattern $A(T)$ is
still arbitrary at this level and $\bar{k}_c$ is the first
admissible unstable mode. The vector $\ro$ is defined up to a
constant and the normalization is made as follows:
\begin{equation}\ro=\begin{pmatrix}1\\M\end{pmatrix} \quad \mbox{with} \quad M=\frac{-\bar{k}_c^2-\gamma\frac{2b}{a+b}}{\gamma(a+b)^2+\bar{k}_c^2d_c}.\label{ro}\end{equation}
Substituting the above result in \eqref{sequence_2} and requiring to eliminate secular terms, we assume
$T_1=0$ and $d^{(1)}=0$ and therefore the solution ${\bf w}_2$ of \eqref{sequence_2} can be straightforwardly obtained as a function of $A$. The source term $\mathbf{G}$ of the
linear problem \eqref{sequence_3} results in:
\begin{equation}\label{G}
{\bf G}=\left(\displaystyle\frac{d A}{d
T}\ro+A {\bf G}_1^{(1)}+A^3 {\bf G}_1^{(3)}
\right)\cos(\bar{k}_c x)+{\bf G}^* ,
\end{equation}
where\\\medskip
${\bf G}_1^{(1)}=\begin{pmatrix}0\\d^{(2)}k_c^2M\end{pmatrix}$\\\bigskip
${\bf G}_1^{(3)}=-\gamma\begin{pmatrix}\frac{2b}{(a+b)^2}+2(a+b)v_1 & 2(a+b)u_1 \\ -\frac{2b}{(a+b)^2}-2(a+b)v_1 & -2(a+b)u_1 \end{pmatrix}\left(\wdz+\displaystyle\frac{1}{2}\wdd \right)-\displaystyle\frac{3}{4}\gamma\begin{bmatrix}M\\-M\end{bmatrix}$\\\bigskip
${\bf G}^*= -\displaystyle\frac{1}{2}\gamma\begin{pmatrix}\frac{2b}{(a+b)^2}+2(a+b)v_1 & 2(a+b)u_1 \\ -\frac{2b}{(a+b)^2}-2(a+b)v_1 & -2(a+b)u_1 \end{pmatrix}\wdd-\displaystyle\frac{1}{4}\gamma\begin{bmatrix}M\\-M\end{bmatrix}$
At $O(\varepsilon^4)$
\begin{align*}
{\bf H}&=2A\frac{\partial A}{\partial T_2}\wdz+A^2{\bf H}_0^{(2)}+A^4{\bf H}_0^{(4)}+\left(2A\frac{\partial A}{\partial T}\wdd+A^2{\bf H}_2^{(2)}+A^4{\bf H}_2^{(4)}\right)\cos(2\bar{k}_cx)\\
&+A^4{\bf H}_4^{(4)}\cos(4\bar{k}_cx),
\end{align*}
where\\
\begin{align*}
{\bf H}_0^{(2)}&= \displaystyle\frac{1}{2}\gamma\begin{pmatrix}\frac{2b}{(a+b)^2}+2(a+b)M & 2(a+b) \\ -\frac{2b}{(a+b)^2}-2(a+b)M & -2(a+b) \end{pmatrix}\wtu\\\\
{\bf H}_0^{(4)}&= \displaystyle\frac{1}{2}\gamma\begin{pmatrix}\frac{2b}{(a+b)^2}+2(a+b)M & 2(a+b) \\ -\frac{2b}{(a+b)^2}-2(a+b)M & -2(a+b) \end{pmatrix}\wtd\\
&+\gamma\begin{pmatrix}\frac{b}{(a+b)^2}\wdz(1)+2(a+b)\wdz(2)+M & \frac{1}{2}\\ -\frac{b}{(a+b)^2}\wdz(1)-2(a+b)\wdz(2)-M & -\frac{1}{2}\end{pmatrix}\wdz\\
&+\displaystyle\frac{1}{2}\gamma\begin{pmatrix}\frac{b}{(a+b)^2}\wdd(1)+2(a+b)\wdd(2)+M & \frac{1}{2}\\ -\frac{b}{(a+b)^2}\wdd(1)-2(a+b)\wdd(2)-M & -\frac{1}{2}\end{pmatrix}\wdd\\\\
{\bf H}_2^{(2)}&=\begin{pmatrix}0&0\\0& 4d^{(2)}k_c^2\end{pmatrix}\wdd+\displaystyle\frac{1}{2}\gamma\begin{pmatrix}\frac{2b}{(a+b)^2}+2(a+b)M & 2(a+b) \\ -\frac{2b}{(a+b)^2}-2(a+b)M & -2(a+b) \end{pmatrix}\wtu\\\\
\end{align*}
\begin{align*}
{\bf H}_2^{(4)}&= \displaystyle\frac{1}{2}\gamma\begin{pmatrix}\frac{2b}{(a+b)^2}+2(a+b)M & 2(a+b) \\ -\frac{2b}{(a+b)^2}-2(a+b)M & -2(a+b) \end{pmatrix}(\wtd+\wtt)\\
&+\gamma\begin{pmatrix}\frac{b}{(a+b)^2}\wdz(1)+2(a+b)\wdz(2)+M & \frac{1}{2}\\ -\frac{b}{(a+b)^2}\wdz(1)-2(a+b)\wdz(2)-M & -\frac{1}{2}\end{pmatrix}\wdd\\
&+\gamma\begin{pmatrix}\frac{b}{(a+b)^2}\wdd(1)+2(a+b)\wdd(2)+M & \frac{1}{2}\\ -\frac{b}{(a+b)^2}\wdd(1)-2(a+b)\wdd(2)-M & -\frac{1}{2}\end{pmatrix}\wdz\\
{\bf H}_4^{(4)}&= \displaystyle\frac{1}{2}\gamma\begin{pmatrix}\frac{2b}{(a+b)^2}+2(a+b)M & 2(a+b) \\ -\frac{2b}{(a+b)^2}-2(a+b)M & -2(a+b) \end{pmatrix}\wtt\\
&+\displaystyle\frac{1}{2}\gamma\begin{pmatrix}\frac{b}{(a+b)^2}\wdd(1)+2(a+b)\wdd(2)+M & \frac{1}{2}\\ -\frac{b}{(a+b)^2}\wdd(1)-2(a+b)\wdd(2)-M & -\frac{1}{2}\end{pmatrix}\wdd
\end{align*}
At $O(\varepsilon^5)$
\begin{align}
{\bf P}&=\left(\frac{\partial A}{\partial T_4}\ro+\frac{\partial A}{\partial T_2}\wtu+3A^2\frac{\partial A}{\partial T_2}\wtd+A{\bf P}_1^{(1)}+A^3{\bf P}_1^{(3)}+A^5{\bf P}_1^{(5)}\right)\cos(\bar{k}_cx)\\
&+\left(3A^2\frac{\partial A}{\partial T_2}\wtt+A^3{\bf P}_3^{(3)}+A^5{\bf P}_3^{(5)}\right)\cos(3\bar{k}_cx)+A^5{\bf P}_5^{(5)}\cos(5\bar{k}_cx),
\label{P}
\end{align}
where\\
\begin{align*}
{\bf P}_1^{(1)}&=\begin{pmatrix}0 & 0 \\ 0 & d^{(2)}k_c^2\end{pmatrix}\wtu+\begin{pmatrix}0\\d^{(4)}k_c^2 M\end{pmatrix}\\ \\
{\bf P}_1^{(3)}&=\begin{pmatrix}0 & 0 \\ 0 & d^{(2)}k_c^2\end{pmatrix}\wtd- \gamma\begin{pmatrix}\frac{2b}{(a+b)^2}+2(a+b)M & 2(a+b) \\ -\frac{2b}{(a+b)^2}-2(a+b)M & -2(a+b) \end{pmatrix}\left(\wqz+\displaystyle\frac{1}{2}\wqd\right)\\
&-\gamma\begin{pmatrix} \frac{2b}{(a+b)^2}\wdz(1)+2(a+b)\wdz(2) & 2(a+b)\wdz(1)+\frac{1}{2}\\ -\frac{2b}{(a+b)^2}\wdz(1)-2(a+b)\wdz(2) &  -2(a+b)\wdz(1)-\frac{1}{2}\end{pmatrix}\wtu\\
&-\displaystyle\frac{1}{2}\gamma\begin{pmatrix} \frac{2b}{(a+b)^2}\wdd(1)+2(a+b)\wdd(2) & 2(a+b)\wdd(1)+\frac{1}{2}\\ -\frac{2b}{(a+b)^2}\wdd(1)-2(a+b)\wdd(2) &  -2(a+b)\wdd(1)-\frac{1}{2}\end{pmatrix}\wtu
\end{align*}
\begin{align*}
{\bf P}_1^{(5)}&=-\gamma\begin{pmatrix}\frac{2b}{(a+b)^2}+2(a+b)M & 2(a+b) \\ -\frac{2b}{(a+b)^2}-2(a+b)M & -2(a+b) \end{pmatrix} \left(\wqu+\displaystyle\frac{1}{2}\wqt\right)\\
&-\gamma\begin{pmatrix} \frac{2b}{(a+b)^2}\wdz(1)+2(a+b)\wdz(2) & 2(a+b)\wdz(1)+\frac{1}{2}\\ -\frac{2b}{(a+b)^2}\wdz(1)-2(a+b)\wdz(2) &  -2(a+b)\wdz(1)-\frac{1}{2}\end{pmatrix}\wtd\\
&-\displaystyle\gamma\begin{pmatrix} \frac{2b}{(a+b)^2}\wdz(1)+2(a+b)\wdz(2) & 2(a+b)\wdz(1)+\frac{1}{2}\\ -\frac{2b}{(a+b)^2}\wdz(1)-2(a+b)\wdz(2) &  -2(a+b)\wdz(1)-\frac{1}{2}\end{pmatrix}(\wtd+\wtt)\\
&-\gamma\begin{pmatrix}\left(\wdz(1)+\frac{1}{2}\wdd(1)\right)M & 2\wdz(1)+\wdd(1)\\-\left(\wdz(1)+\frac{1}{2}\wdd(1)\right)M & -2\wdz(1)-\wdd(1)\end{pmatrix}\wdz\\
&-\displaystyle\frac{1}{2}\gamma\begin{pmatrix}\Big(\wdz(1)+\wdd(1)\Big)M & 2\Big(\wdz(1)+\wdd(1)\Big)\\-\Big(\wdz(1)+\wdd(1)\Big)M & -2\Big(\wdz(1)+\wdd(1)\Big)\end{pmatrix}\wdd\\\\
{\bf P}_3^{(3)}&=\begin{pmatrix}0 & 0 \\ 0 & 9d^{(2)}k_c^2\end{pmatrix}\wtt-\displaystyle\frac{1}{2}\gamma\begin{pmatrix}\frac{2b}{(a+b)^2}+2(a+b)M & 2(a+b) \\ -\frac{2b}{(a+b)^2}-2(a+b)M & -2(a+b) \end{pmatrix} \wqd\\
&-\displaystyle\gamma\begin{pmatrix} \frac{2b}{(a+b)^2}\wdz(1)+2(a+b)\wdz(2) & 2(a+b)\wdz(1)+\frac{1}{2}\\ -\frac{2b}{(a+b)^2}\wdz(1)-2(a+b)\wdz(2) &  -2(a+b)\wdz(1)-\frac{1}{2}\end{pmatrix}\wtu
\end{align*}
\begin{align*}
{\bf P}_3^{(5)}&= -\displaystyle\frac{1}{2}\gamma\begin{pmatrix}\frac{2b}{(a+b)^2}+2(a+b)M & 2(a+b) \\ -\frac{2b}{(a+b)^2}-2(a+b)M & -2(a+b) \end{pmatrix}(\wqt+\wqq)\\
&-\gamma\begin{pmatrix} \frac{2b}{(a+b)^2}\wdz(1)+2(a+b)\wdz(2) & 2(a+b)\wdz(1)+\frac{1}{2}\\ -\frac{2b}{(a+b)^2}\wdz(1)-2(a+b)\wdz(2) &  -2(a+b)\wdz(1)-\frac{1}{2}\end{pmatrix}\wtt\\
&-\displaystyle\frac{1}{2}\gamma\begin{pmatrix} \frac{2b}{(a+b)^2}\wdz(1)+2(a+b)\wdz(2) & 2(a+b)\wdz(1)+\frac{1}{2}\\ -\frac{2b}{(a+b)^2}\wdz(1)-2(a+b)\wdz(2) &  -2(a+b)\wdz(1)-\frac{1}{2}\end{pmatrix}\wtd\\
&-\displaystyle\frac{1}{2}\gamma\begin{pmatrix}M\wdz(1) & 2\wdz(1)\\ -M\wdz(1) & -2\wdz(1)\end{pmatrix}\wdd\\
&-\displaystyle\frac{1}{2}\gamma\begin{pmatrix} M\wdd(1) & 2\wdd(1) \\ -M\wdd(1) & -2\wdd(1)\end{pmatrix}\left(\wdz+\displaystyle\frac{1}{2}\wdd\right)\\\\
{\bf P}_5^{(5)}&= -\displaystyle\frac{1}{2}\gamma\begin{pmatrix}\frac{2b}{(a+b)^2}+2(a+b)M & 2(a+b) \\ -\frac{2b}{(a+b)^2}-2(a+b)M & -2(a+b) \end{pmatrix}\wqq\\
&-\displaystyle\frac{1}{2}\gamma\begin{pmatrix} \frac{2b}{(a+b)^2}\wdz(1)+2(a+b)\wdz(2) & 2(a+b)\wdz(1)+\frac{1}{2}\\ -\frac{2b}{(a+b)^2}\wdz(1)-2(a+b)\wdz(2) &  -2(a+b)\wdz(1)-\frac{1}{2}\end{pmatrix}\wtt\\
&-\displaystyle\frac{1}{4}\gamma\begin{pmatrix} M\wdd(1) & 2\wdd(1) \\ -M\wdd(1) & -2\wdd(1)\end{pmatrix}\wdd
\end{align*}

Putting
\begin{equation}\label{sigmaL}
\sigma=-\frac{<{\bf G}_1^{(1)}, \bfpsi>}{<\ro,
\bfpsi>},\qquad  L=\frac{<{\bf G}_1^{(3)}, \bfpsi>}{<\ro,
\bfpsi>},
\end{equation}
where ${\bfpsi} \in
Ker\left\{\left(J -\bar{k}_c^2D^{d_c}
\right)^\dag\right\}$, the solvability condition $<{\bf G}, \bfpsi>$ for the equation \eqref{G} leads to \eqref{SL3}.\\
Putting
\begin{equation}\label{sigmaLR}
\tilde{\sigma}=-\frac{<{\bf P}_1^{(1)}, \bfpsi>}{<\ro,
\bfpsi>},\qquad  L=\frac{<{3\sigma\wtd-L\wtu+\bf P}_1^{(3)}, \bfpsi>}{<\ro,
\bfpsi>},\qquad R=\frac{<{3L\wtd+\bf P}_1^{(5)}, \bfpsi>}{<\ro,
\bfpsi>}
\end{equation}
the Fredholm alternative $<{\bf P}, \bfpsi>$ for the equation \eqref{P} leads to
\begin{equation}
\frac{\partial A}{\partial T_4}=\tilde{\sigma}A-\tilde{L}A^3+\tilde{R}A^5.
\label{SL_tilde}
\end{equation}
Adding up \eqref{SL_tilde} to \eqref{SL3} one gets \eqref{SL5},
with
$$\bar{\sigma}=\sigma+\varepsilon^2\tilde{\sigma}, \qquad \bar{L}=L+\varepsilon^2\tilde{L}, \qquad \bar{R}=\varepsilon^2\tilde{R}.$$

\section{Appendix}
\label{AppB}
We define the following operators:\\
\begin{align*}
L_{ij}^1&=J-(i^2\phi_1^2+j^2\psi_1^2)D^{dc}\\
L_{ij}^2&=J-(i^2\phi_2^2+j^2\psi_2^2)D^{dc}\\
L_{mn}&=J-\big((\phi_1+m\phi_2)^2+(\psi_1+n\psi_2^2)\big)D^{dc}\\
\end{align*}
\begin{itemize}
\item At $O(\varepsilon^2)$, again, assuming that $T_1=0$ and $d^{(1)}=0$ to eliminate secular terms, one obtains:\\
\begin{equation*}
\begin{split}
L^{d_c}\wdue = {\bf F}^{(2)}\equiv &-\displaystyle\frac{1}{4}\gamma\begin{pmatrix}2(a+b)M+\frac{b}{(a+b)^2}\\-2(a+b)M-\frac{b}{(a+b)^2}\end{pmatrix} A_1^2\sum_{i,j=0,2}\cos(i\phi_1x)\cos(j\psi_1y)\\
&-\displaystyle\frac{1}{4}\gamma\begin{pmatrix}2(a+b)M+\frac{b}{(a+b)^2}\\-2(a+b)M-\frac{b}{(a+b)^2}\end{pmatrix} A_2^2\sum_{i,j=0,2}\cos(i\phi_1x)\cos(j\psi_1y)\\
&-\displaystyle\frac{1}{2}\gamma\begin{pmatrix}2(a+b)M+\frac{b}{(a+b)^2}\\-2(a+b)M-\frac{b}{(a+b)^2}\end{pmatrix} A_1A_2\sum_{m,n=-1,1}\cos((\phi_1+m\phi_2)x)\cos((\psi_1+n\psi_2)y).
\end{split}
\end{equation*}
All these terms identically satisfy the compatibility conditions and the solution is:
\begin{equation*}
\begin{split}
\wdue &=A_1^2\sum_{i,j=0,2}\ww_{2ij}^1\cos(i\phi_1 x)\cos(j\psi_1 y)+A_2^2\sum_{i,j=0,2}\ww_{2ij}^2\cos(i\phi_2 x)\cos(j\psi_2 y)\\
&+A_1A_2\sum_{m,n=-1,1}\ww_{2mn}\cos((\phi_1+m\phi_2)x)\cos((\psi_1+n\psi_2)y),
\label{sol_h2(i)}
\end{split}
\end{equation*}
where the vectors $\ww_{2ij}^l$ $(l=1,2)$ and $\ww_{m,n}$ are the solutions of the following linear systems:
$$
L^1_{ij}\ww_{2ij}^1=-\quart \gamma\begin{pmatrix}2(a+b)M+\frac{b}{(a+b)^2}\\-2(a+b)M-\frac{b}{(a+b)^2}\end{pmatrix}\quad L^2_{ij}\ww_{2ij}^2=-\quart\gamma\begin{pmatrix}2(a+b)M+\frac{b}{(a+b)^2}\\-2(a+b)M-\frac{b}{(a+b)^2}\end{pmatrix}$$
$$L_{mn}\ww_{2mn}= -\half\gamma\begin{pmatrix}2(a+b)M+\frac{b}{(a+b)^2}\\-2(a+b)M-\frac{b}{(a+b)^2}\end{pmatrix}.$$
\item $O(\varepsilon^3)$\\
\begin{equation*}
\begin{split}
L^{d_c}\wtre={\bf G}^{(2)}\equiv &\left[\frac{dA_1}{dT_2}\ro+A_1 {\bf G}_1^{(21)}-A_1^3 {\bf G}_2^{(21)}-A_1A_2^2{\bf G}_3^{(21)}\right]\cos(\phi_1x)\cos(\psi_1y)\\
&+\left[\frac{dA_2}{dT_2}\ro+A_2 {\bf G}_1^{(22)}-A_2^3 {\bf G}_2^{(22)}-A_1^2A_2 {\bf G}_3^{(22)}\right]\cos(\phi_2x)\cos(\psi_2y)\\
&+\bar{\bf G}^{(2)},
\label{h3G(i)}
\end{split}
\end{equation*}
where
\begin{align*}
{\bf G}_1^{(21)}&={\bf G}_1^{(22)}=\ddue\kk\vet\\\\
{\bf G}_2^{(21)}&=\gamma\begin{pmatrix}\frac{2b}{(a+b)^2}+2(a+b)M & 2(a+b) \\ -\frac{2b}{(a+b)^2}-2(a+b)M & -2(a+b) \end{pmatrix}\left(\ww_{200}^1+\half\ww_{220}^1+\half\ww_{202}^1+\quart\ww_{222}^1\right)+\frac{9}{16}\gamma\begin{bmatrix}M\\-M\end{bmatrix}\\\\
{\bf G}_2^{(22)}&=\gamma\begin{pmatrix}\frac{2b}{(a+b)^2}+2(a+b)M & 2(a+b) \\ -\frac{2b}{(a+b)^2}-2(a+b)M & -2(a+b) \end{pmatrix}\left(\ww_{200}^2+\half\ww_{220}^2+\half\ww_{202}^2+\quart\ww_{222}^2\right)
+\gamma\begin{bmatrix}M\\-M\end{bmatrix}\\\\
{\bf G}_3^{(21)}&=\gamma\begin{pmatrix}\frac{2b}{(a+b)^2}+2(a+b)M & 2(a+b) \\ -\frac{2b}{(a+b)^2}-2(a+b)M & -2(a+b) \end{pmatrix}\left(\ww_{200}^2+\quart\ww_{211}+\quart\sum_{m,n=-1,1}\ww_{2mn}\right)
+\gamma\begin{bmatrix}M\\-M\end{bmatrix}\\
\end{align*}
\begin{align*}
{\bf G}_3^{(22)}&=\gamma\begin{pmatrix}\frac{2b}{(a+b)^2}+2(a+b)M & 2(a+b) \\ -\frac{2b}{(a+b)^2}-2(a+b)M & -2(a+b) \end{pmatrix}\left(\ww_{200}^1+\quart\sum_{m,n=-1,1}\ww_{2mn}\right)
+\frac{3}{4}\gamma\begin{bmatrix}M\\-M\end{bmatrix}\\
\end{align*}
As regards the vector $\bar{\bf G}^{(2)}$, it needs to consider three different cases:
\begin{itemize}
\item[(i)] only one of the relations $\phi_1=\psi_2=0$ and $\phi_2=\psi_1=0$ holds (we shall assume, without loss of generality,  $\phi_1=\psi_2=0$).\\
The explicit expression for $\bar{\bf G}^{(2)}$ is:\\
\begin{align*}
\bar{\bf G^{(2)}}&=\bar{\bf G}_2^{(21)}A_1^3\cos(\psi_1y)+\bar{\bf G}_3^{(22)}A_1^2A_2\cos(\phi_2x)\\
&+\bar{\bf G}_2^{(22)}A_2^3\cos(\phi_2x)+\bar{\bf G}_3^{(21)}A_1A_2^2\cos(\psi_1y)+\bar{\bf G}^{(2)*},
\end{align*}\\
where $\bar{\bf G}^{(2)*}$ contains only terms orthogonal to ${\bm \psi}$ and\\
\begin{align*}
\bar{\bf G}_2^{(21)}&=\half\gamma\begin{pmatrix}\frac{2b}{(a+b)^2}+2(a+b)M & 2(a+b) \\ -\frac{2b}{(a+b)^2}-2(a+b)M & -2(a+b) \end{pmatrix}\left(\ww_{220}^1+\half\ww_{222}^1\right)+\displaystyle\frac{3}{16}\gamma\begin{bmatrix}M\\-M\end{bmatrix}\\ \\
\bar{\bf G}_2^{(22)}&=\half\gamma\begin{pmatrix}\frac{2b}{(a+b)^2}+2(a+b)M & 2(a+b) \\ -\frac{2b}{(a+b)^2}-2(a+b)M & -2(a+b) \end{pmatrix}\left(\ww_{202}^2+\half\ww_{222}^2\right)+\frac{3}{16}\begin{bmatrix}M\\-M\end{bmatrix}\\ \\
\bar{\bf G}_3^{(21)}&=\gamma\begin{pmatrix}\frac{2b}{(a+b)^2}+2(a+b)M & 2(a+b) \\ -\frac{2b}{(a+b)^2}-2(a+b)M & -2(a+b) \end{pmatrix}\left(\ww_{202}^2+\quart\sum_{m,n=-1,1}\ww_{2mn}\right)+\frac{3}{4}\begin{bmatrix}M\\-M\end{bmatrix}\\\\
\bar{\bf G}_3^{(22)}&=\gamma\begin{pmatrix}\frac{2b}{(a+b)^2}+2(a+b)M & 2(a+b) \\ -\frac{2b}{(a+b)^2}-2(a+b)M & -2(a+b) \end{pmatrix}\left(\ww_{220}^1+\quart\sum_{m,n=-1,1}\ww_{2mn}\right)+\frac{3}{4}\begin{bmatrix}M\\-M\end{bmatrix}.\\
\end{align*}
So, the expression for the parameters in \eqref{4.19} are
\begin{equation*}
\sigma=\sigma_l=-\frac{<{\bf G}_1^{(21)},{\bm\psi}>}{<\ro,{\bm\psi}>}\qquad L_l=-\frac{<{\bf G}_2^{(2l)},{\bm\psi}>}{<\ro,{\bm\psi}>}+\bar{L}_l\qquad
R_l=\frac{<{\bf G}_3^{(2l)},{\bm\psi}>}{<\ro,{\bm\psi}>}+\bar{R}_l,
\end{equation*}
where
\begin{equation*}
\bar{L}_l=-\frac{<\bar{\bf G}_2^{(2l)},\tilde{\psi}>}{<\ro,{\bm\psi}>},\qquad
\bar{R}_l=\frac{<\bar{\bf G}_3^{(2l)},\tilde{\psi}>}{<\ro,{\bm\psi}>}.
\end{equation*}
\item [(ii)] only one of $\phi_l$ and $\psi_l$ vanishes (we shall assume, without loss of generality, $\phi_1=0$).\\
In this case, one has
$$\bar{\bf G}_2^{(22)}=\bar{\bf G}_3^{(21)}=0 \qquad \mbox{from which follows} \qquad  \bar{L}_2=\bar{R}_1=0,$$
while \, $\bar{\bf G}_2^{(21)}$,\,  $\bar{\bf G}_3^{(22)}$,\, $\bar{L}_1$ \, and \, $\bar{R}_2$ \, are as in case (i).
\item [(iii)] $\phi_l$ and $\psi_l$ are both different from zero.\\
In this last case, $\bar{\bf G}^{(2)}$ involves only terms orthogonal to ${\bm {\psi}}$ and so $\bar{L}_l$ and $\bar{R}_l$ are all zero.
\end{itemize}
\end{itemize}

\section{Appendix}
\label{AppC}
\begin{itemize}
\item $O(\varepsilon^2)$\\
\begin{equation}
\begin{split}
L^{d_c}\wdue={\bf F}^{(3)}&\equiv\Biggl(\frac{dA_1}{dT_1}\ro+\kk\duno\vet A_1-\gamma\begin{pmatrix}2(a+b)M+b\frac{b}{(a+b)^2}\\-2(a+b)M-b\frac{b}{(a+b)^2}\end{pmatrix}A_1A_2\Biggr)\cos(\phi_1x)\cos(\psi_1y)\\
&+\Biggl(\frac{dA_2}{dT_1}\ro+\kk\duno\vet A_2-\quart \gamma \begin{pmatrix}2(a+b)M+b\frac{b}{(a+b)^2}\\-2(a+b)M-b\frac{b}{(a+b)^2}\end{pmatrix}A_1^2\Biggr)\cos(\phi_2x)+{\bf F}^{(3)*},
\label{ris}
\end{split}
\end{equation}
where
\begin{equation*}
\begin{split}
\bar{\bf F}^{(3)*}=& -\quart\gamma\begin{pmatrix}2(a+b)M+b\frac{b}{(a+b)^2}\\-2(a+b)M-b\frac{b}{(a+b)^2}\end{pmatrix} A_1^2-\half\gamma\begin{pmatrix}2(a+b)M+b\frac{b}{(a+b)^2}\\-2(a+b)M-b\frac{b}{(a+b)^2}\end{pmatrix}A_2^2\\
&-\quart \gamma\begin{pmatrix}2(a+b)M+b\frac{b}{(a+b)^2}\\-2(a+b)M-b\frac{b}{(a+b)^2}\end{pmatrix}A_1^2\cos(2\psi_1y)\\
&-\half\gamma\begin{pmatrix}2(a+b)M+b\frac{b}{(a+b)^2}\\-2(a+b)M-b\frac{b}{(a+b)^2}\end{pmatrix}A_1^2\cos(2\phi_2x)\\
&-\gamma\begin{pmatrix}2(a+b)M+b\frac{b}{(a+b)^2}\\-2(a+b)M-b\frac{b}{(a+b)^2}\end{pmatrix}A_1A_2\cos(3\phi_1x)\cos(\psi_1 y)\\
&-\quart \gamma\begin{pmatrix}2(a+b)M+b\frac{b}{(a+b)^2}\\-2(a+b)M-b\frac{b}{(a+b)^2}\end{pmatrix}A_1^2\cos(2\phi_1x)\cos(2\psi_1y)
\end{split}
\end{equation*}
is orthogonal to $\bm \psi$.\\
By imposing the solvability condition, one has
\begin{equation}
\begin{split}
\frac{\partial A_1}{\partial T_1}&=\sigma A_1-LA_1A_2\\
\frac{\partial A_2}{\partial T_1}&=\sigma A_2-\frac{L}{4}A_1^2,
\label{sistema_sec_h2}
\end{split}
\end{equation}
where
$$\sigma=-\frac{<\begin{pmatrix}0\\\kk\duno M\end{pmatrix},{\bm\psi}>}{<\ro,{\bm\psi}>},\qquad L=-\frac{<\gamma\begin{pmatrix}2(a+b)M+b\frac{b}{(a+b)^2}\\-2(a+b)M-b\frac{b}{(a+b)^2}\end{pmatrix},{\bm\psi}>}{<\ro,{\bm\psi}>}.$$

This system, as it is easy to prove, does not admit any stable stationary solution, so the weakly nonlinear analysis, at this order, is not able to predict the amplitude of the pattern.\\
Therefore, using the Landau equations \eqref{sistema_sec_h2} and the Fredholm alternative, Eq. \eqref{ris} can be solved and its solution $\wdue$ has the following form:
\begin{equation*}
\begin{split}
\wdue&=\Big(A_1\ww_{211}^{1(1)}+A_1A_2\ww_{211}^{1(2)}\Big)\cos(\phi_1x)\cos(\psi_1y)+\Big(A_2\ww_{211}^{2(1)}+A_1^2\ww_{211}^{2(2)}\Big)\cos(\phi_2x)\\
&+A_1^2\Big(\ww_{200}^1+\ww_{202}^1\cos(2\psi_1y)+\ww_{222}^1\cos(2\phi_1x)\cos(2\psi_1y)\Big)\\
&+A_2^2\Big(\ww_{200}^2+\ww_{220}^2\cos(2\phi_2x)\Big)+A_1A_2\ww_{231}\cos(3\phi_1x)\cos(\psi_1y),
\label{sol_h2(ii)}
\end{split}
\end{equation*}
where the vectors $\ww_{2ij}$ are the solutions of the following linear systems:
\begin{align*}
L_{11}^l\ww_{211}^{l(1)}&=\sigma\ro+\begin{pmatrix}0\\ \kk\duno M\end{pmatrix}\\\\
L_{11}^1\ww_{211}^{1(2)}&=-L\ro-\gamma\begin{pmatrix}2(a+b)M+b\frac{b}{(a+b)^2}\\-2(a+b)M-b\frac{b}{(a+b)^2}\end{pmatrix}\\\\
L_{11}^2\ww_{211}^{2(2)}&=-\frac{L}{4}\ro-\quart\gamma\begin{pmatrix}2(a+b)M+b\frac{b}{(a+b)^2}\\-2(a+b)M-b\frac{b}{(a+b)^2}\end{pmatrix}\\\\
L_{ij}^1\ww_{2ij}^1&=-\quart\gamma\begin{pmatrix}2(a+b)M+b\frac{b}{(a+b)^2}\\-2(a+b)M-b\frac{b}{(a+b)^2}\end{pmatrix}\\\\
L_{i0}^2\ww_{2i0}^2&=-\half\gamma\begin{pmatrix}2(a+b)M+b\frac{b}{(a+b)^2}\\-2(a+b)M-b\frac{b}{(a+b)^2}\end{pmatrix}\\\\
L_{31}^1\ww_{231}&=-\gamma\begin{pmatrix}2(a+b)M+b\frac{b}{(a+b)^2}\\-2(a+b)M-b\frac{b}{(a+b)^2}\end{pmatrix}
\end{align*}
\item $O(\varepsilon^3)$\\
\begin{equation*}
\begin{split}
L^{d_c}\wtre={\bf G}^{(3)}&\equiv\left[\frac{dA_1}{T_2}\ro+A_1 {\bf G}_1^{(31)}+A_1A_2 {\bf G}_2^{(31)}+A_1^3{\bf G}_3^{(31)}+A_1A_2^2{\bf G}_4^{(31)}\right]\cos(\phi_1x)\cos(\psi_1y)\\
&+\left[\frac{dA_2}{T_2}\ro+A_2 {\bf G}_1^{(32)}+A_1^2 {\bf G}_2^{(32)}+A_2^3{\bf G}_3^{(32)}+A_1^2A_2{\bf G}_4^{(32)}\right]\cos(\phi_2x)\cos(\psi_2y)\\
&+{\bf G}^{(3)*},
\label{h3G(ii)}
\end{split}
\end{equation*}
where
\begin{align*}
{\bf G}_1^{(31)}&= \sigma\ww_{211}^{1(1)}+\duno\kk\mat\ww_{211}^{1(1)}+\ddue\kk\vet\\\\
{\bf G}_1^{(32)}&= \sigma\ww_{211}^{2(1)}+\duno\kk\mat\ww_{211}^{2(1)}+\ddue\kk\vet\\\\
{\bf G}_2^{(31)}&= 2\sigma\ww_{211}^{1(2)}-L\ww_{211}^{1(1)}+\duno\kk\mat\ww_{211}^{1(2)}\\
&-\half\gamma \begin{pmatrix}\frac{2b}{(a+b)^2}+2(a+b)M & 2(a+b)\\ -\frac{2b}{(a+b)^2}-2(a+b)M & -2(a+b)\end{pmatrix}\Big(\ww_{211}^{1(1)}+\ww_{211}^{2(1)}\Big)\\\\
\end{align*}
\begin{align*}
{\bf G}_2^{(32)}&= 2\sigma\ww_{211}^{2(2)}-\frac{L}{4}\ww_{211}^{2(1)}+\duno\kk\mat\ww_{211}^{2(2)}\\
&-\quart \gamma \begin{pmatrix}\frac{2b}{(a+b)^2}+2(a+b)M & 2(a+b)\\ -\frac{2b}{(a+b)^2}-2(a+b)M & -2(a+b)\end{pmatrix}\ww_{211}^{1(1)}\\\\
{\bf G}_3^{(31)}&=
-\frac{L}{4}\ww_{211}^{1(2)}-\gamma \begin{pmatrix}\frac{2b}{(a+b)^2}+2(a+b)M & 2(a+b)\\ -\frac{2b}{(a+b)^2}-2(a+b)M & -2(a+b)\end{pmatrix}\left(\half\ww_{211}^{2(2)}+\ww_{200}^1+\half\ww_{202}^1+\quart\ww_{222}^1\right)\\
&-\frac{9}{16}\gamma\begin{bmatrix}M\\-M\end{bmatrix}\\\\
{\bf G}_3^{(32)}&=
-\gamma \begin{pmatrix}\frac{2b}{(a+b)^2}+2(a+b)M & 2(a+b)\\ -\frac{2b}{(a+b)^2}-2(a+b)M & -2(a+b)\end{pmatrix}\left(\ww_{200}^2+\half\ww_{220}^2\right)-\frac{3}{4}\gamma\begin{bmatrix}M\\-M\end{bmatrix}\\\\
{\bf G}_4^{(31)}&=
-L\ww_{211}^{1(2)}-\gamma \begin{pmatrix}\frac{2b}{(a+b)^2}+2(a+b)M & 2(a+b)\\ -\frac{2b}{(a+b)^2}-2(a+b)M & -2(a+b)\end{pmatrix}\left(\ww_{200}^2+\half\ww_{211}^{1(2)}+\half\ww_{231}\right)-\frac{3}{2}\gamma\begin{bmatrix}M\\-M\end{bmatrix}\\\\
{\bf G}_4^{(32)}&=
-2L\ww_{211}^{2(2)}-\gamma \begin{pmatrix}\frac{2b}{(a+b)^2}+2(a+b)M & 2(a+b)\\ -\frac{2b}{(a+b)^2}-2(a+b)M & -2(a+b)\end{pmatrix} \left(\ww_{200}^1+\quart\ww_{211}^{1(2)}+ \quart\ww_{231}\right)-\frac{3}{4}\gamma\begin{bmatrix}M\\-M\end{bmatrix},
\end{align*}
while ${\bf G}^{(3)*}$ does not contain secular terms.\\
By imposing the Fredholm alternative $<{\bf G}^{(3)},{\bm\psi}>=0$ and defining
\begin{equation*}
\tilde{\sigma}_l=-\frac{<{\bf G}_1^{(3l)},{\bm\psi}>}{<\ro,{\bm\psi}>}\qquad
\tilde{L}_l=\frac{<{\bf G}_2^{(3l)},{\bm\psi}>}{<\ro,{\bm\psi}>}\qquad
\tilde{R}_l=-\frac{<{\bf G}_3^{(3l)},{\bm\psi}>}{<\ro,{\bm\psi}>}\qquad
\tilde{S}_l=-\frac{<{\bf G}_4^{(3l)},{\bm\psi}>}{<\ro,{\bm\psi}>}
\end{equation*}
we find
\begin{equation}
\begin{split}
\frac{\partial A_1}{\partial T_2}&=\tilde{\sigma}_1A_1-\tilde{L}_1A_1A_2+\tilde{R}_1A_1^3+\tilde{S}_1A_1A_2^2\\
\frac{\partial A_2}{\partial T_2}&= \tilde{\sigma}_2 A_2-\tilde{L}_2A_1^2+\tilde{R}_2A_2^3+\tilde{S}_2A_1^2A_2.
\label{sistema_sec_h3_part}
\end{split}
\end{equation}
\end{itemize}
\end{small}
Adding up \eqref{sistema_sec_h2} to \eqref{sistema_sec_h3_part} one gets the system \eqref{4.32} for the amplitude $A_1$ and $A_2$, where:
\begin{equation*}
\bar{\sigma}_l=\sigma+\varepsilon\tilde{\sigma}_l,\qquad \bar{L}_1=L+\varepsilon\tilde{L}_1,\qquad\bar{L}_2=\frac{L}{4}+\varepsilon\tilde{L}_2,\qquad
\bar{R}_l=\varepsilon\tilde{R}_l,\qquad
\bar{S}_l=\varepsilon\tilde{S}_l.
\end{equation*}
\section*{Conclusions}\label{concl}
In the present paper we have investigated the Turing mechanism induced by linear cross-diffusion for a two variable \SS reaction-diffusion system. We have determined the parameter space distinguishing, each time, the supercritical region from the subcritical one. In particular we have found that the subcritical region increase as $d_v$ (the cross-diffusion term for the activator) increases, while it decrease as $d_u$ (the cross-diffusion term for the inhibitor) increases.\\
By performing a weakly nonlinear analysis, we have predicted the amplitude and the shape of the pattern, deriving the Stuart-Landau equation. The same analysis has been carried out for two-dimensional domains and in this case we have obtained different forms of patterns, such as rolls, squares, hexagons and other mixed-mode structures.\\
The analysis of the amplitude equations has allowed to investigate the occurrence of multiple branches of stable solutions, leading to hysteresis, that we have observed both in 1D and 2D cases. Of particular relevance is the transition from rolls to hexagons when the bifurcation parameter is varied.\\
Other aspects of the problem could be examined, as the Turing and Hopf bifurcation interactions and the arising oscillating pattern and the spatio-temporal chaos in the complex Ginzburg-Landau equation which describes the amplitude of the homogeneous oscillatory solution \cite{BP06, BB12}.

\section*{Acknowledgments}\label{ack}
The work of G.G. and S.L. was partially supported by
GNFM-INdAM through a Progetto Giovani grant.

\bibliographystyle{unsrt}
\bibliography{pattern2}

\end{document}